\documentclass[11pt]{article}
\pdfoutput=1
\usepackage{jheppub}
\usepackage[...]{youngtab}
\usepackage{xfrac,appendix,amsfonts,physics,tabularx,multirow,amsmath,mathalfa,color,colortbl,ccaption,mathtools,bbold,kantlipsum,slashbox,changepage,titlesec,graphicx}
\usepackage[usenames,dvipsnames]{xcolor}
\usepackage[framemethod=TikZ]{mdframed}
\usepackage{longtable,caption}

\usepackage{array}
\newcolumntype{L}[1]{>{\raggedright\let\newline\\\arraybackslash\hspace{0pt}}m{#1}}
\newcolumntype{C}[1]{>{\centering\let\newline\\\arraybackslash\hspace{0pt}}m{#1}}
\newcolumntype{R}[1]{>{\raggedleft\let\newline\\\arraybackslash\hspace{0pt}}m{#1}}

\definecolor{Grayy}{gray}{0.9}
\definecolor{MyBlue}{rgb}{0.0,0.0,0.9}
\definecolor{MyColor}{rgb}{0.5, 0.1, 0.6}
\definecolor{MyRed}{rgb}{0.0,0.9,0.0}
\colorlet{Bluee}{MyBlue!6}
\colorlet{Redd}{MyRed!1}
\allowdisplaybreaks

\mdfdefinestyle{sid}{%
    linecolor=black,
    outerlinewidth=1.0pt,
    roundcorner=7pt,
    innerrightmargin=15pt,
    innerleftmargin=15pt,
    backgroundcolor=gray!30
		}	

\definecolor{Red}{rgb}{1,0,0}
\makeatletter

\newcommand*\bigcdot{\mathpalette\bigcdot@{.5}}
\newcommand*\bigcdot@[2]{\mathbin{\vcenter{\hbox{\scalebox{#2}{$\m@th#1\bullet$}}}}}
\makeatother

\usepackage{tikz}
\usetikzlibrary{decorations.pathmorphing}
\usetikzlibrary{decorations.pathreplacing}
\usetikzlibrary{shapes, shapes.geometric, shapes.symbols, shapes.arrows, shapes.multipart, shapes.callouts, shapes.misc}
\tikzset{snake it/.style={decorate, decoration=snake}}
\tikzset{7brane/.style={circle, draw=black, fill=black,ultra thick,inner sep=1.5 pt, minimum size=1 pt,}, c/.default={4pt}}

\tikzset{big7brane/.style={circle, draw=black, fill=black,ultra thick,inner sep=2.5 pt, minimum size=1 pt,}, c/.default={4pt}}
\tikzset{u/.style={circle, draw=black, fill=white, thick,inner sep=2 pt, minimum size=2 pt,},f/.style={square, draw=black, fill=white,ultra thick,inner sep=4 pt, minimum size=2 pt,}}

\tikzset{so/.style={circle, draw=black, fill=red, thick,inner sep=2 pt, minimum size=2 pt,},f/.style={square, draw=black, fill=white,ultra thick,inner sep=4 pt, minimum size=2 pt,}}

\tikzset{sp/.style={circle, draw=black, fill=blue,thick,inner sep=2 pt, minimum size=2 pt,},f/.style={square, draw=black, fill=white,ultra thick,inner sep=4 pt, minimum size=2 pt,}}

\tikzset{uf/.style={rectangle, draw=black, fill=white, thick,inner sep=2.5 pt, minimum size=4 pt,}}

\tikzset{spf/.style={rectangle, draw=black, fill=blue, thick,inner sep=2.5 pt, minimum size=4 pt,}}

\tikzset{sof/.style={rectangle, draw=black, fill=red, thick,inner sep=2.5 pt, minimum size=4 pt,}}
\usetikzlibrary{positioning}
\usetikzlibrary{arrows}

\title{\boldmath Topological entanglement and hyperbolic volume}

\author[a]{Aditya Dwivedi}
\author[b]{, Siddharth Dwivedi}
\author[a]{, Bhabani Prasad Mandal}
\author[c]{, Pichai Ramadevi}
\author[d]{, Vivek Kumar Singh}

\affiliation[a]{Department of Physics, Institute of Science, Banaras Hindu University,\\ Varanasi, 221005, India}
\affiliation[b]{Center for Theoretical Physics, College of Physical Science and Technology, Sichuan University,\\
Chengdu, 610064, China}
\affiliation[c]{Department of Physics, Indian Institute of Technology Bombay, \\ Powai, Mumbai, 400076, India }
\affiliation[d]{Department of Mathematics, Indian Institute of Science Education and Research, Pune, \\ Pashan, Pune, 411008, India}

\emailAdd{aditya.dwivedi13@bhu.ac.in, sdwivedi@scu.edu.cn, bhabani@bhu.ac.in, ramadevi@phy.iitb.ac.in, vivek.singh@fuw.edu.pl}

\abstract{The entanglement entropy of many quantum systems is difficult to compute in general. They are obtained as a limiting case of the R\'enyi  entropy of index $m$, which captures the higher moments of the reduced density matrix. In this work, we study pure bipartite states associated with $S^3$ complements of a two-component link which is a connected sum of a knot $\mathcal{K}$ and the Hopf link. For this class of links, the Chern-Simons theory provides the necessary setting to visualise the $m$-moment of the reduced density matrix as a three-manifold invariant $Z(M_{\mathcal{K}_m})$, which is the partition function of $M_{\mathcal{K}_m}$. Here $M_{\mathcal{K}_m}$ is a closed 3-manifold associated with the knot $\mathcal K_m$, where $\mathcal K_m$ is a connected sum of $m$-copies of  $\mathcal{K}$ (i.e., $\mathcal{K}\#\mathcal{K}\ldots\#\mathcal{K}$) which mimics the well-known replica method. We analyse the partition functions $Z(M_{\mathcal{K}_m})$ for SU(2) and SO(3) gauge groups, in the limit of the large Chern-Simons coupling $k$. For SU(2) group, we show that $Z(M_{\mathcal{K}_m})$ can grow at most polynomially in $k$. On the contrary, we conjecture that $Z(M_{\mathcal{K}_m})$ for SO(3) group shows an exponential growth in $k$, where the leading term of $\ln Z(M_{\mathcal{K}_m})$ is the hyperbolic volume of the knot complement $S^3\backslash \mathcal{K}_m$. We further propose that the R\'enyi entropies associated with SO(3) group converge to a finite value in the large $k$ limit. We present some examples to validate our conjecture and proposal.}

\usepackage[normalem]{ulem}  

\begin{document} 
			\maketitle
		\flushbottom

\section{Introduction} \label{sec1}
The study of quantum entanglement and finding the possible patterns of entanglement that can emerge in a quantum field theory (QFT)  is generally an important question in quantum mechanics and quantum information theory. However, due to large degrees of freedom, it is a difficult exercise to analyse the entanglement structures in a generic QFT. Nevertheless, a class of QFTs known as   `topological quantum field theories' (TQFT's) provides a tractable system to analyse entanglement structures. 

The three-dimensional Chern-Simons theory is one such TQFT \cite{Witten:1988hf} which provides a natural framework to study invariants of knots, links, and three-manifolds. Interestingly, the Chern-Simons path integral on a three-manifold $M$ with boundary $\partial M= \Sigma$ is given by a state belonging to a finite-dimensional Hilbert space $\mathcal{H}_{\Sigma}$.  Such a finite-dimensional Hilbert space is a key behind explicitly computing the entanglement measures. 

\begin{figure}[htbp]
	\centering
		\includegraphics[width=0.70\textwidth]{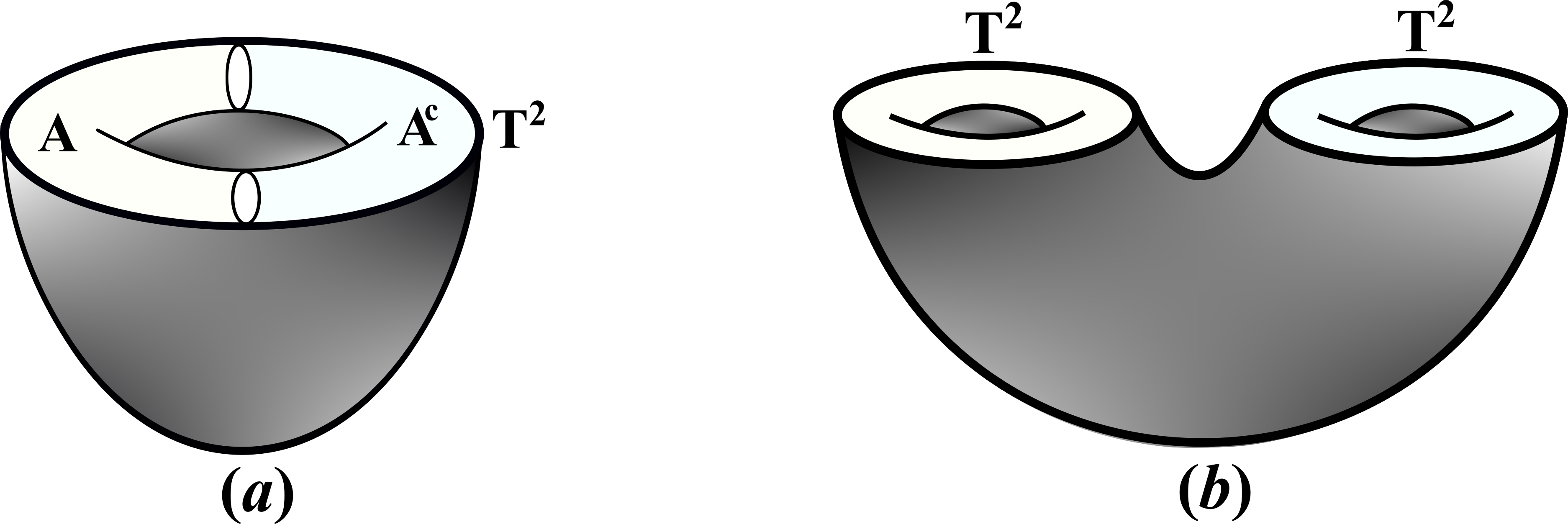}
	\caption{The two approaches to study the topological entanglement: the manifold in ($a$) has a single torus boundary, which is bi-partitioned into spatially connected regions $A$ and $A^c$. The manifold in ($b$) has two disjoint torus boundaries.}
	\label{T2partition}
\end{figure}

There are two different topological set-ups to obtain the entanglement entropy in Chern-Simons theory as shown in figure \ref{T2partition}. In figure \ref{T2partition}($a$), the manifold has a single torus boundary which is bi-partitioned into spatially connected regions $A$ and its complement $A^c$.  
When we trace out the region $A^c$, we obtain the reduced density matrix $\rho_A$ and the corresponding `topological entanglement entropy'. Such an entropy  is  independent of the length or the area of the region $A$ or $A^c$ and have been studied in \cite{Kitaev:2005dm,Levin:2006zz,Dong:2008ft}. Another approach to study the topological entanglement was given in \cite{Balasubramanian:2016sro}, where the three-manifold is actually a link complement\footnote{Given a link $\mathcal{L}$ embedded in $S^3$, the link complement $S^3 \backslash \mathcal{L}$ is a three-dimensional manifold which is obtained by removing a tubular neighborhood around $\mathcal{L}$ from $S^3$, i.e $S^3 \backslash \mathcal{L} \equiv S^3 - \text{interior}(\mathcal{L}_{\text{tub}})$.}
of $S^3$ with two or more disjoint torus boundaries as shown in figure \ref{T2partition}($b$). The topological entanglement structure can be obtained by tracing out one of the boundary components, which is often termed as `multi-boundary entanglement'. We refer the interested readers to \cite{Balasubramanian:2016sro,Dwivedi:2017rnj,Balasubramanian:2018por,Hung:2018rhg,Melnikov:2018zfn,Camilo:2019bbl,Dwivedi:2019bzh,Buican:2019evc,Zhou:2019ezk,Dwivedi:2020jyx,Dwivedi:2020rlo} for the recent developments in this study. For the three-manifold $M$ whose boundary consists of multiple disjoint  torus boundaries ($\partial M = \Sigma_1 \sqcup \Sigma_2 \sqcup \ldots \sqcup \Sigma_n$), the associated Hilbert space is
\begin{equation}
\mathcal{H}_{\partial M} = \mathcal{H}_{\Sigma_1} \otimes \mathcal{H}_{\Sigma_2} \otimes \ldots \otimes \mathcal{H}_{\Sigma_n} ~.
\end{equation}
When $M$ is a link complement ($M=S^3 \backslash \mathcal{L}$), the probability amplitudes of the associated state $\ket{\mathcal{L}}$ are the Chern-Simons partition functions $Z(S^3; \mathcal L)$  of $S^3$ in the presence of the link $\mathcal{L}$ (see \cite{Balasubramanian:2016sro}), which are  proportional to the link invariants \cite{Witten:1988hf}. Therefore, the entanglement measures for such a state can be written in terms of the link invariants of $\mathcal{L}$. 

In this work, we analyse the semiclassical (large $k$) asymptotics of the trace of the unnormalised reduced density matrices ($\text{Tr}[\sigma(\mathcal{L})]$) of two-party states associated with the link complement $S^3 \backslash \mathcal{L}$ for a class of two-component links $\mathcal L$ viewed as connected sum of prime knots $\mathcal{K}$ with the Hopf link $T_{2,2}$:
\begin{equation}
\mathcal{L} = \mathcal{K} \# T_{2,2}~.
\label{KConnectedHopfLink}
\end{equation}
Qualitatively, $\text{Tr}[\sigma(\mathcal{K} \# T_{2,2})]$ describes the Chern-Simons partition function of some closed three-manifold $M_{\mathcal{K}}$ (although the topology of $M_{\mathcal {K}}$ is not known to us) \cite{Melnikov:2018zfn}. For the class of links \eqref{KConnectedHopfLink}, this trace is given as \cite{Balasubramanian:2016sro}:
\begin{equation}
\text{Tr}[\sigma(\mathcal{K}\# T_{2,2})] = Z(M_{\mathcal{K}}) = \sum_R 
\abs{H_R(\mathcal{K};q=e^{\frac{2 \pi i}{k+y}})}^2 ~,
\label{traceKconnectT22}
\end{equation}
where $H_R(\mathcal K;q)$ are the reduced quantum invariants of knot $\mathcal K$ with $y$ being the dual Coxeter number of the gauge group (see the details in section \ref{sec3}).  Such a trace \eqref{traceKconnectT22} for SU(2) gauge group  almost resembles the Turaev-Viro invariant  $\text{TV}_r(S^3 \backslash \mathcal{K}\,;t=e^{\frac{\pi i}{r}})$ for the knot complement $S^3 \backslash \mathcal{K}$ with $t^2=q$ and $r=(k+2)$ \cite{zbMATH00167790, detcherry2018turaev}. In fact, this resemblance to $\text{TV}_r(S^3 \backslash \mathcal{K}\,;t )$ motivates us  to investigate the large $k$ asymptotics of \eqref{traceKconnectT22} for the class of links $\mathcal{K}\# T_{2,2}$ within the Chern-Simons theory.

For SU(2) gauge group, we observe  the trace \eqref{traceKconnectT22} follows the same polynomial growth in $k$  as that of Turaev-Viro invariant $\text{TV}_r(S^3 \backslash \mathcal{K}\,;t =e^{\frac{\pi i}{r}})$ with $r=(k+2)$. Incidentally, the large $k$ asymptotics of Turaev-Viro invariant \cite{detcherry2018turaev} evaluated at a different root of unity $t=e^{\frac{2 \pi i}{r'}}$, with $r'=(k+1)$ being an odd integer, captures the hyperbolic volume of the knot complement $S^3 \backslash \mathcal{K}$. Such a change of variable naturally happens when we study the knot invariants in SO(3) Chern-Simons theory (we refer to section \ref{sec3} for more details).

Hence our focus in this work is to investigate the trace \eqref{traceKconnectT22} for the links of type $\mathcal{K}\# T_{2,2}$ within the context of SO(3) Chern-Simons theory. Interestingly, our numerical analysis, combined with the Kashaev's conjecture \cite{kashaev1997hyperbolic}, is instrumental in extracting the geometrical features of $S^3 \backslash \mathcal{K}$ from \eqref{traceKconnectT22}. Specifically, we conjecture an exponential growth of $\text{Tr}[\sigma(\mathcal{K}\# T_{2,2})]=Z(M_{\mathcal{K}})$ with $k$, where the growth rate is determined by the hyperbolic volume of $S^3 \backslash \mathcal{K}$.

We further propose that the R\'enyi entropies for SO(3) gauge group associated with the state $\ket{\mathcal{K} \# T_{2,2}}$ converge to a finite value as $k \to \infty$, although currently we do not have a geometric or topological interpretation of this limit.

The paper is organized as follows. In section \ref{sec2}, we discuss the preliminaries, including a brief discussion of our set-up. In section \ref{sec3}, we begin by analysing the reduced density matrices and their traces associated with the links of type $\mathcal{K}\# T_{2,2}$. We further analyse their large $k$ asymptotics for two gauge groups: SU(2) and SO(3), respectively. In section \ref{sec4}, we present several examples where we do explicit numerical computations to verify the results conjectured and proposed in section \ref{sec3} for the SO(3) group. We finally conclude in section \ref{sec5}.
\section{Preliminaries} \label{sec2}
\subsection{Chern-Simons theory and multi-boundary states}
Three-dimensional Chern-Simons theory based on a compact gauge group $G$ and coupling $k \in \mathbb{Z}$ is described by the following metric independent action:
\begin{equation}
S(A) = \frac{k}{4\pi} \int_M \text{Tr}\left(A \wedge dA + \frac{2}{3} A \wedge A \wedge A \right) ~.
\label{CSaction}
\end{equation}
Here $A = A_{\mu}dx^{\mu}$ is a matrix-valued gauge field, and $M$ denotes a three-manifold. The gauge-invariant operators are the Wilson loops. Given a knot $\mathcal{K}$ embedded in $M$, the Wilson loop operator is defined by taking the trace of the holonomy of $A$ around $\mathcal{K}$ :
\begin{equation}
W_{R}(\mathcal{K}) = \text{Tr}_R \,P\exp\left(i\oint_{\mathcal{K}} A \right) ~.
\end{equation} 
Note that the knot $\mathcal K$ carries a representation $R$ of the gauge group $G$, and the trace is over the matrix corresponding to the representation $R$.  Chern-Simons partition function, which encodes the topological information of the three-manifold $M$, is given by 
\begin{equation}
Z(M) = \int e^{i S(A)} \mathcal DA ~,
\end{equation}
where $\mathcal D A$ denotes integration over all the gauge-invariant classes of connections. In fact, $Z(M)$ is referred to as a three-manifold invariant (up to an overall normalisation) in the literature.

\begin{figure}[htbp]
	\centering
		\includegraphics[width=1.00\textwidth]{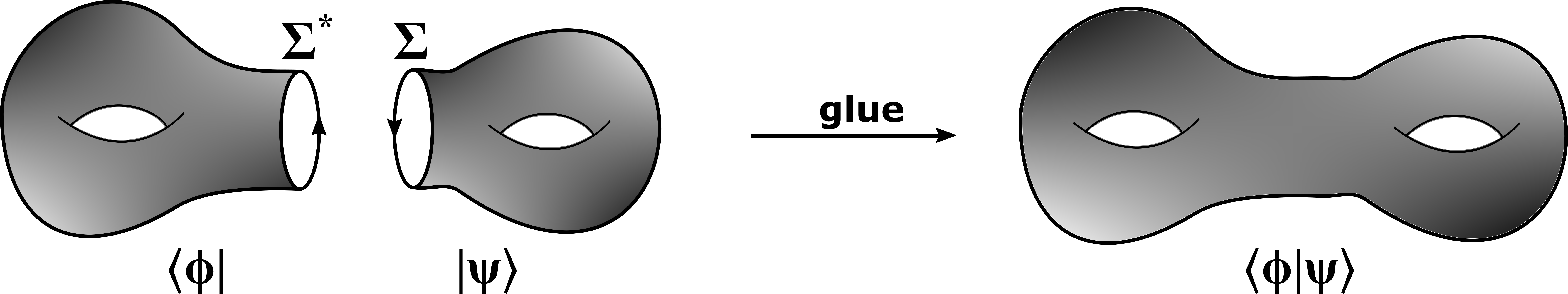}
	\caption{Two manifolds on left with same boundary but opposite orientation. The path integral on these manifolds gives states $\bra{\phi} \in \mathcal{H}_{\Sigma^*}$ and $\ket{\psi} \in \mathcal{H}_{\Sigma}$. The inner product $\bra{\phi}\ket{\psi}$ will be the partition function of a closed three-manifold (shown in right) obtained by gluing.}
	\label{gluing}
\end{figure}

The partition function of $M$ in the presence of knots and links is obtained by inserting the appropriate Wilson loop operators in the integral. For example, consider a  link $\mathcal{L}$ made of component knots  $\mathcal{K}_1$,  $\mathcal{K}_2$ given by Wilson link operator $W_{R1,R_2}(\mathcal L)=W_{R_1}(\mathcal K_1) W_{R_2} (\mathcal K_2)$.  Note that the representations $R_1,R_2$ are placed on the component knots $\mathcal K_1, \mathcal K_2$ respectively. The partition function in the presence of such a link is given by 
 \begin{equation}
Z(M; \mathcal L[R_1,R_2] ) = \int e^{i S(A)}\, W_{R_1}(\mathcal{K}_1) W_{R_2}(\mathcal{K}_2) \, \mathcal D A ~. \label{bdy3manifold}
\end{equation}  
Suppose the manifold is closed with no boundary, the link invariants $V_{R_1,R_2}(\mathcal L)$ 
with the two components carrying representation $R_1, R_2$ are given by the expectation value of the Wilson link operator:
\begin{equation}
V_{R_1,R_2}(\mathcal L) =\frac{1}{Z(M)}  \int e^{i S(A)}\, W_{R_1}(\mathcal{K}_1) W_{R_2}(\mathcal{K}_2) \, \mathcal D A ~.
\end{equation}
When $M$ has a boundary $\Sigma$, the path integral of the theory on $M$ with the Wilson link 
$\mathcal L$  insertion is interpreted as a  state $\ket {\Psi} \equiv \ket {\mathcal L} \in \mathcal H_{\Sigma}$.
Further note that if we reverse the orientation of the boundary, the associated Hilbert space becomes the dual of the original Hilbert space: 
\begin{equation}
\mathcal{H}_{\Sigma^*} = \mathcal{H}^*_{\Sigma} ~.
\end{equation} 
As a result, there exists a natural pairing, the inner product $\langle \Phi| \Psi \rangle$ for any two states $\ket{\Psi} \in \mathcal{H}_{\Sigma}$ and $\bra{\Phi} \in \mathcal{H}_{\Sigma^*}$. This technique can be used to compute the partition functions $Z(M)$ of complicated closed manifolds by gluing two disconnected pieces along their oppositely oriented boundary as shown in the figure \ref{gluing}.
When the boundary of the manifold $M$ consists of disjoint components, i.e., $\Sigma = \Sigma_1 \sqcup \Sigma_2$, the Hilbert space associated with $\Sigma$ is the tensor product of Hilbert spaces associated with each component, i.e.
\begin{equation}
\mathcal{H}_{\Sigma} = \mathcal{H}_{\Sigma_1} \otimes \mathcal{H}_{\Sigma_2} ~.
\end{equation} 
For the states $\ket{\Psi}\in \mathcal{H}_{\Sigma}$, we can study  the entanglement structure by  tracing out the Hilbert space  $\mathcal H_{\Sigma_1} ~{\rm or} ~ \mathcal H_{\Sigma_2}$. In this work, we will consider the link complement manifold 
\begin{equation}
M=S^3 \backslash \mathcal{L} ~,
\end{equation} 
whose boundary is $\Sigma= T^2 \sqcup T^2$  for  any  two-component link $\mathcal{L}$. We denote the state associated with $S^3 \backslash \mathcal{L}$ as $\ket{\mathcal{L}} \in \mathcal H_{T^2} \times \mathcal H_{T^2}$. We will now review the essential steps \cite{Balasubramanian:2016sro}  of computing such states. 
 As discussed in  Ref.\cite{Witten:1988hf}, the basis states of the Hilbert space  $\mathcal{H}_{T^2}$  
 are in one-to-one correspondence with the integrable representations of the affine Lie algebra 
 $\hat{\mathfrak{g}}_k$  at level $k$.  Here $\mathfrak{g}$ represents the Lie algebra associated with the group $G$. The Chern-Simons path integral on a  solid torus with a Wilson loop  along the non-contractible cycle carrying  integrable representation $\alpha$  is given by  the basis state $\ket{e_{\alpha}}$:
\begin{equation}
\begin{array}{c}
\includegraphics[width=0.25\linewidth]{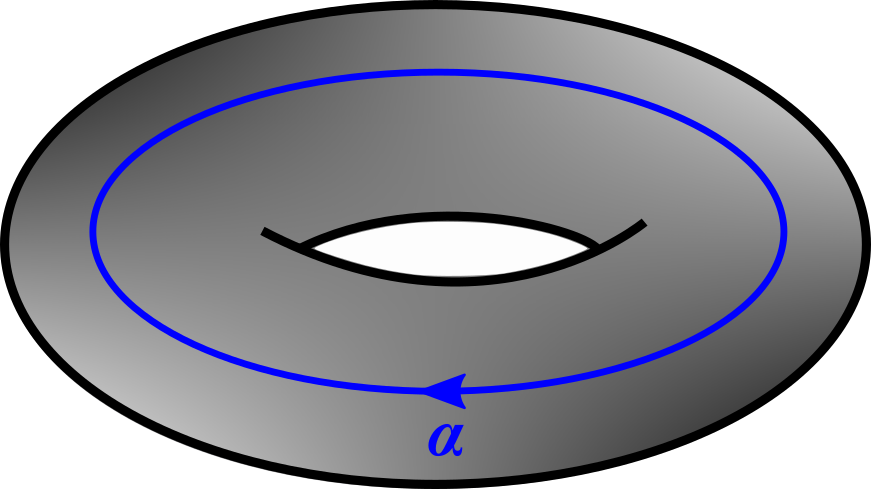}
\end{array} = \ket{e_{\alpha}} ~.
\end{equation}
Using  such basis states, we can expand the bipartite state $\ket{\mathcal{L}}$ as:
\begin{equation}
\ket{\mathcal{L}} = \sum_{\alpha} \sum_{\beta} C_{\alpha \beta} \ket{e_{\alpha}} \otimes \ket{e_{\beta}} = \sum_{\alpha} \sum_{\beta} C_{\alpha \beta} \ket{e_{\alpha},e_{\beta}} ~,
\end{equation}
where $C_{\alpha \beta}$ are the  complex coefficients given by partition function \eqref{bdy3manifold} \cite{Balasubramanian:2016sro}:
\begin{equation}
C_{\alpha \beta} = \bra{e_{\alpha},e_{\beta}}\ket{\mathcal{L}} = Z(S^3; \mathcal{L}[\alpha,\beta]) ~.
\end{equation}
Hence the two-component link state can be written as
\begin{equation}
\ket{\mathcal{L}} = \sum_{\alpha} \sum_{\beta} Z(S^3; \mathcal{L}[\alpha,\beta]) \ket{e_{\alpha},e_{\beta}} ~.
\label{linkstate}
\end{equation}
To study its entanglement properties, we require a reduced density matrix by partial tracing one of the two Hilbert spaces. We will now present these entanglement features in the following subsection.
\subsection{Reduced density matrix and R\'enyi entropy}
Given the link state $\ket{\mathcal{L}}\in  \mathcal{H}_1 \otimes \mathcal{H}_2$ in \eqref{linkstate}, the corresponding density matrix operator  is 
\begin{equation}
\hat{\rho}(\mathcal{L}) = \ket{\mathcal{L}} \bra{\mathcal{L}} ~.
\end{equation}
The reduced density matrix is computed by tracing out one of the Hilbert space, say $\mathcal{H}_{2}$:
\begin{equation}
\rho(\mathcal{L}) = \frac{\text{Tr}_{\mathcal{H}_2}[\hat{\rho}(\mathcal{L})]}{\braket{\mathcal{L}}} = \frac{1}{\braket{\mathcal{L}}}\sum_{\alpha} \expval{\hat{\rho}(\mathcal{L})}{e_\alpha} ~,
\end{equation}
where the factor $\braket{\mathcal{L}}$ in the denominator ensures that the trace of $\rho(\mathcal{L})$ is unity.  For convenience, we will work with  the unnormalised version 
of the reduced density matrix as follows:
\begin{equation}
\sigma(\mathcal{L}) = \text{Tr}_{\mathcal{H}_2}[\hat{\rho}(\mathcal{L})] ~.
\label{reduceden}
\end{equation} 
The entanglement measures associated with $\ket{\mathcal{L}}$ can be computed from the matrix $\sigma(\mathcal{L})$. For example, the R\'enyi entropy of index $m$ can be written in terms of $m$-moments or the trace of the $m^{\text{th}}$ power of $\sigma(\mathcal{L})$ as following:
\begin{equation}
\mathcal{R}_m = \frac{1}{1-m} \ln\left(\frac{\text{Tr}[\sigma^m(\mathcal{L})]}{(\text{Tr}[\sigma(\mathcal{L})])^m} \right) ~.
\label{Renyi-conclusion}
\end{equation}
The entanglement entropy is given as,
\begin{equation}
\mathcal{E} = \lim_{m \to 1} \mathcal{R}_m ~.
\label{EEm1limit}
\end{equation}
There also exists a minimum entropy which is controlled by the maximum eigenvalue of the reduced density matrix. Further, in the large $m$ limit,  the corresponding R\'enyi entropy is minimum:
\begin{equation}
\mathcal{R}_{\text{min}} \equiv \mathcal{R}_{\infty} = \lim_{m \to \infty} \mathcal{R}_m ~.
\label{REminfylimit}
\end{equation}
In the next section, we will study some of the properties of the density matrices and the entanglement measures associated with a class of two-component links $\mathcal L$ which is a connected sum of a prime knot  $\mathcal{K}$ with  Hopf link.
\section{Bi-partite state: Connected sum of a knot $\mathcal{K}$ and Hopf link} \label{sec3}
\begin{figure}[htbp]
	\centering
		\includegraphics[width=0.43\textwidth]{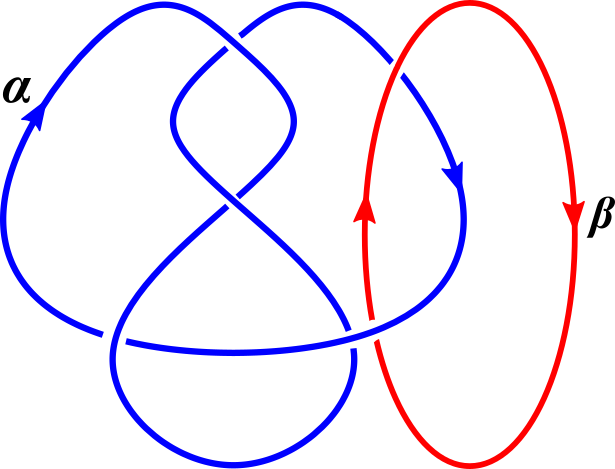}
	\caption{The two component link $4_1 \# T_{2,2}$ which is the connected sum of figure-eight knot and Hopf link.}
	\label{Fig8andHopf}
\end{figure}

In this work, we consider the two-component link $\mathcal{L} = \mathcal{K} \# T_{2,2}$ which is a connected sum of a prime knot $\mathcal{K}$ and the Hopf link $T_{2,2}$. A typical example of such a link is shown in the figure \ref{Fig8andHopf} when $\mathcal{K}$ is the figure-eight knot. 
In this section, we will closely analyse the bi-partite state $\ket{\mathcal{K} \# T_{2,2}}$ and its properties. In particular, our focus will be on the traces of the unnormalised reduced density matrices. As mentioned in the introduction, we will see that these traces have a close resemblance to the Turaev-Viro invariant \cite{zbMATH00167790, detcherry2018turaev}, motivating us to focus on the Chern-Simons theory based on SU(2) as well as SO(3) gauge groups. 

We will first start with a generic discussion on the density matrices associated with $\ket{\mathcal{K} \# T_{2,2}}$ in the following subsection.
\subsection{Reduced density matrices}
Following the prescription given in the previous section, we can write the state associated with the link $\mathcal{K} \# T_{2,2}$ as
\begin{equation}
\ket{\mathcal{K} \#  T_{2,2}} = \sum_{\alpha,\, \beta} Z(S^3, \mathcal{K}_{\alpha}\, \# \, T_{2,2}[\alpha,\beta]) \ket{e_{\alpha},e_{\beta}} ~,
\end{equation}
where the probability amplitudes are the Chern-Simons partition functions of $S^3$ in presence of the link $\mathcal{K}\, \# \, T_{2,2}$. The labels $\alpha$ and $\beta$ are the representations carried by $\mathcal{K}$ and unknot components of the link respectively (as shown in the example in figure \ref{Fig8andHopf}). This partition function can be computed following Ref.\cite{Witten:1988hf} and is given as,
\begin{equation}
Z(S^3, \mathcal{K}_\alpha\, \# \, T_{2,2}[\alpha,\beta]) =  \frac{Z(S^3, \mathcal{K}_\alpha) \, Z(S^3, T_{2,2}[\alpha,\beta])}{Z(S^3, U_\alpha)} = Z(S^3, \mathcal{K}_\alpha) \frac{\mathcal{S}_{\alpha \beta}}{\mathcal{S}_{0\alpha}} ~,
\end{equation} 
where $U_\alpha$ is the unknot with representation $\alpha$. In the last equality, the partition functions of Hopf link and unknot are given in terms of the elements of the modular transformation matrix\footnote{The operator $\mathcal{S}$ is one of the generators of the modular group SL(2,\,$\mathbb{Z}$). In the above discussion, the operator $\mathcal{S}$ is written in a matrix form with element $\mathcal{S}_{\alpha \beta}$, where $\alpha$ and $\beta$ label the integrable representations of affine Lie algebra $\mathfrak{g}_k$ at level $k$. For $\mathfrak{su}(2)_k$ and $\mathfrak{so}(3)_k$ algebras, the matrix $\mathcal{S}$ is symmetric, unitary and satisfies $\mathcal{S}^{*}=\mathcal{S}$.} $\mathcal{S}$. 
Thus we can write the state as:
\begin{equation}
\ket{\mathcal{K}\# T_{2,2}} = \sum_{\alpha,\, \beta} Z(S^3, \mathcal{K}_\alpha) \frac{\mathcal{S}_{\alpha \beta}}{\mathcal{S}_{0\alpha}} \, \ket{e_\alpha,e_\beta} ~.
\label {stat}
\end{equation}
As the entanglement properties of a state do not change under a local unitary change of the basis, we simplify the above state by performing the following transformation:
\begin{equation}
\ket{e_\beta} = \sum_{x} \mathcal{S}_{\beta x}^{*} \, \ket{e_x} ~.
\label{unitarychange}
\end{equation} 
Substituting such a transformation in \eqref{stat} and using the unitary property of the $\mathcal{S}$ matrix, we get
\begin{equation}
\ket{\mathcal{K}\# T_{2,2}} = \sum_{\alpha} \frac{Z(S^3, \mathcal{K}_\alpha)}{\mathcal{S}_{0\alpha}} \, \ket{e_\alpha,e_\alpha} ~.
\label{stateKconnectT22}
\end{equation}
Interestingly, the probability amplitudes appearing in the above state are precisely the reduced quantum invariants\footnote{The reduced invariants are normalised such that the unknot invariant is unity, i.e. $H_{\alpha}(U;q)=1$.} of the knot $\mathcal{K}$ evaluated at a specific root of unity \cite{Witten:1988hf}:
\begin{equation} 
\frac{Z(S^3, \mathcal{K}_{\alpha})}{\mathcal{S}_{0 \alpha}} = \frac{Z(S^3, \mathcal{K}_{\alpha})/Z(S^3)}{\mathcal{S}_{0 \alpha}/Z(S^3)} =  \frac{V_{\alpha}(\mathcal{K})}{V_{\alpha}(U)} = H_{\alpha}(\mathcal K;q=e^{\frac{2\pi i}{k+y}}) ~,
\end{equation}
where $H_{\alpha}(\mathcal{K};q)$ are the reduced invariants of knot $\mathcal{K}$ evaluated at $q=e^{\frac{2\pi i}{k+y}}$ with $y$ being the dual Coxeter number of the gauge group. Thus we arrive at the following state:
\begin{equation}
\ket{\mathcal{K}\# T_{2,2}} = \sum_{\alpha} H_{\alpha}(\mathcal K;q=e^{\frac{2\pi i}{k+y}}) \, \ket{e_\alpha,e_\alpha} ~.
\label{state-final}
\end{equation}
The unnormalised reduced density matrix \eqref{reduceden} for this state is a diagonal matrix:
\begin{equation}
\sigma(\mathcal{K}\# T_{2,2}) = \text{diag}\left\{\abs{H_{\alpha}(\mathcal K;q=e^{\frac{2\pi i}{k+y}})}^2 \quad;\quad \alpha \in \text{Integrable representations}\right\} ~,
\label{rhoGeneral}
\end{equation} 
where the entries are labeled by various integrable representations $\alpha$.
The $m$-moment or the trace of the $m^{\text{th}}$ power of this matrix is:
\begin{equation}
\text{Tr}[\sigma^m(\mathcal{K}\# T_{2,2})] = \sum_{\alpha} \abs{H_{\alpha}(\mathcal K;q=e^{\frac{2\pi i}{k+y}})}^{2m} ~.
\label{traceSO3K}
\end{equation}

We also know that the trace of the reduced density matrix $\sigma(\mathcal{K}\# T_{2,2})$ can be viewed as gluing oppositely oriented torus boundaries resulting in a closed
three-manifold without any boundary. Some examples have been qualitatively discussed in Ref.\cite{Melnikov:2018zfn}. For clarity, let us look at the simplest link $\mathcal{L} = U \# T_{2,2}$ involving the unknot $U$ whose Jones polynomial is 1. Clearly,
\begin{equation}
\text{Tr}[\sigma(U\# T_{2,2})] = \sum_{\alpha}(1)  = {\rm dim} ~\mathcal H_{T^2} = Z(T^2 \times S^1)~, \label{unknotinv}
\end{equation}
giving the Chern-Simons partition function of $M_U \equiv T^2 \times S^1$. Replacing $U$ by a non-trivial prime knot $\mathcal{K}$ 
 would eventually give us:
\begin{equation}
\text{Tr}[\sigma(\mathcal{K}\# T_{2,2})] = \sum_{\alpha} \abs{H_{\alpha}(\mathcal K;q=e^{\frac{2\pi i}{k+y}})}^{2} = Z(M_{\mathcal{K}}) \label{MK-manifold}
\end{equation}
for some closed three-manifold $M_{\mathcal{K}}$, though we cannot specify the explicit topology of $M_{\mathcal{K}}$. Further, the $m$-moment of the unnormalised reduced density matrix $\sigma(\mathcal{K}\# T_{2,2})$ can be viewed as joining $m$-copies of the oppositely oriented $T^2$ boundaries. This is called the replica method \cite{Calabrese:2004eu} and it again results in a closed three-manifold. In the following subsection, we will present the 
knot theoretic interpretation  of the replica method and their relations to $m$-moments of $\sigma(\mathcal{K}\# T_{2,2})$ defining three-manifold invariants. We will show that the R\'enyi entropies of index $m$ can be written in terms of these three-manifold invariants. 
\subsection{R\'enyi entropies and three-manifold invariants} 
The R\'enyi entropy of index $m$ associated with the reduced density matrix of state $\ket{\mathcal{K}\# T_{2,2}}$ can be computed as:
\begin{equation}
\mathcal{R}_m = \frac{1}{1-m} \ln \left[ \frac{\text{Tr}[\sigma^m(\mathcal{K}\# T_{2,2})]}{(\text{Tr}[\sigma(\mathcal{K}\# T_{2,2})])^m}\right] = \frac{1}{1-m} \ln \left[ \frac{\text{Tr}[\sigma^m(\mathcal{K}\# T_{2,2})]}{Z(M_{\mathcal{K}})^m}\right] ~.
\end{equation}
We aim to write the numerator of the logarithmic term also as a three-manifold invariant. Just like our discussion that $\text{Tr}[\sigma(\mathcal{K}\# T_{2,2})] = Z(M_{\mathcal{K}})$, a similar topological attribution is also true for the higher moments of the reduced density matrix. We will now elaborate the knot theoretic picture of the replica trick to obtain $\text{Tr}[\sigma^m(\mathcal{K}\# T_{2,2})]$. 

Replica method involving $m$-copies is achievable by replacing $\mathcal{K}$ by the following connected sum of knots:
\begin{equation}
\mathcal{K}_m \equiv \underbrace{\mathcal{K}\, \#\, \mathcal{K}\, \#\, \ldots \#\, \mathcal{K}}_m ~.
\end{equation}
Incidentally, the reduced knot invariants for connected sum of two knots $\mathcal K_1 \# \mathcal K_2$  is equal to  the product of the reduced knot invariants for knots $\mathcal K_1$ and $\mathcal K_2$. Using this fact and \eqref{traceSO3K}, we will have:
\begin{equation}
\text{Tr}[\sigma(\mathcal{K}_m \# T_{2,2})] = \sum_{\alpha} \abs{H_{\alpha}(\mathcal{K}_m;q)}^{2} = \sum_{\alpha} \abs{H_{\alpha}(\mathcal K;q)}^{2m} = \text{Tr}[\sigma^m(\mathcal{K} \# T_{2,2})] ~,
\label{traceconnectsum}
\end{equation}
with $q=e^{\frac{2\pi i}{k+y}}$.
Thus using \eqref{MK-manifold} in \eqref{traceconnectsum}, we realize:
\begin{equation}
\text{Tr}[\sigma^m(\mathcal{K}\# T_{2,2})] = Z(M_{\mathcal{K}_m}) 
\label{ZMKm}
\end{equation}
and hence we can write the R\'enyi entropies in terms of three-manifold invariants:
\begin{equation}
\boxed{\mathcal{R}_m = \frac{1}{1-m} \ln \left[ \frac{Z(M_{\mathcal{K}_m})}{Z(M_{\mathcal{K}})^m}\right]} ~.
\label{RenyiActualValues}
\end{equation}

So far, our discussion has been for a generic gauge group. In the following subsections, we consider the cases when the gauge group is SU(2) and SO(3, respectively. In particular, we focus on the large $k$ asymptotics of the traces $\text{Tr}[\sigma^m(\mathcal{K}\# T_{2,2})] = Z(M_{\mathcal{K}_m})$, which almost resembles the Turaev-Viro invariants evaluated at certain roots of unity \cite{detcherry2018turaev}. We start with the SU(2) case in the following subsection. 
\subsection{SU(2) gauge group \& $m$-moments of density matrix}
The integrable representations for the affine algebra $\mathfrak{su}(2)_k$ are given as:
\begin{equation}
 \left\{\,\underbrace{\tiny\yng(6)}_{\alpha} \,\,:\,\, 0 \leq \alpha \leq k \right\}~.
\end{equation} 
Hence the basis states of $\mathcal{H}_{T^2}$ can be accordingly labeled:
\begin{equation}
\text{basis}(\mathcal{H}_{T^2}) = \left\{\ket{e_{\alpha}} \,\,:\,\, \alpha \equiv  \underbrace{\tiny\yng(6)}_{\alpha} \right\} ~.
\end{equation}
Using \eqref{state-final} and noting that the reduced invariants $H_{\alpha}$ are the colored Jones polynomials, the state is written as:
\begin{equation}
\ket{\mathcal{K} \# T_{2,2}} = \sum_{\alpha=0}^k J_{\alpha}(\mathcal{K}\,;e^{\frac{2\pi i}{k+2}}) \, \ket{e_\alpha,e_\alpha}    
\end{equation}
and the $m$-moment or the trace of the $m^{\text{th}}$ power of the unnormalised reduced density matrix associated with $\ket{\mathcal{K} \# T_{2,2}}$ will be:
\begin{equation}
\text{Tr}[\sigma^m(\mathcal{K} \# T_{2,2})] = \sum_{\alpha=0}^k \abs{J_{\alpha}(\mathcal{K}\,;e^{\frac{2\pi i}{k+2}})}^{2m} ~.  
\label{traceSU2appB}
\end{equation}
We will now show that the large $k$ asymptotics of \eqref{traceSU2appB} is the same as that of Turaev-Viro invariant \cite{zbMATH00167790} of $S^3 \backslash \mathcal{K}_m$ at a specific root of unity.
\subsubsection{Large $k$ asymptotics: The polynomial growth}
Following \cite{detcherry2018turaev}, the Turaev-Viro invariant of $S^3 \backslash \mathcal{K}_m$ evaluated at $e^{\frac{\pi i}{k+2}}$ with integer $k \geq 1$ can be written in terms of values of its colored Jones polynomials:\footnote{The colored Jones $J_{\alpha}(\mathcal K;q)$ here are the reduced invariants which are normalised such that $J_{\alpha}(U;q)$ is unity as mentioned in footnote 3. The invariants appearing in \cite{detcherry2018turaev} follow a different normalisation where $J_{\alpha}(U;q)$ is equal to the $q$-number $[\alpha+1]$. The Turaev-Viro invariant written in \eqref{TVappB} takes care of this.}
\begin{align}
\text{TV}_{k+2}(S^3 \backslash \mathcal{K}_m\,; e^{\frac{\pi i}{k+2}}) &= \frac{2}{k+2}\, \sum_{\alpha=0}^k \sin^2 \left(\frac{\pi  \alpha+\pi}{k+2}\right) \abs{J_{\alpha}(\mathcal{K}_m\,;e^{\frac{2\pi i}{k+2}})}^2 \nonumber \\
&= \frac{2}{k+2}\, \sum_{\alpha=0}^k \sin^2 \left(\frac{\pi  \alpha+\pi}{k+2}\right) \abs{J_{\alpha}(\mathcal{K}\,;e^{\frac{2\pi i}{k+2}})}^{2m} ~.
\label{TVappB}
\end{align} 
We also have the following inequality which is valid for $0 \leq \alpha \leq k$ :
\begin{equation}
\frac{1}{(k+2)^2} \leq \sin^2 \left(\frac{\pi  \alpha+\pi}{k+2}\right) \leq 1 ~.
\label{ineq1}
\end{equation}
Using \eqref{traceSU2appB} and \eqref{TVappB}, and applying \eqref{ineq1}, we obtain the following inequalities:
\begin{equation}
\frac{2}{(k+2)^3}\, \text{Tr}[\sigma^m(\mathcal{K} \# T_{2,2})] \leq \text{TV}_{k+2}(S^3 \backslash \mathcal{K}_m\,; e^{\frac{\pi i}{k+2}}) \leq \frac{2}{k+2}\, \text{Tr}[\sigma^m(\mathcal{K} \# T_{2,2})] ~.
\label{ineq2}
\end{equation}
Further, as mentioned in \cite{detcherry2018turaev}, the growth of $\text{TV}_{r}(M\,; e^{\frac{\pi i}{r}})$ is expected to be a polynomial in $r$ for 3-manifolds $M$. Thus the above inequality tells that $\text{Tr}[\sigma^m(\mathcal{K}\# T_{2,2})]$  can grow at most polynomially in $k$. In particular, taking log, dividing by $k$ and letting $k \to \infty$ in the expression \eqref{ineq2}, we get:
\begin{equation}
\lim_{k \to \infty} \frac{\ln\, \text{Tr}[\sigma^m(\mathcal{K} \# T_{2,2})]}{k} = \lim_{k \to \infty} \frac{\ln \text{TV}_{k+2}(S^3 \backslash \mathcal{K}_m\,; e^{\frac{\pi i}{k+2}})}{k} = 0 ~.
\label{SU2largekTV}
\end{equation}

Since the trace $\text{Tr}[\sigma^m(\mathcal{K} \# T_{2,2})]$ in \eqref{traceSU2appB} can be viewed as the SU(2) partition function $Z(M_{\mathcal{K}_m})$, we have the following limit:
\begin{equation}
\lim_{k \to \infty} \frac{\ln\, Z(M_{\mathcal{K}_m})}{k} = \lim_{k \to \infty} \frac{\ln \text{TV}_{k+2}(S^3 \backslash \mathcal{K}_m\,; e^{\frac{\pi i}{k+2}})}{k} = 0 ~.
\end{equation}

In the following section, we investigate the large $k$ asymptotics for SO(3) gauge group. 
\subsection{SO(3) gauge group \& $m$-moments of density matrix}
Here, we will confine to Chern-Simons theory based on SO(3) gauge group where the level $k$ is a positive even integer. Since the group SO(3) has rank one, the integrable representations are given by the Dynkin labels $\alpha$ of the highest weights in the following range: 
\begin{align}
0 \leq \alpha \leq k \quad;\quad \alpha \in \mathbb{Z},\,\, k \in 2\mathbb{Z} ~.
\end{align}
Using the group theoretical fact that the representation of SO(3) with Dynkin label $\alpha$ is isomorphic to the representation of SU(2) with Dynkin label $2\alpha$, we can write the list of integrable representations of $\mathfrak{so}(3)_k$ in terms of following Young diagrams:
\begin{equation}
\left\{\,\underbrace{\tiny\yng(6)}_{2\alpha} \,\,:\,\, 0 \leq \alpha \leq k \right\}~.
\end{equation} 
Hence the basis states of $\mathcal{H}_{T^2}$ can be accordingly labeled:
\begin{equation}
\text{basis}(\mathcal{H}_{T^2}) = \left\{\ket{e_{\alpha}} \,\,:\,\, \alpha \equiv  \underbrace{\tiny\yng(6)}_{2\alpha} \right\} ~.
\end{equation}
Thus, the state \eqref{state-final} associated with the link $\mathcal{K} \# T_{2,2}$ can be written as
\begin{equation}
\ket{\mathcal{K} \# T_{2,2}} = \sum_{\alpha=0}^k W_{\alpha}(\mathcal{K}\,;e^{\frac{2\pi i}{k+1}}) \, \ket{e_\alpha,e_\alpha} ~, 
\label{stateSO3}
\end{equation}
where $W_{\alpha}(\mathcal{K}\,;q)$ are polynomials in $q$ and are the colored knot invariants for SO(3) group.\footnote{The reduced Chern-Simons invariants of knots for SO($N$) gauge group are called the colored Kauffman polynomials: $\text{Kauff}_{\alpha}(\mathcal{K};a,q)$, where the two variables are $a=q^{N-1}$ and $q=e^{\frac{2\pi i}{k+N-2}}$. For the group SO(3), we define $W_{\alpha}(\mathcal{K}\,;q) \equiv \text{Kauff}_{\alpha}(\mathcal{K};q^2,q)$ with $q=e^{\frac{2\pi i}{k+1}}$ which enter as probability amplitudes in \eqref{stateSO3}.} Using the isomorphism between the SO(3) and SU(2) homologies, we can convert these SO(3) invariants into the colored Jones invariants. The two invariants are related as (we refer to \cite{Nawata:2013mzx} for some explicit examples):
\begin{equation}
W_{\alpha}(\mathcal{K}\,;q) = J_{2\alpha}(\mathcal{K}\,;q^2) ~,
\end{equation}
where $J_{2\alpha}$ is the colored Jones invariant evaluated for the spin $\alpha$ representation of SU(2) group. Thus we can rewrite the state in terms of the colored Jones invariants as:
\begin{equation}
\ket{\mathcal{K}\# T_{2,2}} = \sum_{\alpha=0}^k J_{2\alpha}(\mathcal{K}\,;e^{\frac{4\pi i}{k+1}}) \, \ket{e_\alpha,e_\alpha} ~.
\label{state-SO3-final}
\end{equation}
The $m$-moment or the trace of the $m^{\text{th}}$ power of the unnormalised reduced density matrix associated with $\ket{\mathcal{K} \# T_{2,2}}$ will be:
\begin{equation}
\text{Tr}[\sigma^m(\mathcal{K} \# T_{2,2})] = \sum_{\alpha=0}^k \abs{J_{2\alpha}(\mathcal{K}\,;e^{\frac{4\pi i}{k+1}})}^{2m} ~.  
\label{traceSO3k}
\end{equation}
One of the key questions we plan to investigate is the large $k$ behaviour of these $m$-
moments. We believe that they must capture geometrical features of the three-manifolds.
This will be the focus in the following subsection.
\subsubsection{Large $k$ asymptotics: The exponential growth}
We numerically checked that the colored Jones $J_{2\alpha}(\mathcal{K}\,;e^{\frac{4\pi i}{k+1}})$ has a maximum value when the
color is $2\alpha = k$ (note that $k$ is a positive even integer). Even though this observation was validated for some knots, we do not have
an analytic argument.

For the SO(3) Chern-Simons theory where coupling $k\in 2\mathbb{Z}$ and $\sigma(\mathcal{K}\# T_{2,2})$ is the unnormalised
reduced density matrix associated with the state $\ket{\mathcal{K}\# T_{2,2}}$ ($\mathcal{K}$ is a prime knot and $T_{2,2}$ is the
Hopf link), we put forth the following conjecture. \\ \\
\textbf{Conjecture 1.} \emph{The large $k$ asymptotics of the trace of the $m^{\text{th}}$ power of the matrix $\sigma(\mathcal{K}\# T_{2,2})$ captures the hyperbolic volume of $S^3 \backslash \mathcal{K}$:} 
\begin{equation}
\lim_{k \to \infty} \frac{\ln\, \text{Tr}[\sigma^m(\mathcal{K}\# T_{2,2})]}{k} = m\frac{\text{Vol}(S^3 \backslash \mathcal{K})}{2\pi} ~.
\label{SO3tracelimit}
\end{equation} \\
\emph{Proof.} Using our numerical result that the colored Jones $J_{k}(\mathcal{K}\,;e^{\frac{4\pi i}{k+1}})$ has maximum value, we can write the following inequality for \eqref{traceSO3k}:
\begin{equation}
\abs{J_{k}(\mathcal{K}\,;e^{\frac{4\pi i}{k+1}})}^{2m} \leq \text{Tr}[\sigma^m(\mathcal{K}\# T_{2,2})] \leq (k+1)\abs{J_{k}(\mathcal{K}\,;e^{\frac{4\pi i}{k+1}})}^{2m} ~.
\end{equation}
Taking log, dividing by $k$ and then taking $k \to \infty$ limit will give
\begin{equation}
\lim_{k \to \infty} \frac{\ln\, \text{Tr}[\sigma^m(\mathcal{K}\# T_{2,2})]}{k} = \lim_{k \to \infty} \frac{\ln\, \abs{J_{k}(\mathcal{K}\,;e^{\frac{4\pi i}{k+1}})}^{2m} }{k} ~.
\label{limitintermediate}
\end{equation}
To write the RHS of the above equation as volume, we will use the following relation between the colored Jones invariant and the Kashaev's invariant \cite{murakami2001colored}:
\begin{equation}
\abs{\text{Kas}_{n}(\mathcal{K}\,;q)} = \abs{J_{n-1}(\mathcal{K}\,;q)} \quad,\quad q = \exp(\frac{2 \pi i\, a}{n}) \quad,\quad \text{gcd}(a,n)=1 ~,
\label{Kas=Jones}
\end{equation}
where $\text{Kas}_{n}(\mathcal{K}\,;q)$ is the Kashaev's invariant of the knot $\mathcal{K}$ and the variable $q$ is taken to be a primitive $n^{\text{th}}$ root of unity. In the large $n$ limit, the above Kashaev's invariant grows exponentially obeying the following conjecture \cite{kashaev1997hyperbolic}:
\begin{equation}
\lim_{n \to \infty} \frac{\ln\, \abs{\text{Kas}_{n}(\mathcal{K}\,;e^{\frac{2\pi i a}{n}})} }{n} = \frac{\text{Vol}(S^3 \backslash \mathcal{K})}{2\pi a} ~.
\label{Kashaveconj}
\end{equation}
That is, its growth rate is determined by the hyperbolic volume of $S^3 \backslash \mathcal{K}$. This conjecture has been verified for $a=1$ case in Ref.\cite{kashaev1997hyperbolic,hikami2003volume} for many hyperbolic knots. For the SO(3) invariants we are studying, we have $n=(k+1)$, which is odd. Hence $a=2$ is also allowed in \eqref{Kas=Jones} and \eqref{Kashaveconj}. We have checked the $a=2$ case of \eqref{Kashaveconj} in appendix \ref{appA} for some torus and non-torus knots. From the above discussion, we can now state that the $m$-moment of the matrix $\sigma$ for the class of two-component links $\mathcal{K} \# T_{2,2}$ must obey the following in the large $k$ limit:
\begin{equation}
\lim_{k \to \infty} \frac{\ln\, \text{Tr}[\sigma^m(\mathcal{K}\# T_{2,2})]}{k} = \lim_{k \to \infty} \frac{\ln\, \abs{J_{k}(\mathcal{K}\,;e^{\frac{4\pi i}{k+1}})}^{2m} }{k} = 2m\frac{\text{Vol}(S^3 \backslash \mathcal{K})}{4\pi} ~,
\end{equation}
which gives our conjectured result \eqref{SO3tracelimit}.

Further, following our earlier arguments that $\text{Tr}[\sigma(\mathcal{K}\# T_{2,2})]$ is the partition function $Z(M_{\mathcal{K}})$, the conjecture \eqref{SO3tracelimit} leads to the following corollary:\\ \\
\textbf{Corollary 1.} \emph{For any prime knot $\mathcal K$, there exists a closed three-manifold $M_{\mathcal K}$ such that the large $k$ asymptotics of  the corresponding SO(3) Chern-Simons partition function $Z(M_{\mathcal K})$ determines the hyperbolic volume of  
$S^3 \backslash \mathcal{K}$:}
\begin{equation}
\lim_{k \to \infty} \frac{\ln\, Z(M_{\mathcal{K}})}{k} = \frac{\text{Vol}(S^3 \backslash \mathcal{K})}{2\pi} ~.
\label{corollary-eq}
\end{equation} 
Thus, even though the explicit topology of $M_{\mathcal{K}}$ is not visualisable, the leading large $k$ behaviour of the SO(3) Chern-Simons partition function $Z(M_{\mathcal{K}})$ captures the hyperbolic volume of $S^3 \backslash \mathcal{K}$. 

Moreover, as we have shown earlier in \eqref{RenyiActualValues} that the R\'enyi entropies can be written in terms of the partition functions $Z(M_{\mathcal{K}_m})$, it will be important to study the large $k$ asymptotics of the R\'enyi entropies. This will be the content of the following subsection. 
\subsubsection{Large $k$ limits of R\'enyi entropies}
We have already discussed that the large $k$ asymptotics of $\text{Tr}[\sigma^m(\mathcal{K}\# T_{2,2})]$ or $Z(M_{\mathcal{K}_m})$ is governed by the asymptotics of $\abs{J_{k}(\mathcal{K}\,;e^{\frac{4\pi i}{k+1}})}$. Since $\abs{J_{k}(\mathcal{K}\,;e^{\frac{4\pi i}{k+1}})} = \abs{\text{Kas}_{k+1}(\mathcal{K}\,;e^{\frac{4\pi i}{k+1}})}$
 from \eqref{Kas=Jones}, it suffices to obtain the full asymptotics of $\abs{\text{Kas}_{k+1}(\mathcal{K}\,;e^{\frac{4\pi i}{k+1}})}$. Setting $n=(k+1)$, the following functional form for the Kashaev's invariant in the large $n$ limit was conjectured in \cite{hikami2003volume}:
\begin{equation}
\frac{2 \pi}{n} \ln \abs{\text{Kas}_{n}(\mathcal{K}\,;q=e^{\frac{2\pi i}{n}})} \sim  c_1(\mathcal{K}) + c_2(\mathcal{K}) \frac{2 \pi}{n}\ln n + \frac{c_3(\mathcal{K})}{n} + \frac{c_4(\mathcal{K})}{n^2} ~,
\label{Kas-trialFun}
\end{equation}
where $c_i(\mathcal{K})$ are knot dependent constants such that \cite{hikami2003volume}:
\begin{equation}
c_1(\mathcal{K}) = \text{Vol}(S^3 \backslash \mathcal{K}) \quad;\quad c_2(\mathcal{K}) = \frac{3}{2} ~.
\end{equation}
Doing the numerical analysis for some of the non-trivial prime knots, we find that we can modify \eqref{Kas-trialFun} for $q = e^{\frac{4\pi i}{n}}$ as following:
\begin{equation}
\frac{4 \pi}{n} \ln \abs{\text{Kas}_{n}(\mathcal{K}\,;q=e^{\frac{4\pi i}{n}})} \sim  a_1(\mathcal{K}) + a_2(\mathcal{K}) \frac{4 \pi}{n}\ln n + \frac{a_3(\mathcal{K})}{n} + \frac{a_4(\mathcal{K})}{n^2} ~,
\end{equation} 
where $a_i(\mathcal{K})$ are constants such that:
\begin{equation}
a_1(\mathcal{K}) = \text{Vol}(S^3 \backslash \mathcal{K}) \quad;\quad a_2(\mathcal{K}) = \frac{3}{2} ~.
\end{equation}
As a result, we will have the following asymptotics:
\begin{equation}
\ln \abs{J_{k}(\mathcal{K},\,q=e^{\frac{4\pi i}{k+1}})} \sim  {\frac{\text{Vol}(S^3 \backslash \mathcal{K})}{4\pi}k} + \ln k^{3/2} + \frac{a_3(\mathcal{K})}{4 \pi} + \frac{a_4(\mathcal{K})}{4\pi k} ~.
\label{trialJones}
\end{equation} 
From \eqref{traceSO3K} and \eqref{ZMKm}, we will have:
\begin{equation}
\frac{Z(M_{\mathcal{K}_m})}{\abs{J_{k}(\mathcal{K},\,e^{\frac{4\pi i}{k+1}})}^{2m}} = \sum_{\alpha=0}^k \abs{\frac{J_{2\alpha}(\mathcal{K}\,;e^{\frac{4\pi i}{k+1}})}{J_{k}(\mathcal{K}\,;e^{\frac{4\pi i}{k+1}})}}^{2m} \equiv X_m(\mathcal{K};k) ~.
\end{equation}
Using \eqref{trialJones}, we have following asymptotics for large $k$:
\begin{equation}
Z(M_{\mathcal{K}_m}) \sim k^{3m} \exp\left[\frac{m\,\text{Vol}(S^3 \backslash \mathcal{K})}{2\pi}k\right] \times \left( \exp\left[m\frac{a_3(\mathcal{K})}{2\pi}+m\frac{a_4(\mathcal{K})}{2\pi k}\right] X_m(\mathcal{K};k) \right) ~.
\end{equation} 
From the numerical computations for several knots, we observe that the term in the parentheses is convergent as $k \to \infty$. So we define a knot dependent constant $C_m$ as:
\begin{equation}
C_m(\mathcal{K}) = \lim_{k \to \infty} \exp\left[m\frac{a_3(\mathcal{K})}{2\pi}+m\frac{a_4(\mathcal{K})}{2\pi k}\right] X_m(\mathcal{K};k) ~.
\end{equation} 
Unfortunately, we do not have an analytical way to compute $C_m(\mathcal{K})$ and we restrict to the numerical computations. To compute $C_m(\mathcal{K})$, we define a function
\begin{equation}
F_m(\mathcal{K}) \equiv Z(M_{\mathcal{K}_m}) \exp\left[-\frac{m\,\text{Vol}(S^3 \backslash \mathcal{K})}{2\pi}k\right]~.
\label{Fmactualvalues}
\end{equation}
From the above discussion, we expect the function $F_m(\mathcal{K})$ (for a prime knot $\mathcal{K}$) to grow as $k^{3m}$ for large values of $k$:
\begin{equation}
F_m(\mathcal{K}) \sim C_m(\mathcal{K})\, k^{3m} ~. \label{Fm-Func-Fit}
\end{equation}
We first evaluate the values of $F_m(\mathcal{K})$ using \eqref{Fmactualvalues} for sufficiently large values of $k$ and then use \eqref{Fm-Func-Fit} to obtain the constants $C_m(\mathcal{K})$ numerically using the least-square fitting method in Mathematica. In section \ref{sec4}, we present examples of prime knots up to six crossings, where $C_m(\mathcal{K})$ has been tabulated up to certain decimal places. With all the numerical results in hand, we propose the following: \\ \\
\textbf{Proposition.} \emph{The large $k$ asymptotics of SO(3) Chern-Simons partition function of $M_{\mathcal{K}_m}$ is given as following:} 
\begin{align}
Z(M_{\mathcal{K}_m}) = \text{Tr}[\sigma^m(\mathcal{K}\# T_{2,2})] \sim  C_m(\mathcal{K}) \,  k^{3m} \exp\left(\frac{m\,\text{Vol}(S^3 \backslash \mathcal{K})}{2\pi}k\right) ~,
\label{SO3traceAsymptotic}
\end{align} 
\emph{where $C_m(\mathcal{K})$ is a knot-dependent constant. As a result, the R\'enyi entropies converge to a finite value in the limit of $k \to \infty$ and its value is determined by the constants $C_m(\mathcal{K})$:}
\begin{align}
\lim_{k \to \infty} \mathcal{R}_m =  \frac{1}{1-m} \ln \left[ \frac{C_m(\mathcal{K})}{C_1(\mathcal{K})^m} \right] ~.
\label{limit-Renyi-proposal}
\end{align} 
In the next section, we will present the numerical results validating our conjecture and proposal. We will restrict to prime knots up to six crossings.
\section{Numerical results for SO(3) group} \label{sec4}
In this section, we will present the numerical results for all prime knots up to six crossings. A list of these knots, along with their hyperbolic volumes, are given in table \ref{listKnots}. As discussed in the previous section, we obtain the constants $C_m(\mathcal{K})$ numerically for various values of $m$. We show the predicted variation of the function $C_m(\mathcal{K})\, k^{3m}$ and contrast it with the actual computed values of $F_m(\mathcal{K})$ from \eqref{Fmactualvalues}. We further list the proposed $k \to \infty$ limiting values of R\'enyi entropies obtained using $C_m(\mathcal{K})$ via \eqref{limit-Renyi-proposal}. We justify these predicted limiting values by plotting the variation of R\'enyi entropies (calculated from \eqref{RenyiActualValues}) with $k$ and showing that the plots tend to converge to the proposed values.   
\begin{table}[!htb]
\centering
\begin{tabular}{|ccccccc|} \hline
\rowcolor{Grayy}
$3_1$ & $4_1$ & $5_1$ & $5_2$ & $6_1$ & $6_2$ & $6_3$ \\ 
$\begin{array}{c}\includegraphics[width=0.10\textwidth]{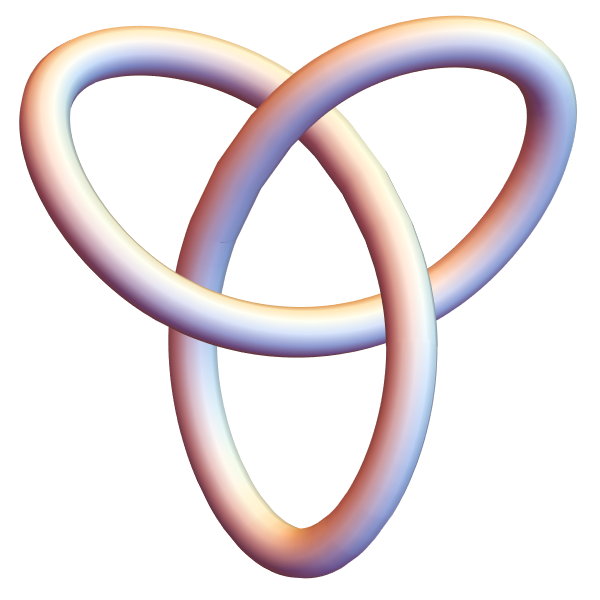}\end{array}$ & $\begin{array}{c}\includegraphics[width=0.10\textwidth]{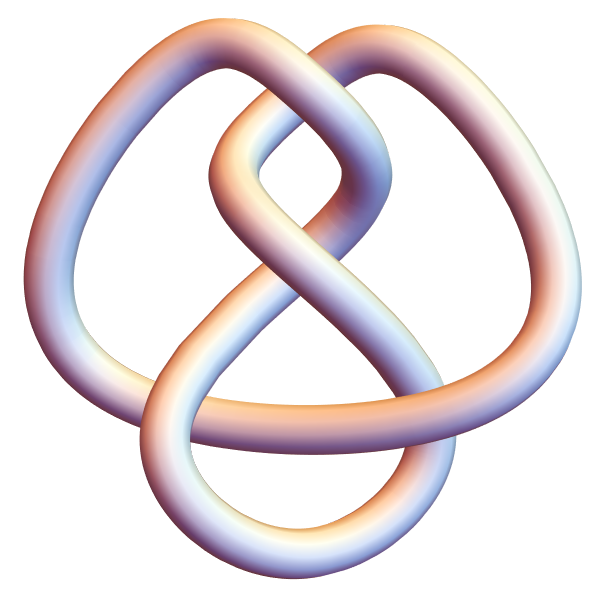}\end{array}$ & $\begin{array}{c}\includegraphics[width=0.10\textwidth]{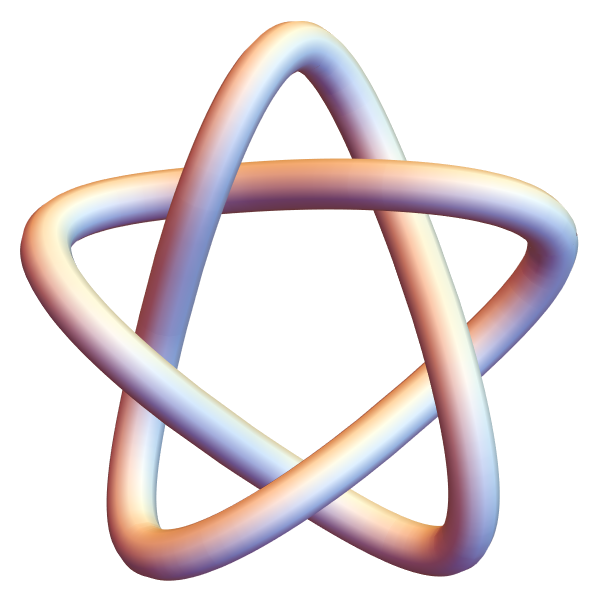}\end{array}$ & $\begin{array}{c}\includegraphics[width=0.10\textwidth]{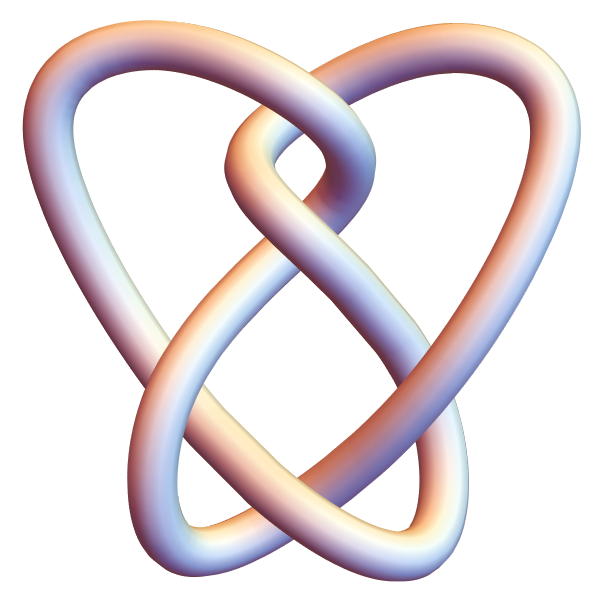}\end{array}$ & $\begin{array}{c}\includegraphics[width=0.10\textwidth]{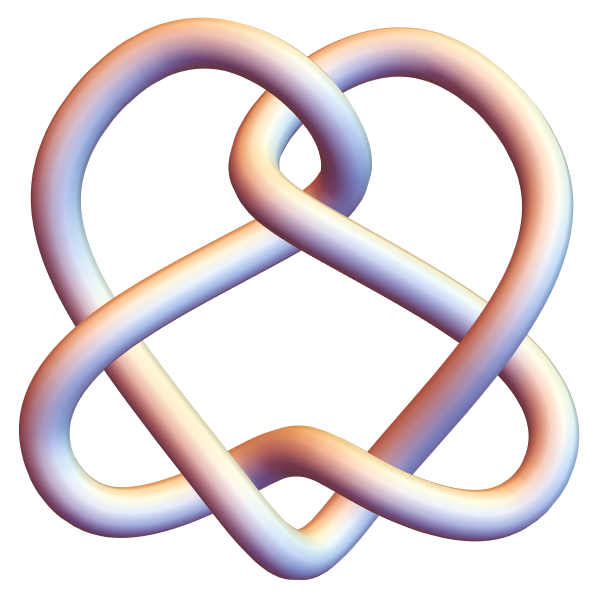}\end{array}$ & $\begin{array}{c}\includegraphics[width=0.10\textwidth]{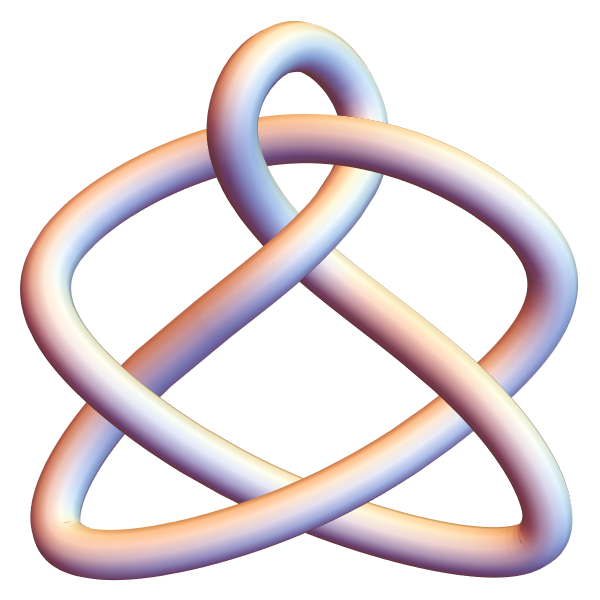}\end{array}$ & $\begin{array}{c}\includegraphics[width=0.10\textwidth]{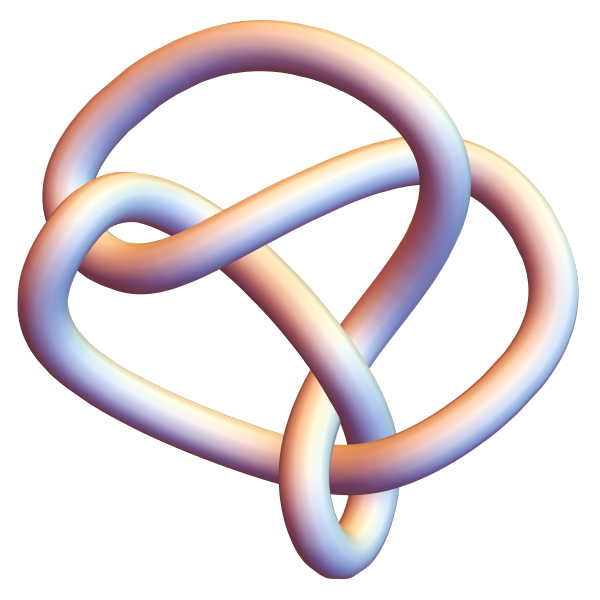}\end{array}$ \\ \rowcolor{Grayy}
0 & 2.02988 & 0 & 2.82812 & 3.16396 & 4.40083 & 5.69302 \\ \hline
\end{tabular}
\caption{List of all non-trivial prime knots up to six crossings. The first row of the table gives the notation of the knots according to the Rolfsen table \cite{rolfsen1990knots}. The second row describes the knot diagram and the third row gives the hyperbolic volume of the knot complement.}
\label{listKnots}
\end{table}
For ease of presentation, a list of plots and tables for various knots are referred in the following: 
\begin{equation*}
\begin{tabular}{|c|c|c|c|c|c|} \hline
\rowcolor{Grayy}
$\mathcal{K}$ & $C_m(\mathcal{K})$ & $C_m(\mathcal{K})$ vs $m$ & $F_m(\mathcal{K})$ vs $k$ & $\lim_{k \to\infty} \mathcal{R}_m$ & $\mathcal{R}_m$ vs $k$ \\
$3_1$ & Table \ref{Cmfor31knot} & Figure \ref{Cvalues31knot} & Figure \ref{traces31knot} & Table \ref{Renyipredicted31knot} & Figure \ref{REvsk31knot}\\
$4_1$ & Table \ref{Cmfor41knot} & Figure \ref{Cvalues41knot} & Figure \ref{traces41knot} & Table \ref{Renyipredicted41knot} & Figure \ref{REvsk41knot}\\
$5_1$ & Table \ref{Cmfor51knot} & Figure \ref{Cvalues51knot} & Figure \ref{traces51knot} & Table \ref{Renyipredicted51knot} & Figure \ref{REvsk51knot}\\
$5_2$ & Table \ref{Cmfor52knot} & Figure \ref{Cvalues52knot} & Figure \ref{traces52knot} & Table \ref{Renyipredicted52knot} & Figure \ref{REvsk52knot}\\
$6_1$ & Table \ref{Cmfor61knot} & Figure \ref{Cvalues61knot} & Figure \ref{traces61knot} & Table \ref{Renyipredicted61knot} & Figure \ref{REvsk61knot}\\
$6_2$ & Table \ref{Cmfor62knot} & Figure \ref{Cvalues62knot} & Figure \ref{traces62knot} & Table \ref{Renyipredicted62knot} & Figure \ref{REvsk62knot}\\
$6_3$ & Table \ref{Cmfor63knot} & Figure \ref{Cvalues63knot} & Figure \ref{traces63knot} & Table \ref{Renyipredicted63knot} & Figure \ref{REvsk63knot} \\ \hline
\end{tabular}
\end{equation*}

We will take up the simplest torus knots in the following section to verify the results of previous section.
\subsection{Torus knots}
\subsubsection{Trefoil knot: $3_1$}
\begin{figure}[htbp]
	\centering
		\includegraphics[width=1.00\textwidth]{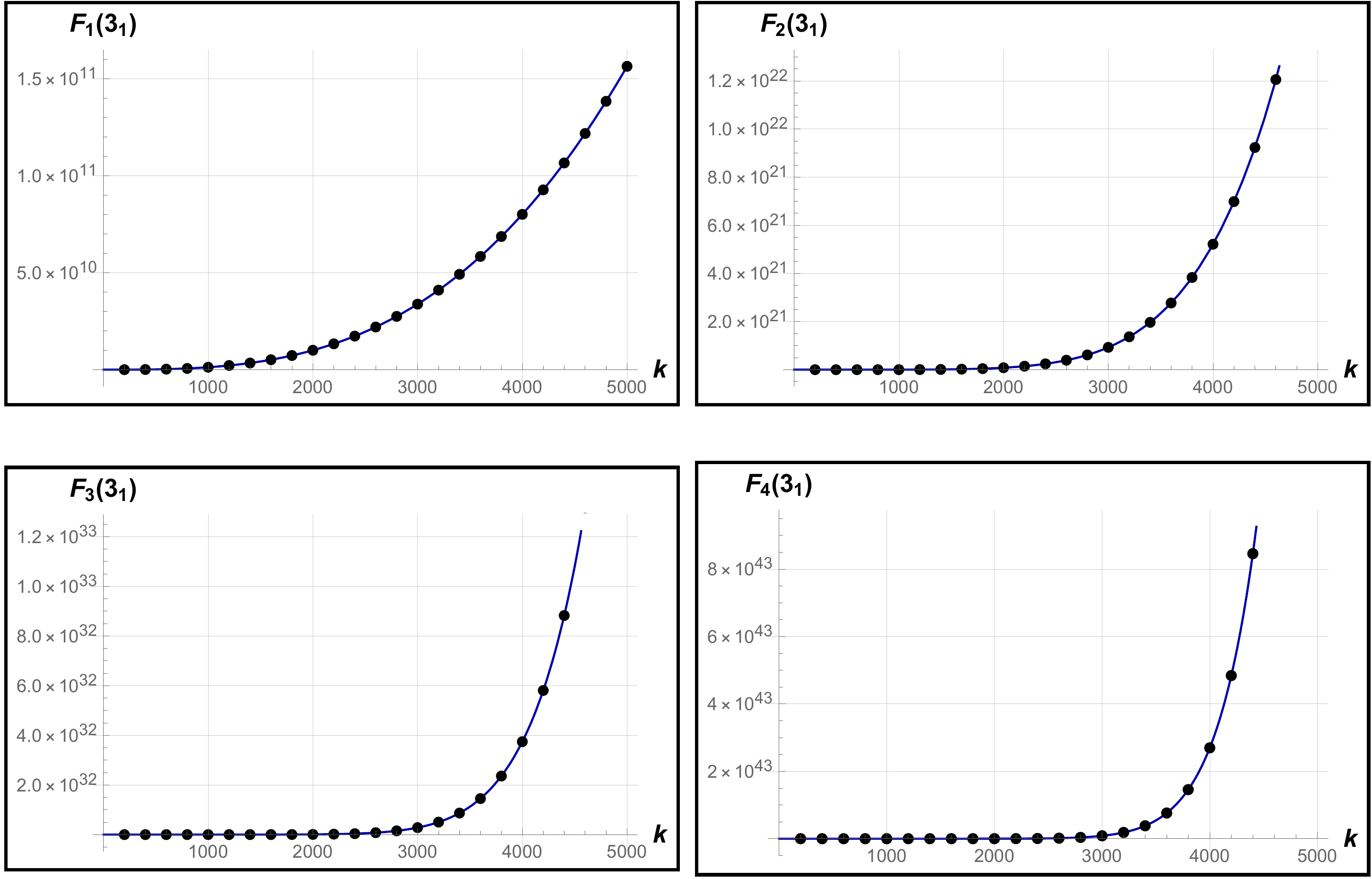}
	\caption{The variation of the function $F_m$ with $k$ for $3_1$ knot. The solid line denotes the function $C_m\, k^{3m}$ with values of $C_m$ given in table \ref{Cmfor31knot}. The $\bullet$ indicates the values of $F_m$ obtained from explicit computations using \eqref{Fmactualvalues}.}
	\label{traces31knot}
\end{figure}

This is a torus knot with vanishing hyperbolic volume: $\text{Vol}(S^3 \backslash 3_1) = 0$. The proposed form for the function $F_m$  
\begin{align}
F_m(3_1) = C_m(3_1) \,  k^{3m}
\label{}
\end{align}
is computed using \eqref{Fmactualvalues} for some sufficiently large values of $k$. In fact, the least-square method in Mathematica enables us to determine the constant $C_m(3_1)$ for various values of $m$. Some of these values are listed in table \ref{Cmfor31knot} and the variation of $\ln C_m(3_1)$ with $m$ is shown in figure \ref{Cvalues31knot}. 
\begin{table}
	\begin{minipage}{0.4\linewidth}
		\centering
		\begin{tabular}{|c|c|} \hline \rowcolor{Grayy}
			$m$ & $C_m(3_1)$  \\ \hline
\footnotesize{1} & \footnotesize{$1.2508395500 \pm 0.0000141876$}  \\
\footnotesize{2} & \footnotesize{$1.2724830787 \pm 0.0000133285$} \\
\footnotesize{3} & \footnotesize{$1.4267959549 \pm 0.0000139704$} \\
\footnotesize{4} & \footnotesize{$1.6058331673 \pm 0.0000151173$} \\
\footnotesize{5} & \footnotesize{$1.8076384699 \pm 0.0000167464$} \\
$\vdots$ & $\vdots$ \\ 
\footnotesize{96} & \footnotesize{$86270.3224528739 \pm 0.3000329135$} \\
\footnotesize{97} & \footnotesize{$97112.8266267165 \pm 0.3310750028$} \\
\footnotesize{98} & \footnotesize{$109318.0234964881 \pm 0.3652932337$} \\
\footnotesize{99} & \footnotesize{$123057.1771676096 \pm 0.4030095261$} \\
\footnotesize{100} & \footnotesize{$138523.0763324435 \pm 0.4445781686$} \\
\hline
		\end{tabular}
		\caption{Values of $C_m(3_1)$}
		\label{Cmfor31knot}
	\end{minipage}\hfill
	\begin{minipage}{0.58\linewidth}
		\centering
		\includegraphics[width=98mm]{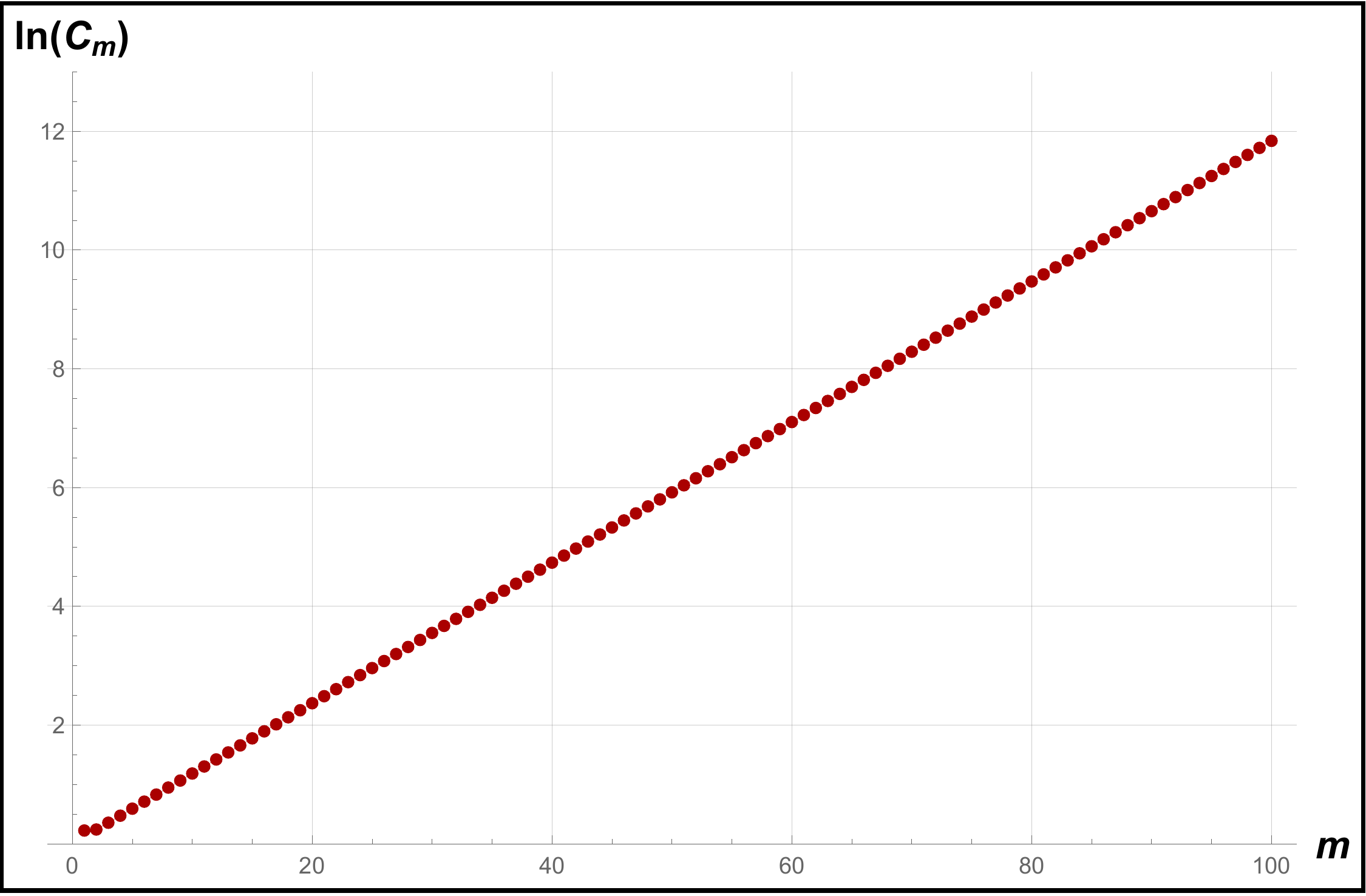}
		\captionof{figure}{Variation of $\ln C_m(3_1)$ with $m$}
		\label{Cvalues31knot}
	\end{minipage}
\end{table}
In figure \ref{traces31knot}, we plot the function $C_m(3_1) \,  k^{3m}$ and compare it with the numerical values of $F_m(3_1)$ obtained explicitly from \eqref{Fmactualvalues}. 
The predicted $k \to \infty$ values of the R\'enyi entropies associated with the state $\ket{3_1 \# T_{2,2}}$ can be computed using \eqref{limit-Renyi-proposal} and we have listed some of these values in table \ref{Renyipredicted31knot}. We also show the variation of the R\'enyi entropies with $k$ in the plots in figure \ref{REvsk31knot}. We see that these plots tend to converge to the predicted $k \to \infty$ values of the R\'enyi entropies.  With this warm-up on trefoil, we present the numerical results for the five crossing torus knot.
\begin{table}
	\begin{minipage}{0.4\linewidth}
		\centering
		\begin{tabular}{|c|c|} \hline \rowcolor{Grayy}
			$m$ & $\lim_{k \to\infty} \mathcal{R}_m$  \\ \hline
\footnotesize{2} & \footnotesize{$0.2066597601 \pm 0.0000249864$}  \\
\footnotesize{3} & \footnotesize{$0.1580067791 \pm 0.0000177041$} \\
\footnotesize{4} & \footnotesize{$0.1405390447 \pm 0.0000154454$} \\
\footnotesize{5} & \footnotesize{$0.1317633872 \pm 0.0000143660$} \\
$\vdots$ & $\vdots$ \\ 
\footnotesize{96} & \footnotesize{$0.1065367978 \pm 0.0000114619$} \\
\footnotesize{97} & \footnotesize{$0.1065252390 \pm 0.0000114607$} \\
\footnotesize{98} & \footnotesize{$0.1065139185 \pm 0.0000114595$} \\
\footnotesize{99} & \footnotesize{$0.1065028290 \pm 0.0000114583$} \\
\footnotesize{100} & \footnotesize{$0.1064919634 \pm 0.0000114571$} \\
\hline
		\end{tabular}
		\caption{The $k \to \infty$ values of the R\'enyi entropies associated with the state $\ket{3_1 \# T_{2,2}}$.}
		\label{Renyipredicted31knot}
	\end{minipage}\hfill
	\begin{minipage}{0.58\linewidth}
		\centering
		\includegraphics[width=95mm]{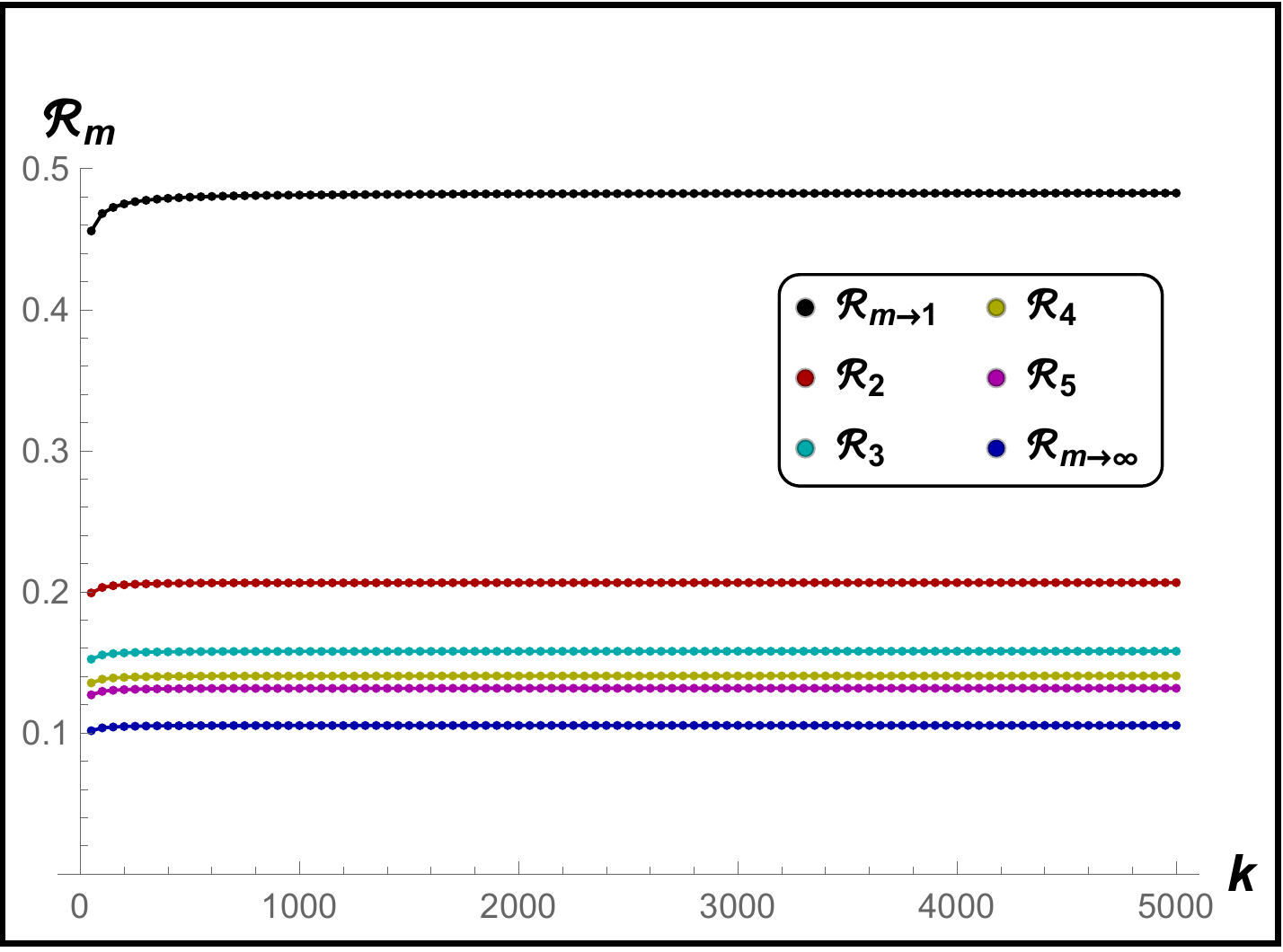}
		\captionof{figure}{Variation of R\'enyi entropies for the state $\ket{3_1 \# T_{2,2}}$ with $k$.}
		\label{REvsk31knot}
	\end{minipage}
\end{table}
\subsubsection{Solomon's Seal knot: $5_1$}
\begin{figure}[htbp]
	\centering
		\includegraphics[width=1.00\textwidth]{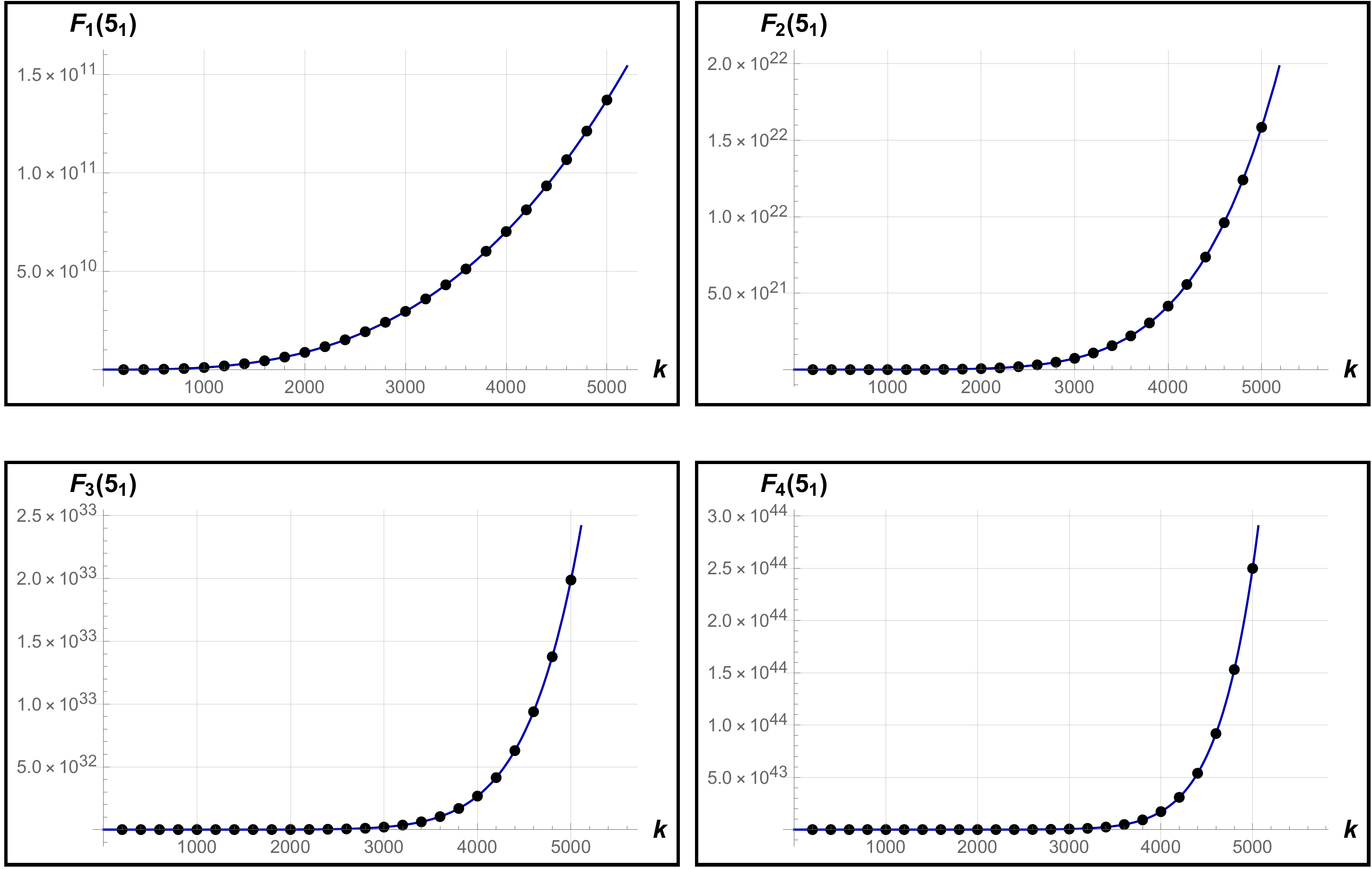}
	\caption{The variation of the function $F_m$ with $k$ for $5_1$ knot. The solid line denotes the function $C_m \,k^{3m}$ with values of $C_m$ given in table \ref{Cmfor51knot}. The $\bullet$ indicates the numerical values of $F_m$ obtained from explicit computations using \eqref{Fmactualvalues}.}
	\label{traces51knot}
\end{figure}

$5_1$ is also a torus knot with vanishing hyperbolic volume: $\text{Vol}(S^3 \backslash 5_1) = 0$.  Similar to our numerical methodology done for trefoil, we compute $F_m(5_1)$ \eqref{Fmactualvalues} and using the proposed large $k$ form  
\begin{align}
F_m(5_1) = C_m(5_1) \,  k^{3m} ~,
\label{}
\end{align}
and least-square method in Mathematica, we obtained the constant $C_m(5_1)$ for various values of $m$. Some of these values are listed in table \ref{Cmfor51knot} and the variation of $C_m(5_1)$ with $m$ is shown in figure \ref{Cvalues51knot}. 
\begin{table}
	\begin{minipage}{0.4\linewidth}
		\centering
		\begin{tabular}{|c|c|} \hline \rowcolor{Grayy}
			$m$ & $C_m(5_1)$  \\ \hline
\footnotesize{1} & \footnotesize{$1.0962198097 \pm 0.0000123719$}  \\
\footnotesize{2} & \footnotesize{$1.0141454043 \pm 0.0000110043$} \\
\footnotesize{3} & \footnotesize{$1.0172196551 \pm 0.0000107028$} \\
\footnotesize{4} & \footnotesize{$1.0228630176 \pm 0.0000105694$} \\
\footnotesize{5} & \footnotesize{$1.0286308590 \pm 0.0000104842$} \\
$\vdots$ & $\vdots$ \\ 
\footnotesize{96} & \footnotesize{$1.7173551256 \pm 0.0000040805$} \\
\footnotesize{97} & \footnotesize{$1.7270567284 \pm 0.0000040224$} \\
\footnotesize{98} & \footnotesize{$1.7368131320 \pm 0.0000039648$} \\
\footnotesize{99} & \footnotesize{$1.7466246464 \pm 0.0000039076$} \\
\footnotesize{100} & \footnotesize{$1.7564915828 \pm 0.0000038509$} \\
\hline
		\end{tabular}
		\caption{Values of $C_m(5_1)$}
		\label{Cmfor51knot}
	\end{minipage}\hfill
	\begin{minipage}{0.58\linewidth}
		\centering
		\includegraphics[width=98mm]{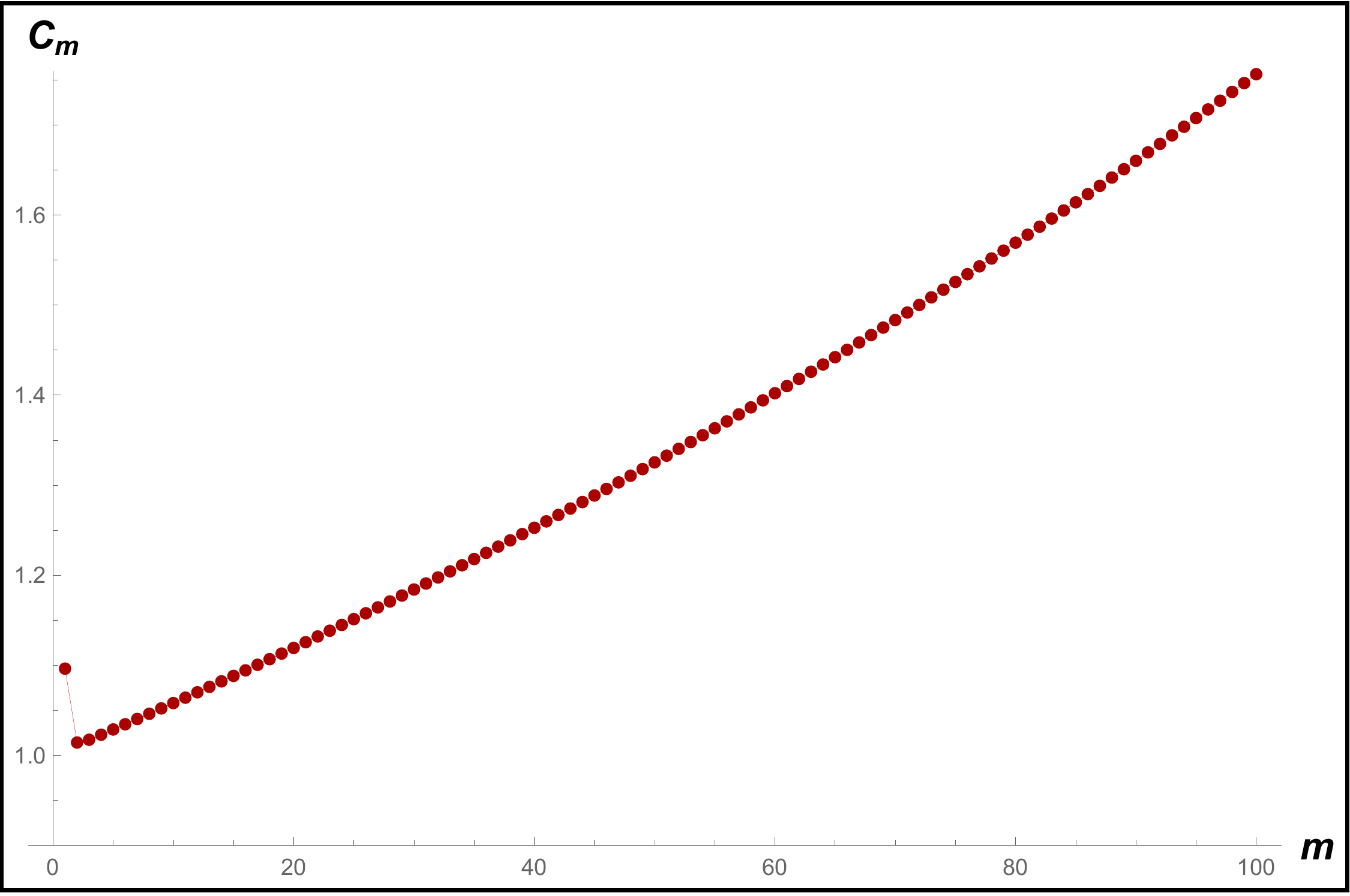}
		\captionof{figure}{Variation of $C_m(5_1)$ with $m$}
		\label{Cvalues51knot}
	\end{minipage}
\end{table}
In figure \ref{traces51knot}, we plot the function $C_m(5_1) \,  k^{3m}$ and compare it with the numerical values of $F_m(5_1)$ obtained explicitly from \eqref{Fmactualvalues}. 
The predicted $k \to \infty$ values of the R\'enyi entropies associated with the state $\ket{5_1 \# T_{2,2}}$ can be computed using \eqref{limit-Renyi-proposal} and we have listed some of these values in table \ref{Renyipredicted51knot}. We also show the variation of the R\'enyi entropies with $k$ in the plots in figure \ref{REvsk51knot}. We see that these plots tend to converge to the predicted $k \to \infty$ values of the R\'enyi entropies. We have also done similar numerical validation for other knots with
non-trivial hyperbolic volume which we present in the following subsection.
\begin{table}
	\begin{minipage}{0.4\linewidth}
		\centering
		\begin{tabular}{|c|c|} \hline \rowcolor{Grayy}
			$m$ & $\lim_{k \to\infty} \mathcal{R}_m$  \\ \hline
\footnotesize{2} & \footnotesize{$0.1696891578 \pm 0.0000250446$}  \\
\footnotesize{3} & \footnotesize{$0.1292650485 \pm 0.0000177275$} \\
\footnotesize{4} & \footnotesize{$0.1149551079 \pm 0.0000154371$} \\
\footnotesize{5} & \footnotesize{$0.1077774922 \pm 0.0000143357$} \\
$\vdots$ & $\vdots$ \\ 
\footnotesize{96} & \footnotesize{$0.0871422756 \pm 0.0000114048$} \\
\footnotesize{97} & \footnotesize{$0.0871328193 \pm 0.0000114036$} \\
\footnotesize{98} & \footnotesize{$0.0871235580 \pm 0.0000114023$} \\
\footnotesize{99} & \footnotesize{$0.0871144857 \pm 0.0000114012$} \\
\footnotesize{100} & \footnotesize{$0.0871055967 \pm 0.0000114000$} \\
\hline
		\end{tabular}
		\caption{The $k \to \infty$ values of the R\'enyi entropies associated with the state $\ket{5_1 \# T_{2,2}}$.}
		\label{Renyipredicted51knot}
	\end{minipage}\hfill
	\begin{minipage}{0.58\linewidth}
		\centering
		\includegraphics[width=95mm]{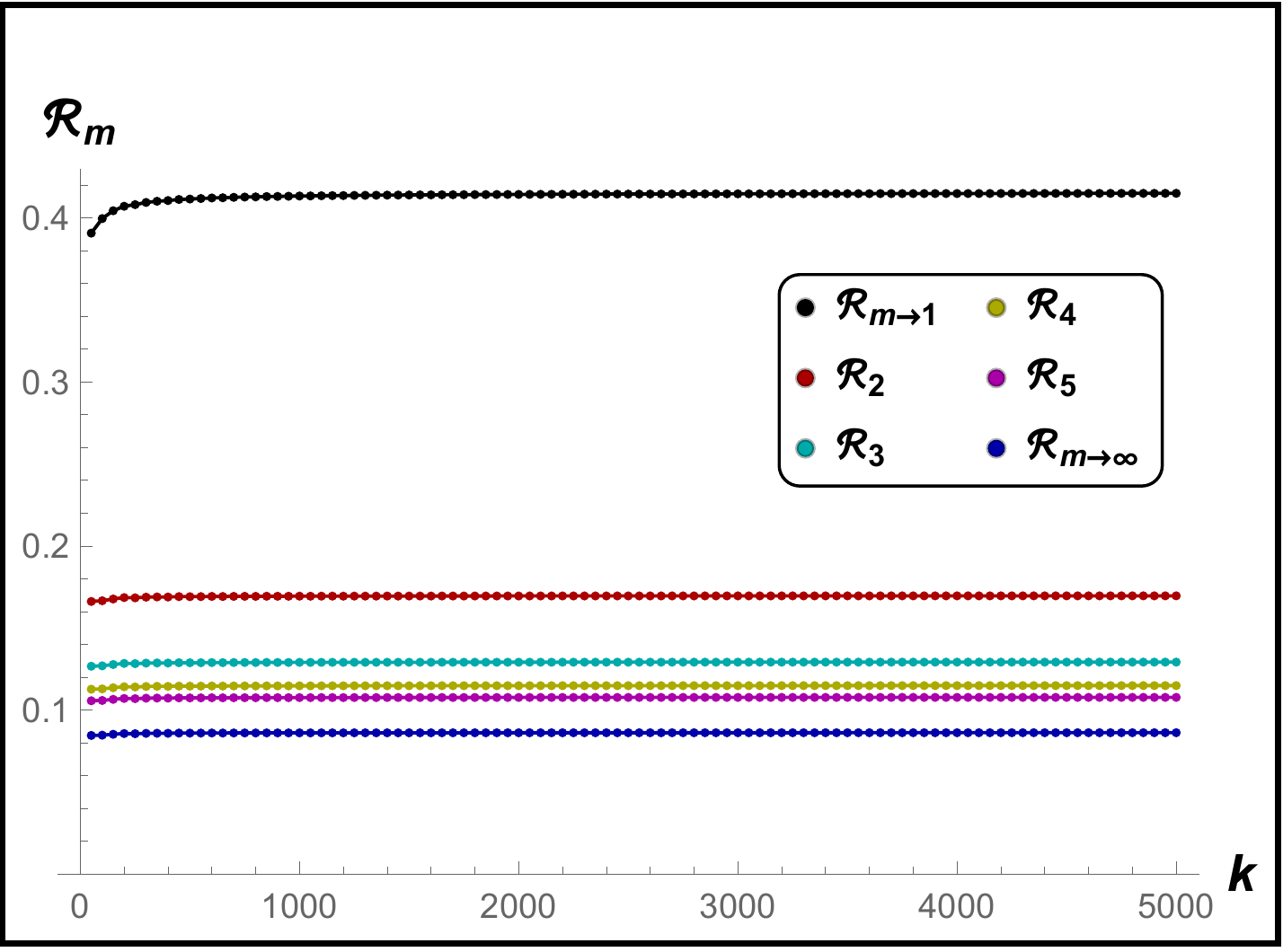}
		\captionof{figure}{Variation of R\'enyi entropies for the state $\ket{5_1 \# T_{2,2}}$ with $k$.}
		\label{REvsk51knot}
	\end{minipage}
\end{table}
\subsection{Non-torus knots}
\subsubsection{Figure-eight knot: $4_1$}
\begin{figure}[htbp]
	\centering
		\includegraphics[width=1.00\textwidth]{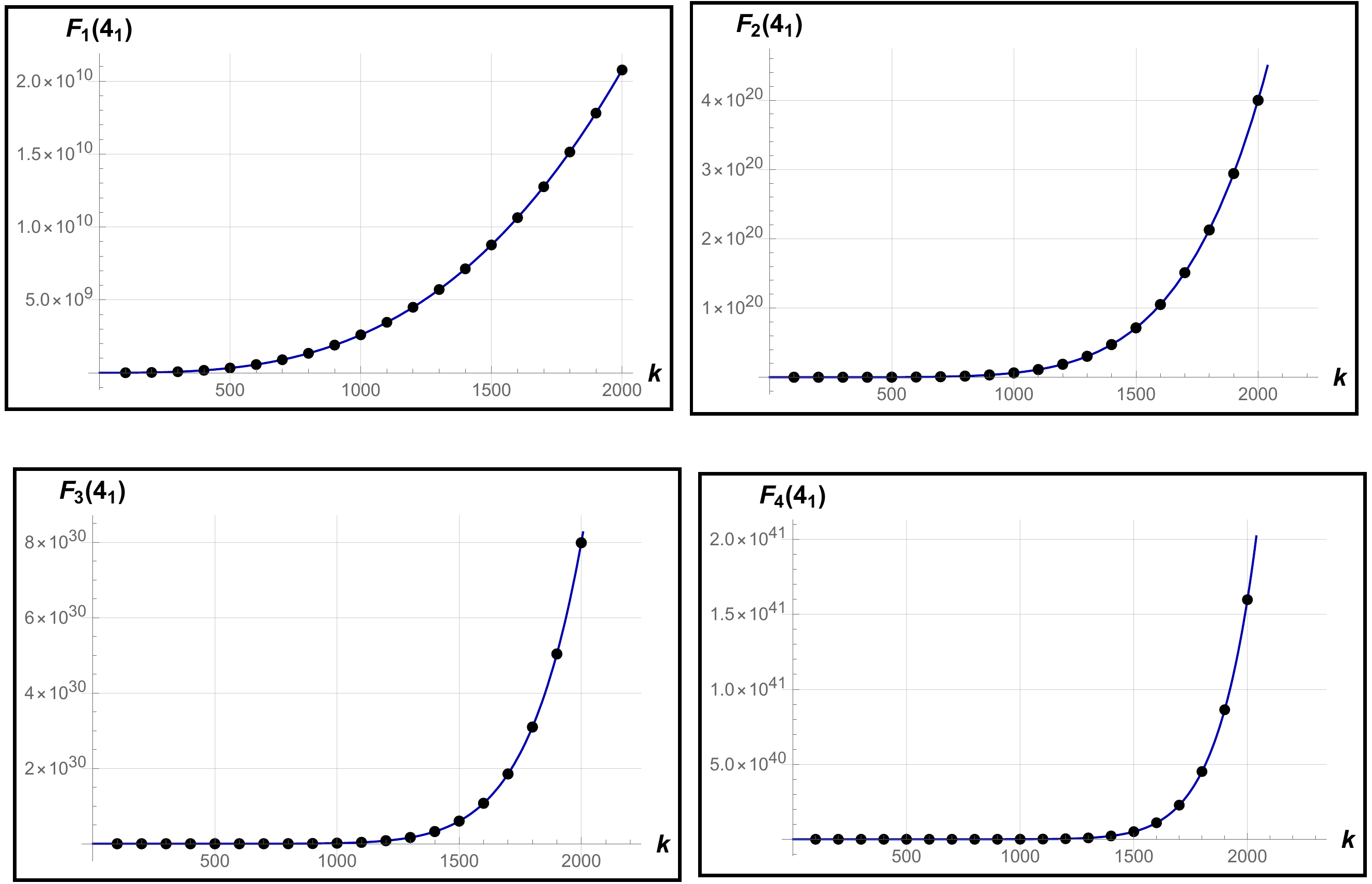}
	\caption{The variation of the function $F_m$ with $k$ for $4_1$ knot. The solid line denotes the function $C_m k^{3m}$ with values of $C_m$ given in table \ref{Cmfor41knot}. The $\bullet$ indicates the numerical values of $F_m$ obtained from explicit computations.}
	\label{traces41knot}
\end{figure}

The lowest crossing non-torus knot is the figure-eight knot $4_1$ with the hyperbolic volume $\text{Vol}(S^3 \backslash 4_1) = 2.029883212819307$. Using $F_m(4_1)$ \eqref{Fmactualvalues} and the following proposed large $k$ form,  
\begin{align}
F_m(4_1) =  C_m(4_1) \,  k^{3m} ~,
\label{}
\end{align}
the constant $C_m(4_1)$ was deduced using the least-square method for various values of $m$. Some of these values are listed in table \ref{Cmfor41knot} and the variation of $\ln C_m(4_1)$ with $m$ is shown in figure \ref{Cvalues41knot}.
\begin{table}
	\begin{minipage}{0.4\linewidth}
		\centering
		\begin{tabular}{|c|c|} \hline \rowcolor{Grayy}
			$m$ & $C_m(4_1)$  \\ \hline
\footnotesize{1} & \footnotesize{$2.5957570692 \pm 0.0001620425$}  \\
\footnotesize{2} & \footnotesize{$6.2495177996 \pm 0.0004659030$} \\
\footnotesize{3} & \footnotesize{$15.6076018833 \pm 0.0011272905$} \\
\footnotesize{4} & \footnotesize{$38.9986179208 \pm 0.0027607882$} \\
\footnotesize{5} & \footnotesize{$97.4465517130 \pm 0.0067868268$} \\
\footnotesize{6} & \footnotesize{$243.4918590302 \pm 0.0167002962$} \\
\footnotesize{7} & \footnotesize{$608.4194021638 \pm 0.0410857092$} \\
\footnotesize{8} & \footnotesize{$1520.2754646304 \pm 0.1009983598$} \\
\footnotesize{9} & \footnotesize{$3798.7623179296 \pm 0.2480046124$} \\
\footnotesize{10} & \footnotesize{$9492.1051075026 \pm 0.6082069793$} \\
\hline
		\end{tabular}
		\caption{Values of $C_m(4_1)$}
		\label{Cmfor41knot}
	\end{minipage}\hfill
	\begin{minipage}{0.58\linewidth}
		\centering
		\includegraphics[width=98mm]{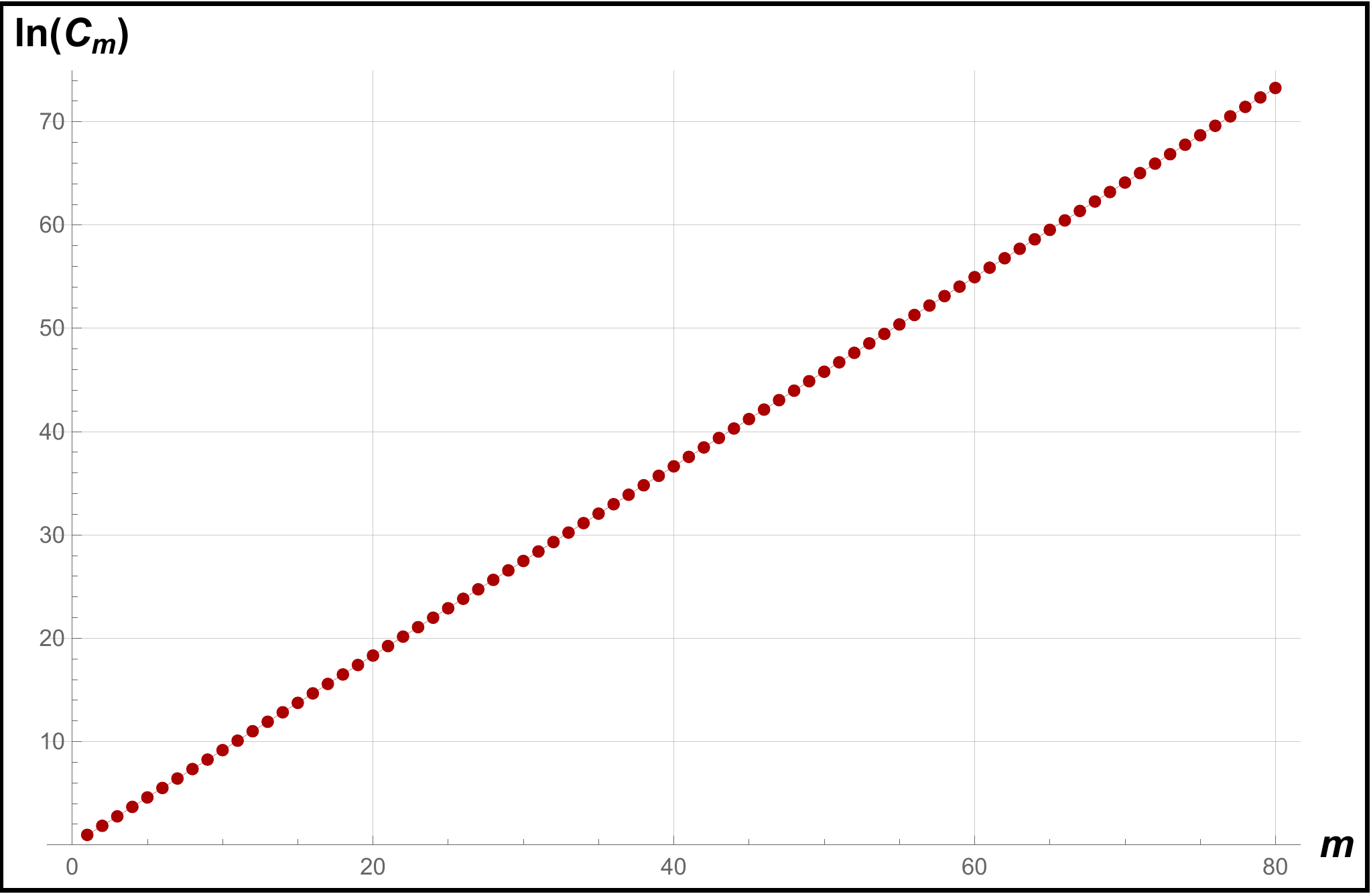}
		\captionof{figure}{Variation of $\ln C_m(4_1)$ with $m$}
		\label{Cvalues41knot}
	\end{minipage}
\end{table}
In figure \ref{traces41knot}, we plot the function $C_m(4_1) \,  k^{3m}$ and compare it with the numerical values of $F_m(4_1)$ obtained explicitly from \eqref{Fmactualvalues}. 
The predicted $k \to \infty$ values of the R\'enyi entropies associated with the state $\ket{4_1 \# T_{2,2}}$ can be computed using \eqref{limit-Renyi-proposal} and we have listed some of these values in table \ref{Renyipredicted41knot}. We also show the variation of the R\'enyi entropies with $k$ in the plots in figure \ref{REvsk41knot}. We see that these plots tend to converge to the predicted $k \to \infty$ values of the R\'enyi entropies. We present similar analysis and our numerical results for the five crossing
non-torus knot in the following section to validate our conjecture and the proposal.
\begin{table}
	\begin{minipage}{0.4\linewidth}
		\centering
		\begin{tabular}{|c|c|} \hline \rowcolor{Grayy}
			$m$ & $\lim_{k \to\infty} \mathcal{R}_m$  \\ \hline
\footnotesize{2} & \footnotesize{$0.0752521224 \pm 0.0001454157$}  \\
\footnotesize{3} & \footnotesize{$0.0569382755 \pm 0.0001003615$}  \\
\footnotesize{4} & \footnotesize{$0.0506622182 \pm 0.0000865149$}  \\
\footnotesize{5} & \footnotesize{$0.0475217594 \pm 0.0000799514$}  \\
\footnotesize{6} & \footnotesize{$0.0456371569 \pm 0.0000761567$}  \\
\footnotesize{7} & \footnotesize{$0.0443805098 \pm 0.0000736947$}  \\
\footnotesize{8} & \footnotesize{$0.0434827000 \pm 0.0000719724$}  \\
\footnotesize{9} & \footnotesize{$0.0428091692 \pm 0.0000707017$}  \\
\footnotesize{10} & \footnotesize{$0.0422851628 \pm 0.0000697266$}  \\
\hline
		\end{tabular}
		\caption{The $k \to \infty$ values of the R\'enyi entropies associated with the state $\ket{4_1 \# T_{2,2}}$.}
		\label{Renyipredicted41knot}
	\end{minipage}\hfill
	\begin{minipage}{0.58\linewidth}
		\centering
		\includegraphics[width=95mm]{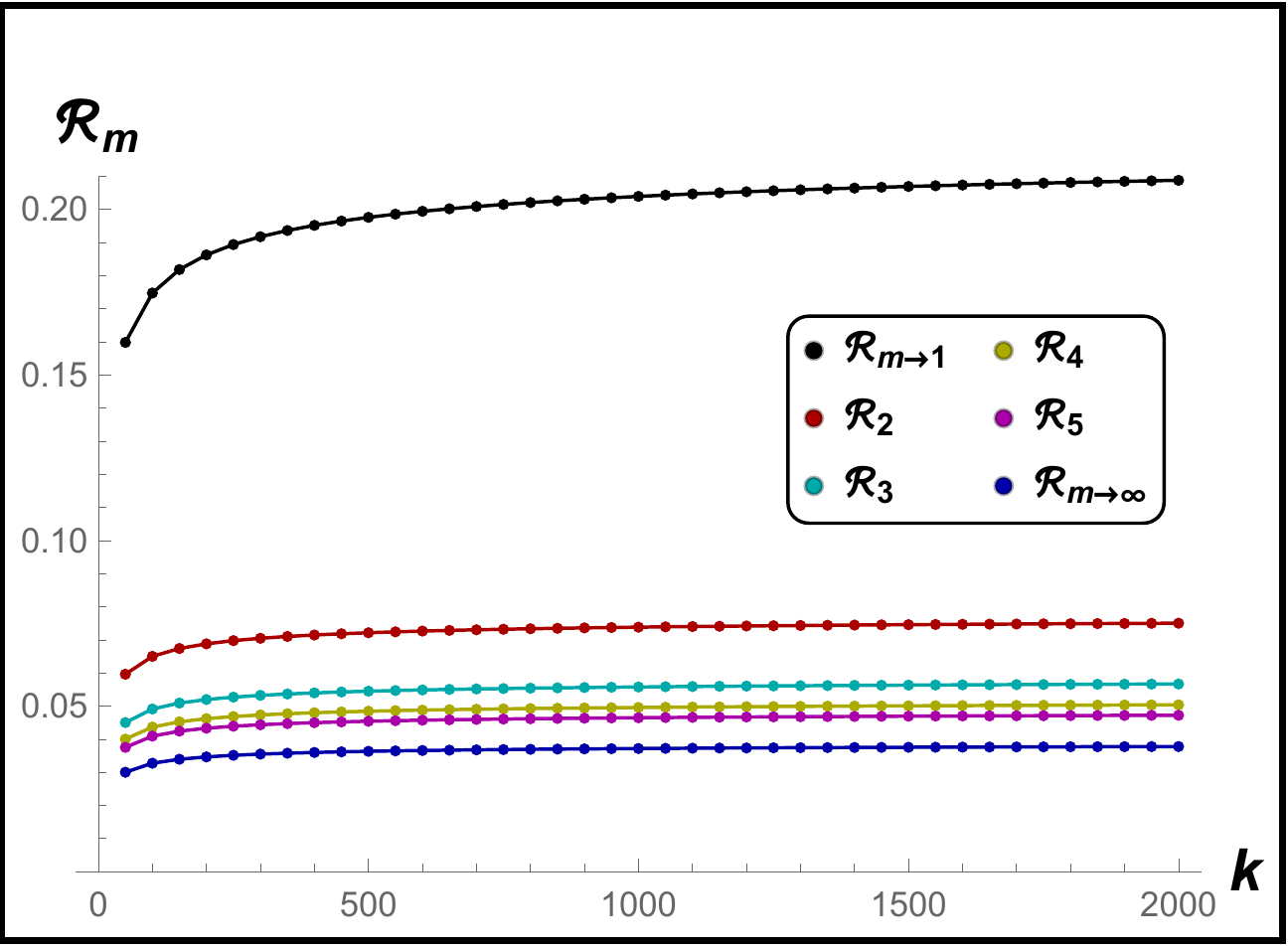}
		\captionof{figure}{Variation of R\'enyi entropies for the state $\ket{4_1 \# T_{2,2}}$ with $k$.}
		\label{REvsk41knot}
	\end{minipage}
\end{table}

\subsubsection{Three-twist knot: $5_2$}
\begin{figure}[htbp]
	\centering
		\includegraphics[width=1.00\textwidth]{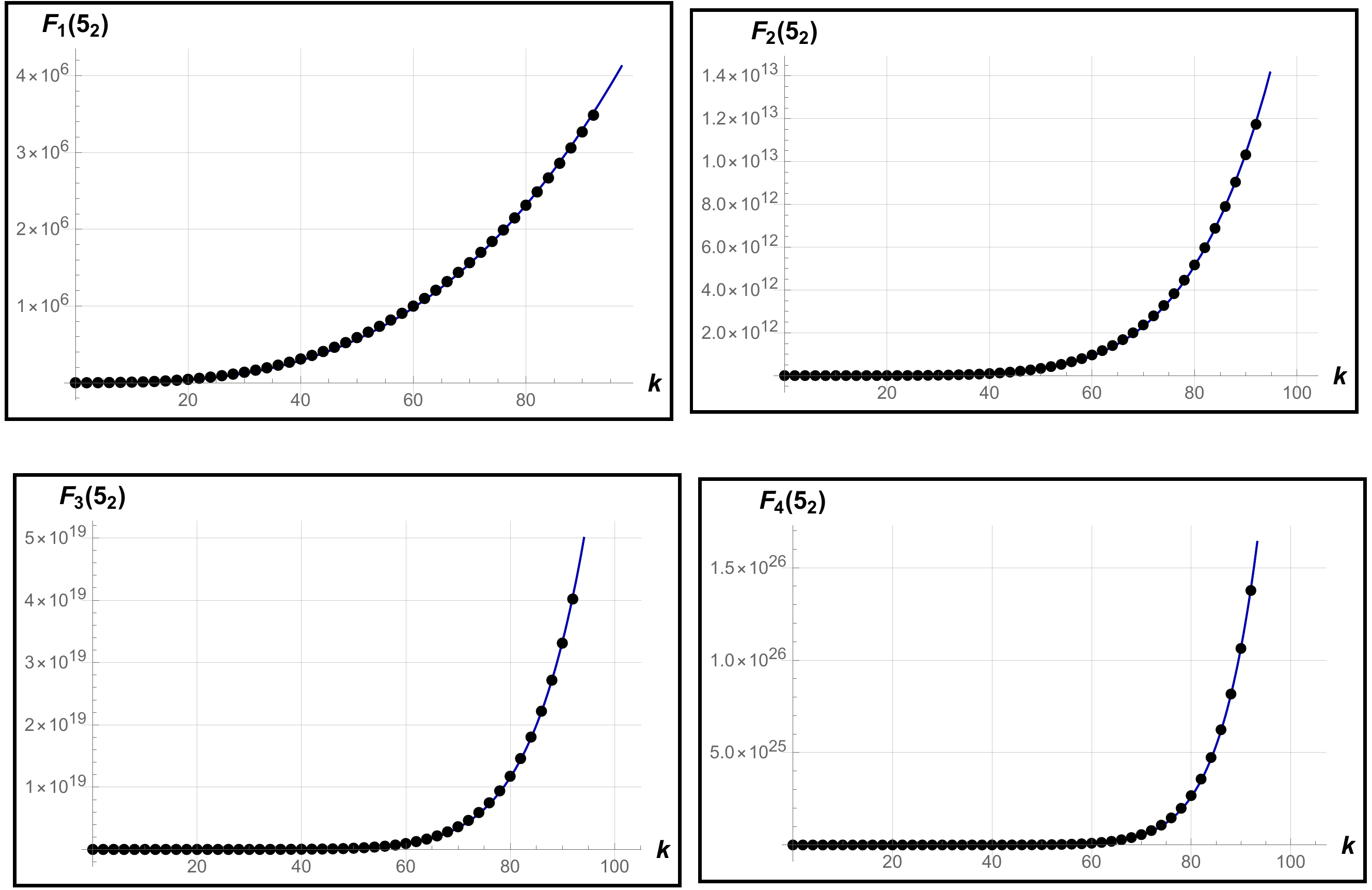}
	\caption{The variation of the function $F_m$ with $k$ for $5_2$ knot. The solid line denotes the function $C_m k^{3m}$ with values of $C_m$ given in table \ref{Cmfor52knot}. The $\bullet$ indicates the numerical values of $F_m$ obtained from explicit computations.}
	\label{traces52knot}
\end{figure}

$F_m(5_2)$ for this five crossing non-torus knot, whose knot complement volume is $\text{Vol}(S^3 \backslash 5_2) = 2.828122088330783$, is computed using \eqref{Fmactualvalues}. Then, from the proposed large $k$ form  
\begin{align}
F_m(5_2) =  C_m(5_2) \,  k^{3m} ~,
\label{}
\end{align}
the constant $C_m(5_2)$ is determined using the least-square method for various values of $m$. Some of these values are listed in table \ref{Cmfor52knot} and the variation of $\ln C_m(5_2)$ with $m$ is shown in figure \ref{Cvalues52knot}.
\begin{table}
	\begin{minipage}{0.4\linewidth}
		\centering
		\begin{tabular}{|c|c|} \hline \rowcolor{Grayy}
			$m$ & $C_m(5_2)$  \\ \hline
\footnotesize{1} & \footnotesize{$4.5156047311 \pm 0.0080192291$}  \\
\footnotesize{2} & \footnotesize{$19.522561281 \pm 0.0323263046$} \\
\footnotesize{3} & \footnotesize{$85.809660629 \pm 0.1375626362$} \\
\footnotesize{4} & \footnotesize{$377.22537837 \pm 0.5933078608$} \\
\footnotesize{5} & \footnotesize{$1658.35734226 \pm 2.5705101499$} \\
\footnotesize{6} & \footnotesize{$7290.66518025 \pm 11.154379776$} \\
\footnotesize{7} & \footnotesize{$32052.9326226 \pm 48.419257460$} \\
\footnotesize{8} & \footnotesize{$140922.2991482 \pm 210.12155723$} \\
\footnotesize{9} & \footnotesize{$619587.6221496 \pm 911.29217860$} \\
\footnotesize{10} & \footnotesize{$2724184.491914 \pm 3949.0717321$} \\
\hline
		\end{tabular}
		\caption{Values of $C_m(5_2)$}
		\label{Cmfor52knot}
	\end{minipage}\hfill
	\begin{minipage}{0.58\linewidth}
		\centering
		\includegraphics[width=98mm]{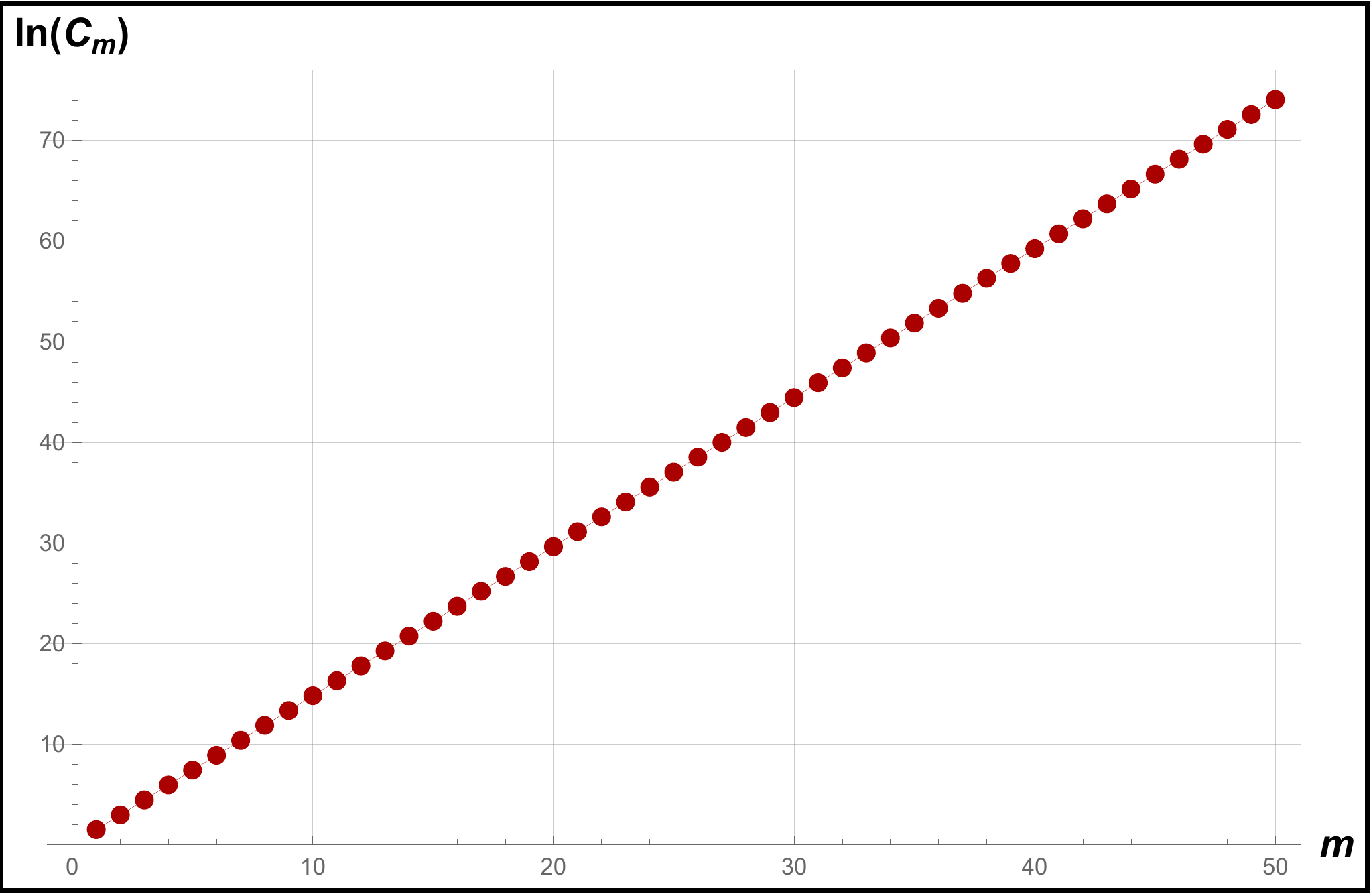}
		\captionof{figure}{Variation of $\ln C_m(5_2)$ with $m$}
		\label{Cvalues52knot}
	\end{minipage}
\end{table}
In figure \ref{traces52knot}, we plot the function $C_m(5_2) \,  k^{3m}$ and compare it with the numerical values of $F_m(5_2)$ obtained explicitly from \eqref{Fmactualvalues}.
The predicted $k \to \infty$ values of the R\'enyi entropies associated with the state $\ket{5_2 \# T_{2,2}}$ can be computed using \eqref{limit-Renyi-proposal} and we have listed some of these values in table \ref{Renyipredicted52knot}. We also show the variation of the R\'enyi entropies with $k$ in the plots in figure \ref{REvsk52knot}. We see that these plots tend to converge to the predicted $k \to \infty$ values of the R\'enyi entropies.  In the rest of this section, we address six crossing non-torus knots to verify
our conjecture and proposal.
\begin{table}
	\begin{minipage}{0.4\linewidth}
		\centering
		\begin{tabular}{|c|c|} \hline \rowcolor{Grayy}
			$m$ & $\lim_{k \to\infty} \mathcal{R}_m$  \\ \hline
\footnotesize{2} & \footnotesize{$0.0435074467 \pm 0.0039188003$}  \\
\footnotesize{3} & \footnotesize{$0.0352428767 \pm 0.0027818216$}  \\
\footnotesize{4} & \footnotesize{$0.0324378784 \pm 0.0024252028$}  \\
\footnotesize{5} & \footnotesize{$0.0310281855 \pm 0.0022534345$}  \\
\footnotesize{6} & \footnotesize{$0.0301769261 \pm 0.0021529270$}  \\
\footnotesize{7} & \footnotesize{$0.0296049755 \pm 0.0020871156$}  \\
\footnotesize{8} & \footnotesize{$0.0291927116 \pm 0.0020407385$}  \\
\footnotesize{9} & \footnotesize{$0.0288803294 \pm 0.0020063206$}  \\
\footnotesize{10} & \footnotesize{$0.0286346099 \pm 0.0019797771$}  \\
\hline
		\end{tabular}
		\caption{The $k \to \infty$ values of the R\'enyi entropies associated with the state $\ket{5_2 \# T_{2,2}}$.}
		\label{Renyipredicted52knot}
	\end{minipage}\hfill
	\begin{minipage}{0.58\linewidth}
		\centering
		\includegraphics[width=95mm]{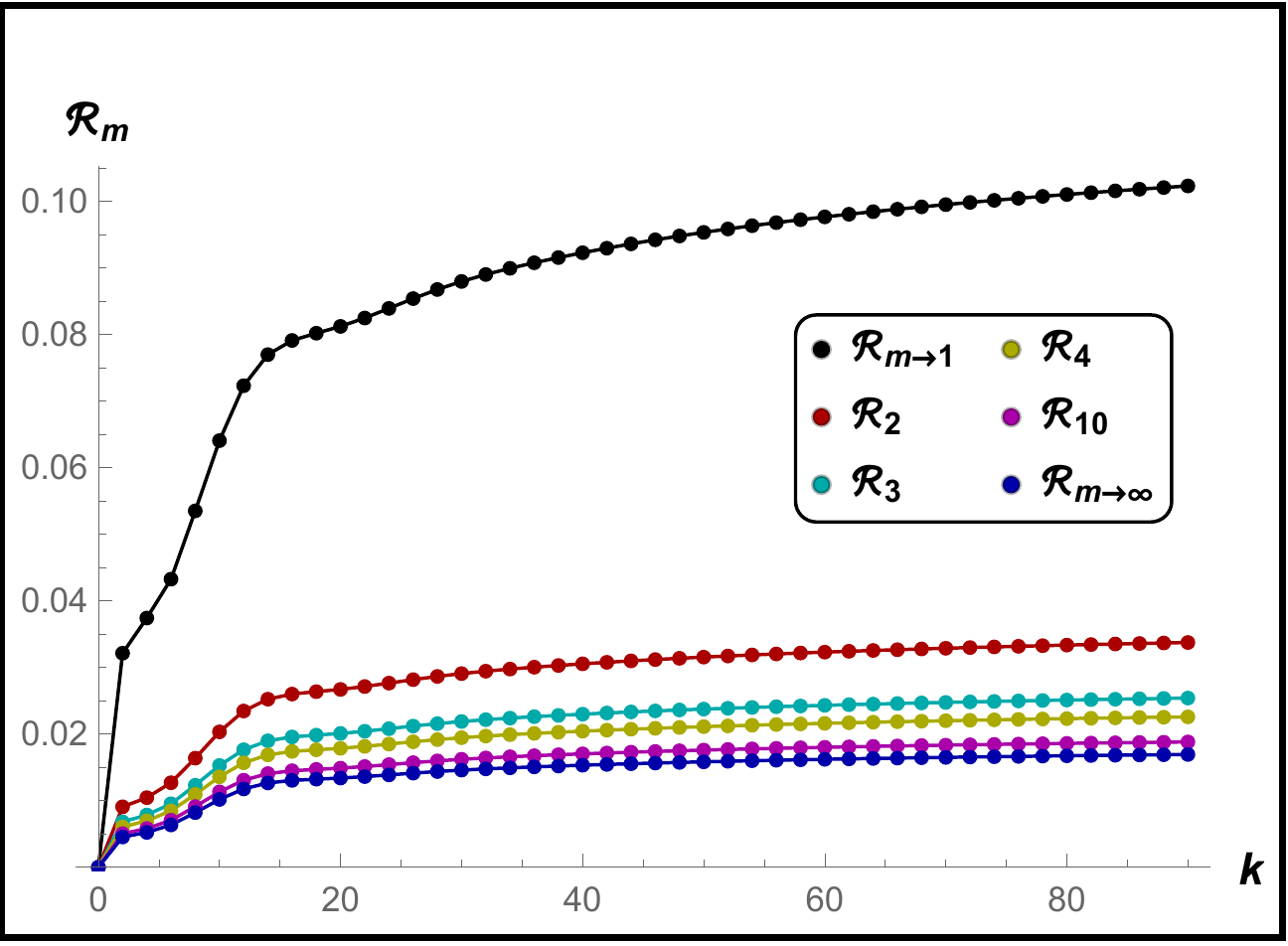}
		\captionof{figure}{Variation of R\'enyi entropies for the state $\ket{5_2 \# T_{2,2}}$ with $k$.}
		\label{REvsk52knot}
	\end{minipage}
\end{table}
\subsubsection{Stevedore knot: $6_1$}
\begin{figure}[htbp]
	\centering
		\includegraphics[width=1.00\textwidth]{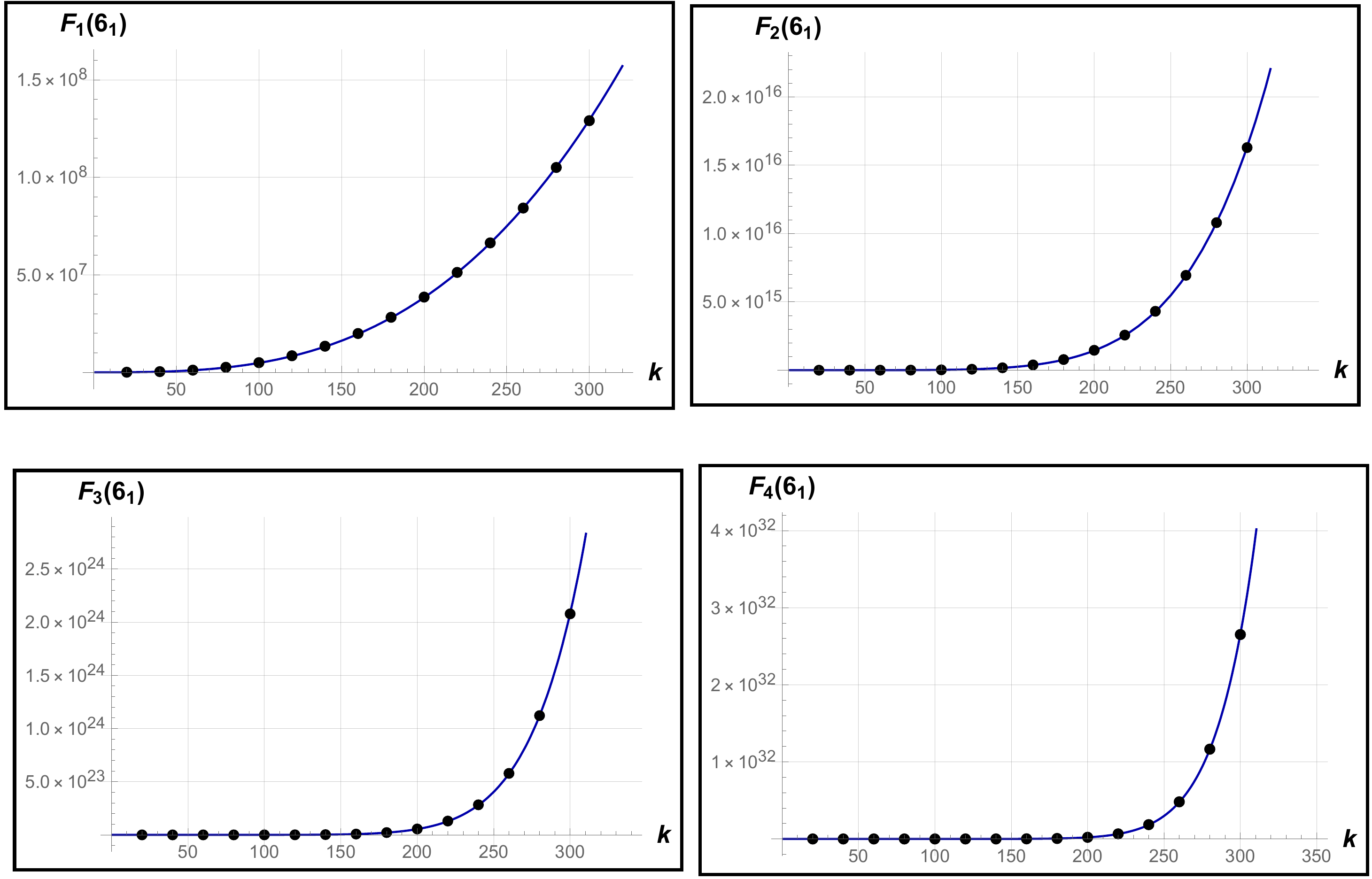}
	\caption{The variation of the function $F_m$ with $k$ for $6_1$ knot. The solid line denotes the function $C_m k^{3m}$ with values of $C_m$ given in table \ref{Cmfor61knot}. The $\bullet$ indicates the numerical values of $F_m$ obtained from explicit computations.}
	\label{traces61knot}
\end{figure}

This is a non-torus knot whose knot complement volume is $\text{Vol}(S^3 \backslash 6_1) =  3.16396322888$. In fact, computing $F_m(6_1)$ \eqref{Fmactualvalues} and relating to the form
\begin{align}
F_m(6_1) \equiv \frac{\text{Tr}[\sigma^m(6_1 \# T_{2,2})]}{\exp\left(\frac{m\,\text{Vol}(S^3 \backslash 6_1)}{2\pi}k\right)} =  C_m(6_1) \,  k^{3m} ~,
\label{}
\end{align}
and using the least-square method, we computed $C_m(6_1)$ for various values of $m$. Some of these values are listed in table \ref{Cmfor61knot} and the variation of $\ln C_m(6_1)$ with $m$ is shown in figure \ref{Cvalues61knot}.
\begin{table}
	\begin{minipage}{0.4\linewidth}
		\centering
		\begin{tabular}{|c|c|} \hline \rowcolor{Grayy}
			$m$ & $C_m(6_1)$  \\ \hline
\footnotesize{1} & \footnotesize{$4.7923232325 \pm 0.0030931604$}  \\
\footnotesize{2} & \footnotesize{$22.3923919843 \pm 0.0135258863$} \\
\footnotesize{3} & \footnotesize{$105.7984752566 \pm 0.0616708792$} \\
\footnotesize{4} & \footnotesize{$499.9052116194 \pm 0.2842710891$} \\
\footnotesize{5} & \footnotesize{$2362.131219566 \pm 1.3135290735$} \\
\footnotesize{6} & \footnotesize{$11161.64299789 \pm 6.0672497245$} \\
\footnotesize{7} & \footnotesize{$52742.37636985 \pm 27.982406758$} \\
\footnotesize{8} & \footnotesize{$249228.8880632 \pm 128.78990106$} \\
\footnotesize{9} & \footnotesize{$1177724.7827384 \pm 591.37837422$} \\
\footnotesize{10} & \footnotesize{$5565389.9109037 \pm 2708.8286274$} \\
\hline
		\end{tabular}
		\caption{Values of $C_m(6_1)$}
		\label{Cmfor61knot}
	\end{minipage}\hfill
	\begin{minipage}{0.58\linewidth}
		\centering
		\includegraphics[width=98mm]{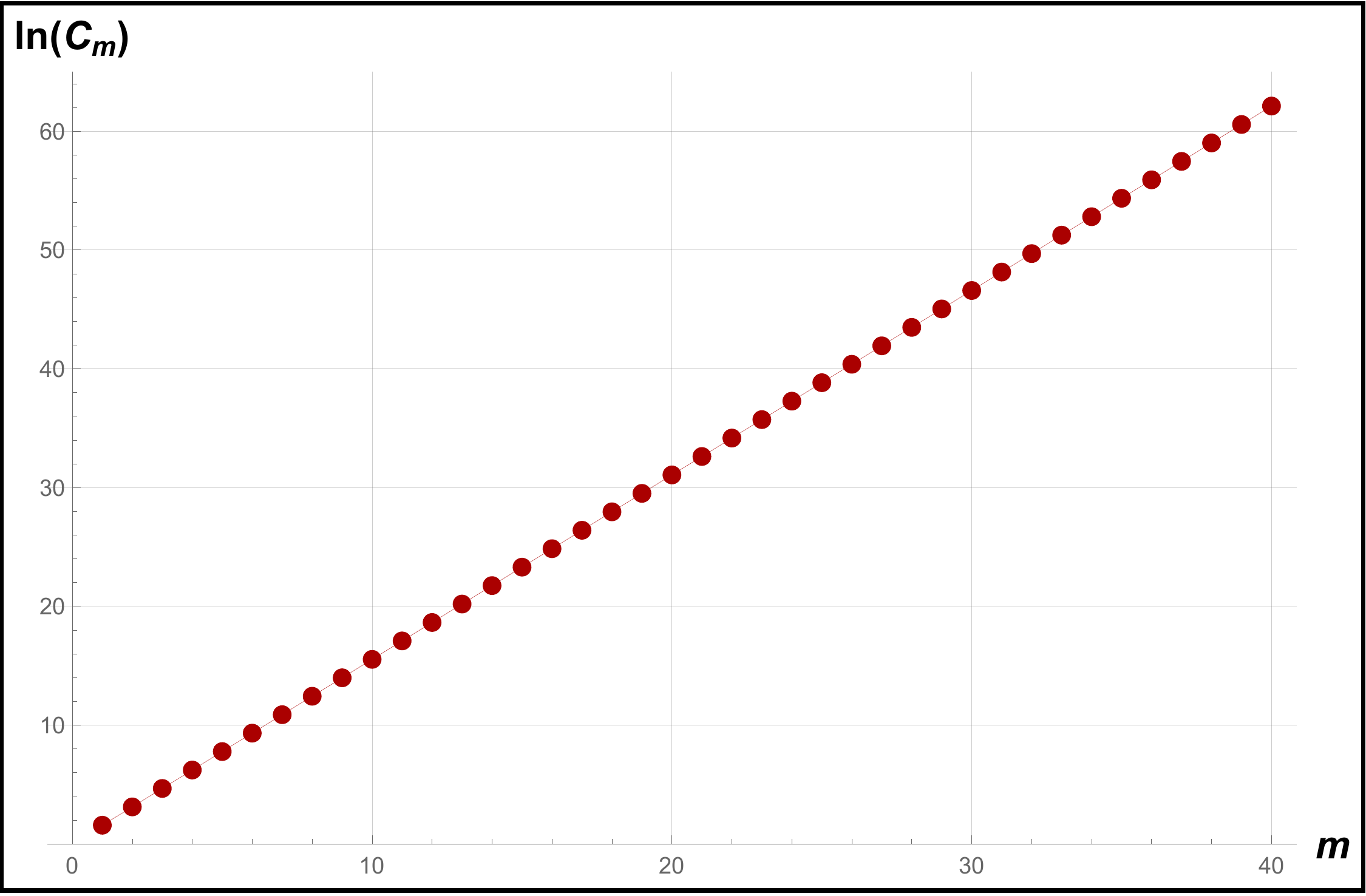}
		\captionof{figure}{Variation of $\ln C_m(6_1)$ with $m$}
		\label{Cvalues61knot}
	\end{minipage}
\end{table}
In figure \ref{traces61knot}, we plot the function $C_m(6_1) \,  k^{3m}$ and compare it with the numerical values of $F_m(6_1)$ obtained explicitly from \eqref{Fmactualvalues}. 
The predicted $k \to \infty$ values of the R\'enyi entropies associated with the state $\ket{6_1 \# T_{2,2}}$ can be computed using \eqref{limit-Renyi-proposal} and we have listed some of these values in table \ref{Renyipredicted61knot}. We also show the variation of the R\'enyi entropies with $k$ in the plots in figure \ref{REvsk61knot}. We see that these plots tend to converge to the predicted $k \to \infty$ values of the R\'enyi entropies.
\begin{table}
	\begin{minipage}{0.4\linewidth}
		\centering
		\begin{tabular}{|c|c|} \hline \rowcolor{Grayy}
			$m$ & $\lim_{k \to\infty} \mathcal{R}_m$  \\ \hline
\footnotesize{2} & \footnotesize{$0.0253093645 \pm 0.0014252152$}  \\
\footnotesize{3} & \footnotesize{$0.0197549127 \pm 0.0010110794$}  \\
\footnotesize{4} & \footnotesize{$0.0178809135 \pm 0.0008812152$}  \\
\footnotesize{5} & \footnotesize{$0.0169392514 \pm 0.0008186904$}  \\
\footnotesize{6} & \footnotesize{$0.0163706838 \pm 0.0007821215$}  \\
\footnotesize{7} & \footnotesize{$0.0159887767 \pm 0.0007581881$}  \\
\footnotesize{8} & \footnotesize{$0.0157136437 \pm 0.0007413313$}  \\
\footnotesize{9} & \footnotesize{$0.0155053518 \pm 0.0007288286$}  \\
\footnotesize{10} & \footnotesize{$0.0153417228 \pm 0.0007191926$}  \\
\hline
		\end{tabular}
		\caption{The $k \to \infty$ values of the R\'enyi entropies associated with the state $\ket{6_1 \# T_{2,2}}$.}
		\label{Renyipredicted61knot}
	\end{minipage}\hfill
	\begin{minipage}{0.58\linewidth}
		\centering
		\includegraphics[width=95mm]{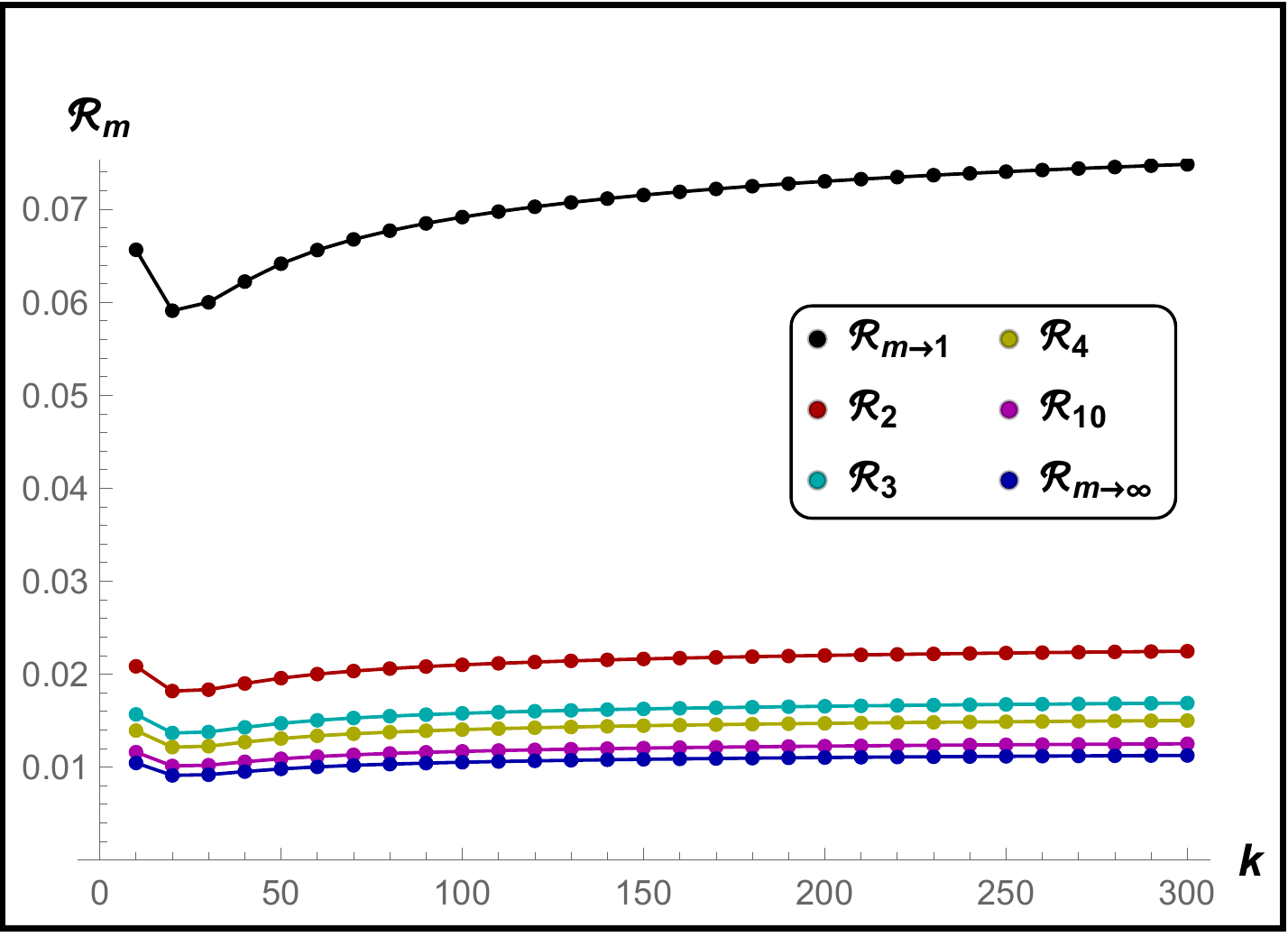}
		\captionof{figure}{Variation of R\'enyi entropies for the state $\ket{6_1 \# T_{2,2}}$ with $k$.}
		\label{REvsk61knot}
	\end{minipage}
\end{table}
\subsubsection{$6_2$ knot}
\begin{figure}[htbp]
	\centering
		\includegraphics[width=1.00\textwidth]{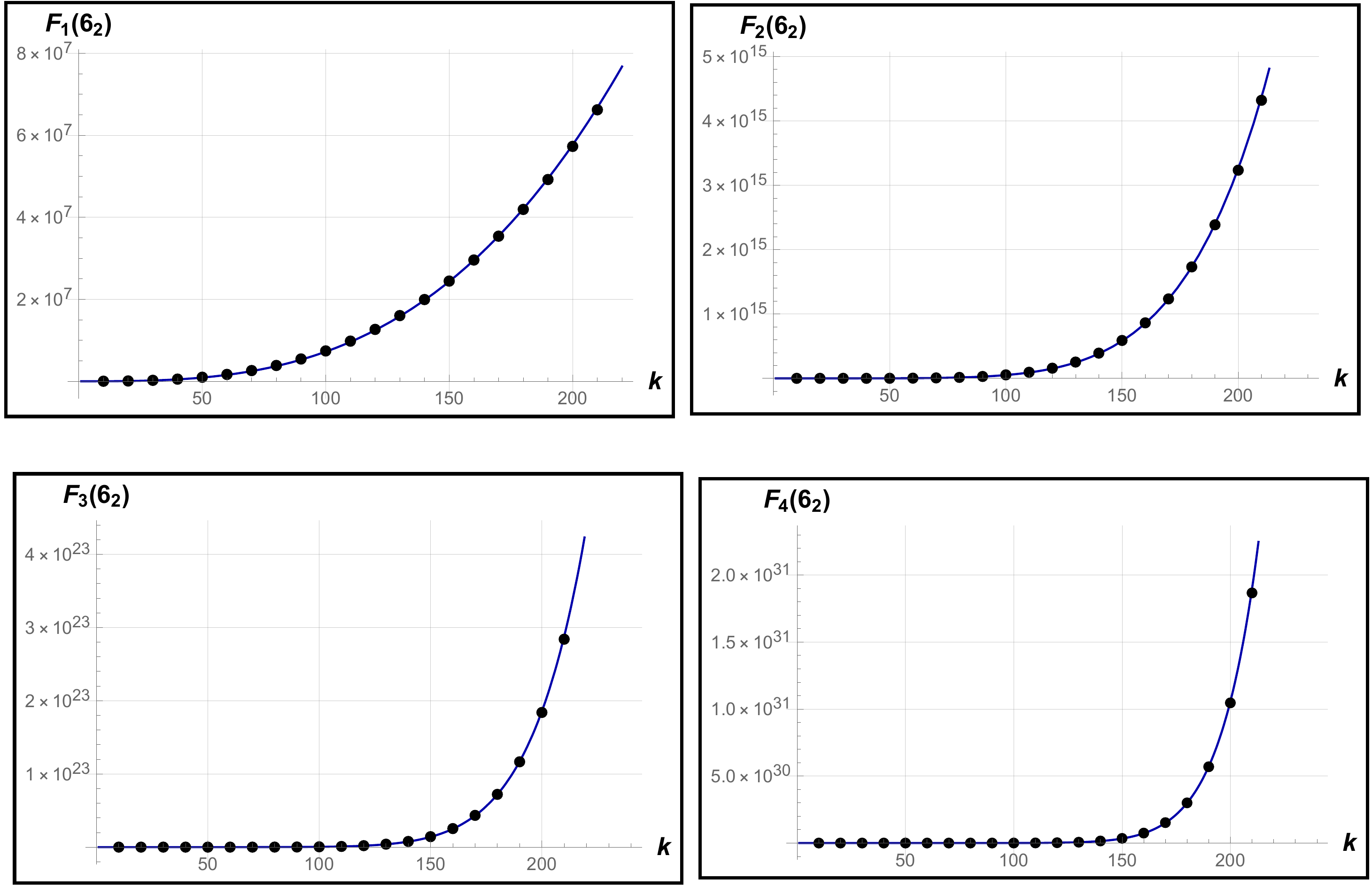}
	\caption{The variation of the function $F_m$ with $k$ for $6_2$ knot. The solid line denotes the function $C_m k^{3m}$ with values of $C_m$ given in table \ref{Cmfor62knot}. The $\bullet$ indicates the numerical values of $F_m$ obtained from explicit computations.}
	\label{traces62knot}
\end{figure}

The hyperbolic volume is $\text{Vol}(S^3 \backslash 6_2) = 4.400832516$ for the complement of the non-torus knot $6_2$. We computed $F_m(6_2)$ \eqref{Fmactualvalues} and compared with the proposed form  
\begin{align}
F_m(6_2) \equiv \frac{\text{Tr}[\sigma^m(6_2 \# T_{2,2})]}{\exp\left(\frac{m\,\text{Vol}(S^3 \backslash 6_2)}{2\pi}k\right)} =  C_m(6_2) \,  k^{3m}
\label{}
\end{align}
to deduce $C_m(6_2)$ for various values of $m$ using least-square method. Some of these values are listed in table \ref{Cmfor62knot} and the variation of $\ln C_m(6_2)$ with $m$ is shown in figure \ref{Cvalues62knot}.
\begin{table}
	\begin{minipage}{0.4\linewidth}
		\centering
		\begin{tabular}{|c|c|} \hline \rowcolor{Grayy}
			$m$ & $C_m(6_2)$  \\ \hline
\footnotesize{1} & \footnotesize{$7.2078662291 \pm 0.0114864102$}  \\
\footnotesize{2} & \footnotesize{$50.9252866533 \pm 0.0745214114$} \\
\footnotesize{3} & \footnotesize{$362.4041343538 \pm 0.5060376066$} \\
\footnotesize{4} & \footnotesize{$2579.2833524324 \pm 3.4585013119$} \\
\footnotesize{5} & \footnotesize{$18358.617056 \pm 23.58889$} \\
\footnotesize{6} & \footnotesize{$130681.2420 \pm 160.1351$} \\
\footnotesize{7} & \footnotesize{$930284.933 \pm 1080.959$} \\
\footnotesize{8} & \footnotesize{$6622851.572 \pm 7253.540$} \\
\footnotesize{9} & \footnotesize{$47151678.319 \pm 48386.251$} \\
\footnotesize{10} & \footnotesize{$335713848.188 \pm 320930.180$} \\
\hline
		\end{tabular}
		\caption{Values of $C_m(6_2)$}
		\label{Cmfor62knot}
	\end{minipage}\hfill
	\begin{minipage}{0.58\linewidth}
		\centering
		\includegraphics[width=98mm]{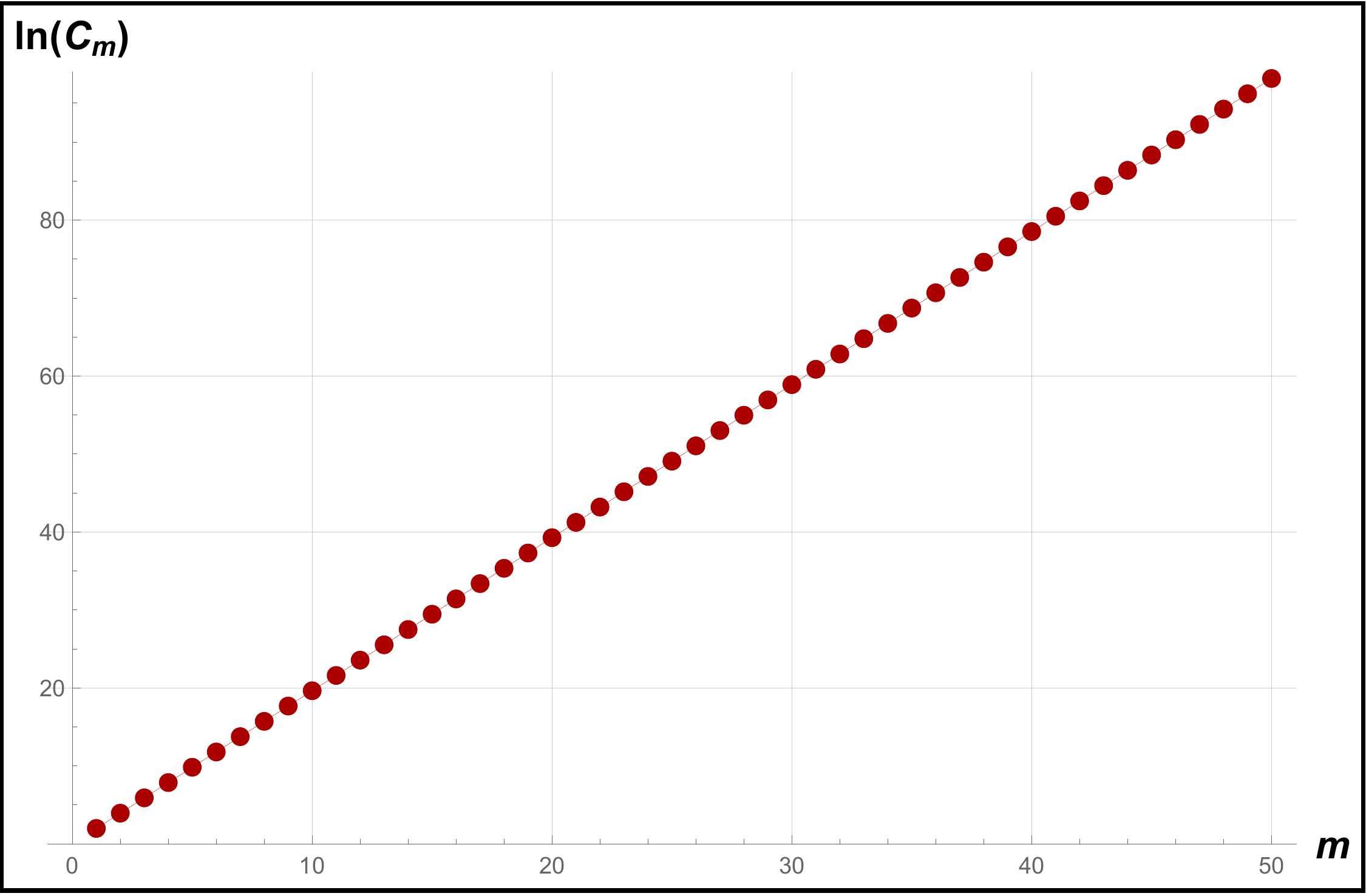}
		\captionof{figure}{Variation of $\ln C_m(6_2)$ with $m$}
		\label{Cvalues62knot}
	\end{minipage}
\end{table}
In figure \ref{traces62knot}, we plot the function $C_m(6_2) \,  k^{3m}$ and compare it with the numerical values of $F_m(6_2)$ obtained explicitly from \eqref{Fmactualvalues}. 
The predicted $k \to \infty$ values of the R\'enyi entropies associated with the state $\ket{6_2 \# T_{2,2}}$ can be computed using \eqref{limit-Renyi-proposal} and we have listed some of these values in table \ref{Renyipredicted62knot}. We also show the variation of the R\'enyi entropies with $k$ in the plots in figure \ref{REvsk62knot}. We see that these plots tend to converge to the predicted $k \to \infty$ values of the R\'enyi entropies.
\begin{table}
	\begin{minipage}{0.4\linewidth}
		\centering
		\begin{tabular}{|c|c|} \hline \rowcolor{Grayy}
			$m$ & $\lim_{k \to\infty} \mathcal{R}_m$  \\ \hline
\footnotesize{2} & \footnotesize{$0.0199863319 \pm 0.0035070715$}  \\
\footnotesize{3} & \footnotesize{$0.0163794510 \pm 0.0024902620$}  \\
\footnotesize{4} & \footnotesize{$0.0151416590 \pm 0.0021712926$}  \\
\footnotesize{5} & \footnotesize{$0.0145026174 \pm 0.0020177258$}  \\
\footnotesize{6} & \footnotesize{$0.0141042798 \pm 0.0019279527$}  \\
\footnotesize{7} & \footnotesize{$0.0138274220 \pm 0.0018692517$}  \\
\footnotesize{8} & \footnotesize{$0.0136210151 \pm 0.0018279583$}  \\
\footnotesize{9} & \footnotesize{$0.0134595614 \pm 0.0017973760$}  \\
\footnotesize{10} & \footnotesize{$0.0133288780 \pm 0.0017738427$}  \\
\hline
		\end{tabular}
		\caption{The $k \to \infty$ values of the R\'enyi entropies associated with the state $\ket{6_2 \# T_{2,2}}$.}
		\label{Renyipredicted62knot}
	\end{minipage}\hfill
	\begin{minipage}{0.58\linewidth}
		\centering
		\includegraphics[width=95mm]{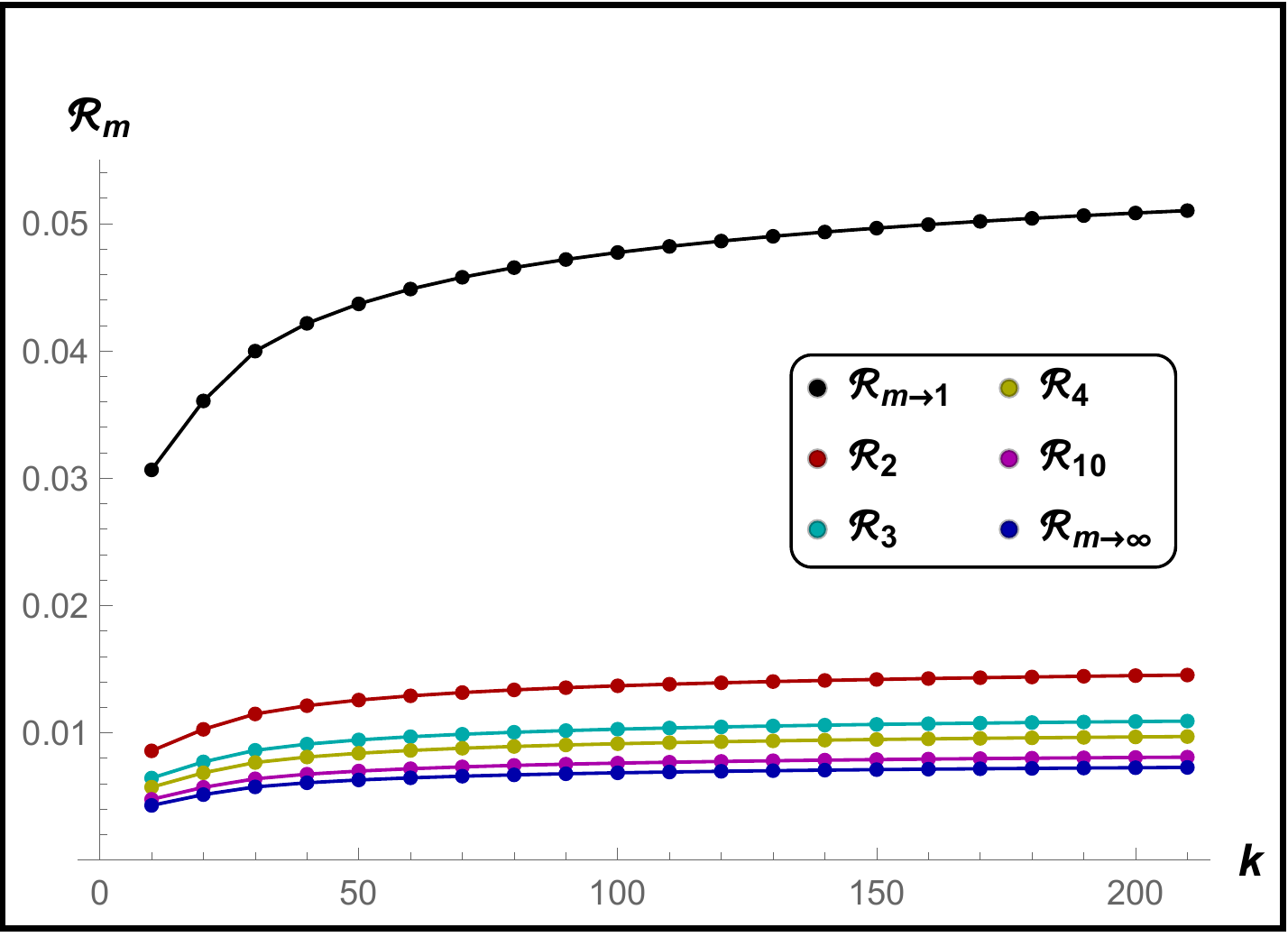}
		\captionof{figure}{Variation of R\'enyi entropies for the state $\ket{6_2 \# T_{2,2}}$ with $k$.}
		\label{REvsk62knot}
	\end{minipage}
\end{table}
\subsubsection{$6_3$ knot}
\begin{figure}[htbp]
	\centering
		\includegraphics[width=1.00\textwidth]{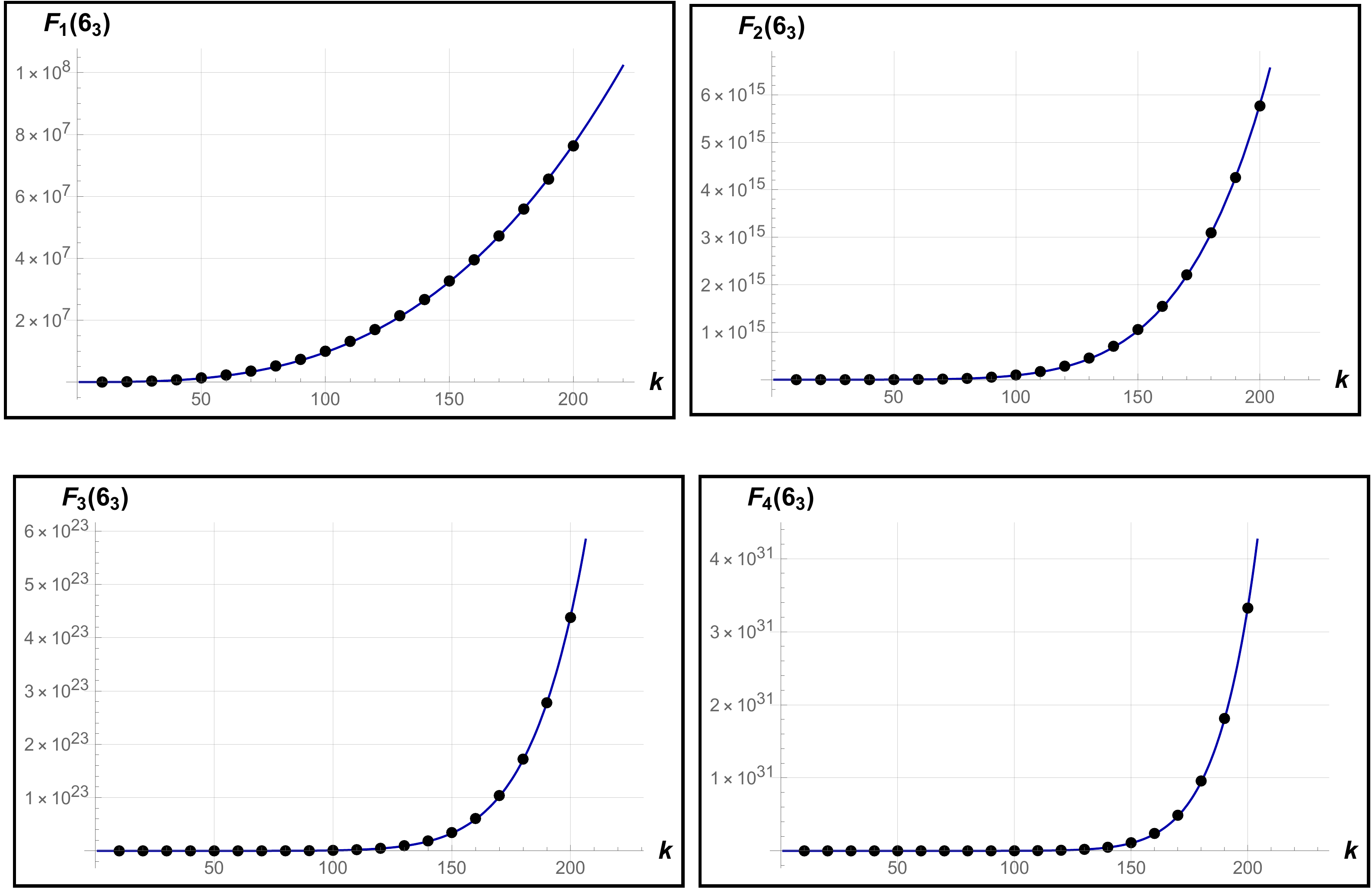}
	\caption{The variation of the function $F_m$ with $k$ for $6_3$ knot. The solid line denotes the function $C_m k^{3m}$ with values of $C_m$ given in table \ref{Cmfor63knot}. The $\bullet$ indicates the numerical values of $F_m$ obtained from explicit computations.}
	\label{traces63knot}
\end{figure}

Finally, we address similar analysis for $6_3$ knot whose knot complement volume is $\text{Vol}(S^3 \backslash 6_3) = 5.69302109$. Thus, working out $F_m(6_3)$ \eqref{Fmactualvalues} and comparing with  
\begin{align}
F_m(6_3) \equiv \frac{\text{Tr}[\sigma^m(6_3 \# T_{2,2})]}{\exp\left(\frac{m\,\text{Vol}(S^3 \backslash 6_3)}{2\pi}k\right)} =  C_m(6_3) \,  k^{3m} ~,
\label{}
\end{align}
the constant $C_m(6_3)$ can be deduced for various values of $m$ using the least-square method. Some of these values are listed in table \ref{Cmfor63knot} and the variation of $\ln C_m(6_3)$ with $m$ is shown in figure \ref{Cvalues63knot}.
\begin{table}
	\begin{minipage}{0.4\linewidth}
		\centering
		\begin{tabular}{|c|c|} \hline \rowcolor{Grayy}
			$m$ & $C_m(6_3)$  \\ \hline
\footnotesize{1} & \footnotesize{$9.5970632893 \pm 0.0185978565$}  \\
\footnotesize{2} & \footnotesize{$90.554839418 \pm 0.160095501$} \\
\footnotesize{3} & \footnotesize{$858.76989699 \pm 1.44999102$} \\
\footnotesize{4} & \footnotesize{$8144.95594531 \pm 13.23524518$} \\
\footnotesize{5} & \footnotesize{$77257.4532157 \pm 120.6783625$} \\
\footnotesize{6} & \footnotesize{$732873.448104 \pm 1096.079889$} \\
\footnotesize{7} & \footnotesize{$6952667.48358 \pm 9906.38029$} \\
\footnotesize{8} & \footnotesize{$65963622.8502 \pm 89061.0909$} \\
\footnotesize{9} & \footnotesize{$625870830.613 \pm 796416.850$} \\
\footnotesize{10} & \footnotesize{$5938663602.54 \pm 7084757.98$} \\
\hline
		\end{tabular}
		\caption{Values of $C_m(6_3)$}
		\label{Cmfor63knot}
	\end{minipage}\hfill
	\begin{minipage}{0.58\linewidth}
		\centering
		\includegraphics[width=98mm]{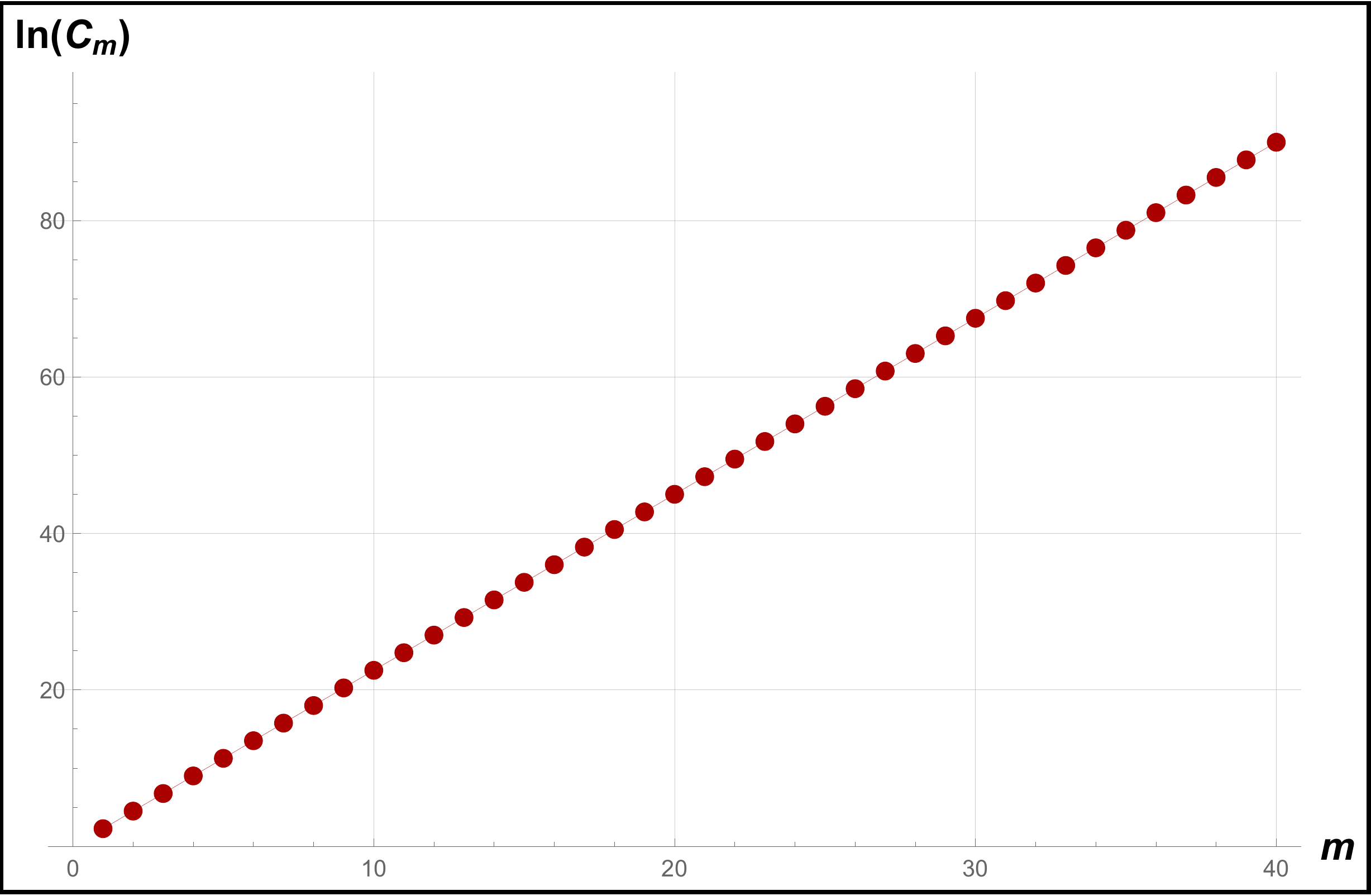}
		\captionof{figure}{Variation of $\ln C_m(6_3)$ with $m$}
		\label{Cvalues63knot}
	\end{minipage}
\end{table}
In figure \ref{traces63knot}, we plot the function $C_m(6_3) \,  k^{3m}$ and compare it with the numerical values of $F_m(6_3)$ obtained explicitly from \eqref{Fmactualvalues}. 
The predicted $k \to \infty$ values of the R\'enyi entropies associated with the state $\ket{6_3 \# T_{2,2}}$ can be computed using \eqref{limit-Renyi-proposal} and we have listed some of these values in table \ref{Renyipredicted63knot}. We also show the variation of the R\'enyi entropies with $k$ in the plots in figure \ref{REvsk63knot}. We see that these plots tend to converge to the predicted $k \to \infty$ values of the R\'enyi entropies.

These numerical analysis and plots for all the prime knots up to six crossings indeed
validate our conjecture and proposal within SO(3) Chern-Simons theory.
\begin{table}
	\begin{minipage}{0.4\linewidth}
		\centering
		\begin{tabular}{|c|c|} \hline \rowcolor{Grayy}
			$m$ & $\lim_{k \to\infty} \mathcal{R}_m$  \\ \hline
\footnotesize{2} & \footnotesize{$0.0169586610 \pm 0.0042599254$}  \\
\footnotesize{3} & \footnotesize{$0.0144352100 \pm 0.0030269172$}  \\
\footnotesize{4} & \footnotesize{$0.0135581550 \pm 0.0026399898$}  \\
\footnotesize{5} & \footnotesize{$0.0130967624 \pm 0.0024536120$}  \\
\footnotesize{6} & \footnotesize{$0.0128029098 \pm 0.0023446021$}  \\
\footnotesize{7} & \footnotesize{$0.0125940093 \pm 0.0022732852$}  \\
\footnotesize{8} & \footnotesize{$0.0124347393 \pm 0.0022230911$}  \\
\footnotesize{9} & \footnotesize{$0.0123074665 \pm 0.0021858981$}  \\
\footnotesize{10} & \footnotesize{$0.0122023867 \pm 0.0021572646$}  \\
\hline
		\end{tabular}
		\caption{The $k \to \infty$ values of the R\'enyi entropies associated with the state $\ket{6_3 \# T_{2,2}}$.}
		\label{Renyipredicted63knot}
	\end{minipage}\hfill
	\begin{minipage}{0.58\linewidth}
		\centering
		\includegraphics[width=95mm]{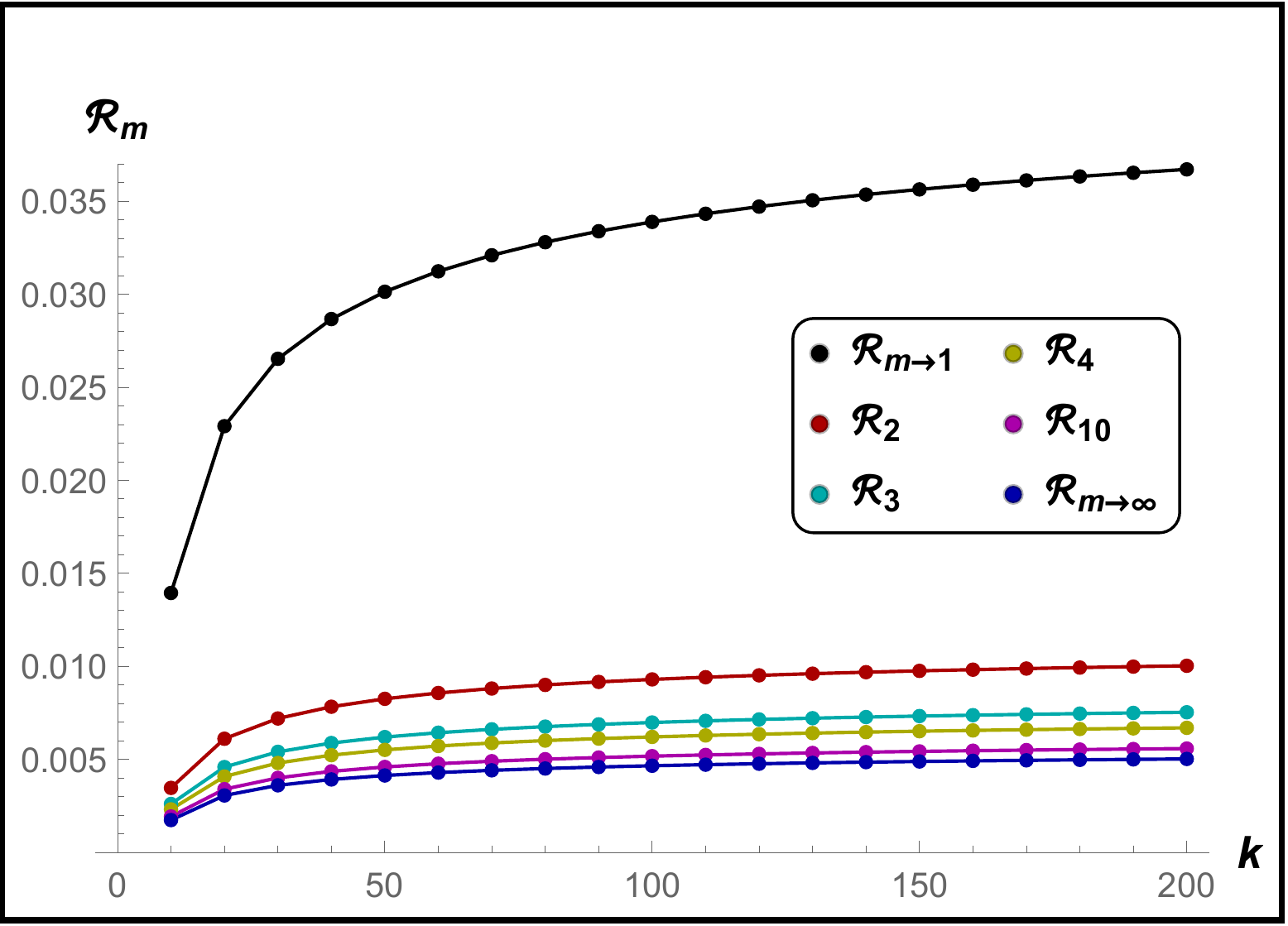}
		\captionof{figure}{Variation of R\'enyi entropies for the state $\ket{6_3 \# T_{2,2}}$ with $k$.}
		\label{REvsk63knot}
	\end{minipage}
\end{table}
\section{Conclusion} \label{sec5}
In this paper, our primary goal was to capture the geometrical features of the three-manifolds from the study of entanglement properties of bi-partite states. Particularly, we focused on the Chern-Simons theory to study the topological entanglement between two torus boundaries of a three-manifold $S^3 \backslash \mathcal L$ (the link complement of a two-component link $\mathcal{L}$). Interestingly, for a class of  two-component links viewed as a connected sum of a prime knot $\mathcal{K}$ and the Hopf link ($\mathcal{L} = \mathcal{K}\, \# \, T_{2,2}$), the probability amplitudes of the state $\ket{\mathcal{K}\, \# \, T_{2,2}}$ are proportional to the quantum invariants of the knot $\mathcal{K}$. Thus the trace of the reduced density matrix \eqref{traceKconnectT22} is almost similar to the Turaev-Viro invariant \cite{zbMATH00167790, detcherry2018turaev} of $S^3 \backslash \mathcal{K}$. This motivated us to study the reduced density matrices for this class of two-component links for both SU(2) and SO(3) gauge groups.

The partial tracing of one of the torus boundary gives the reduced density matrix $\sigma(\mathcal{K} \# T_{2,2})$ \eqref{rhoGeneral}, which provides the starting point for computing various entanglement measures. The R\'enyi entropy with index $m$ can be obtained from the $m$-moment or the trace of the $m^{\text{th}}$ power of the matrix $\sigma(\mathcal{K} \# T_{2,2})$ using \eqref{Renyi-conclusion}. Incidentally, the trace operation on the reduced density matrix can be viewed as gluing the two oppositely oriented torus boundaries resulting in a closed three-manifold $M_{\mathcal K}$  giving the  Chern-Simons partition function  $Z(M_{\mathcal K})=\text{Tr}[\sigma(\mathcal K \# T_{2,2})]$. The subscript on the three-manifold $M_{\mathcal K}$ keeps track of the knot involved in the two-component link $\mathcal L= \mathcal K \# T_{2,2}$. We would like to mention that the authors of the Ref.\cite{Gang:2017cwq} study the perturbative expansion of the SL$(2,\mathbb{C})$ Chern-Simons partition function of closed three-manifolds which are obtained from the Dehn filling of link complements. It would be worthwhile to explore the results of \cite{Gang:2017cwq} in the set-up of our paper. In particular, it will be interesting to see if the manifold $M_{\mathcal K}$ can be viewed as the Dehn filling of certain link complement which can enable us to obtain the subleading terms in the expansion of $Z(M_{\mathcal K})$. We leave this discussion for future work.

We were able to reproduce similar geometrical results for the higher moments of the reduced density matrix. 
Our crucial observation that the replica trick in the entanglement context is quantified by the connected sums of knots led to \eqref{traceconnectsum} and \eqref{ZMKm} respectively. Thus we have shown that all the $m$-moments of the matrix $\sigma(\mathcal{K} \# T_{2,2})$ are the Chern-Simons partition functions $Z(M_{\mathcal K_m})$ where $\mathcal K_m =\mathcal K \# \mathcal K \ldots \# \mathcal K$ denotes the connected sum of $m$-copies of $\mathcal K$. As a result, the R\'enyi entropies with index $m$ can be written in terms of the three-manifold invariants $Z(M_{\mathcal K_m})$ as given in \eqref{RenyiActualValues}.

Even though we cannot visualise the explicit topology of $M_{\mathcal K}$, we were able to address the large $k$ asymptotics of $Z(M_{\mathcal K})$ for SU(2) and SO(3) gauge groups. Using the asymptotics of Turaev-Viro invariant of $S^3 \backslash \mathcal{K}$ evaluated at $e^{\frac{\pi i}{k+2}}$, we were able to show in \eqref{SU2largekTV}, that $\text{Tr}[\sigma(\mathcal K \# T_{2,2})] = Z(M_{\mathcal K})$ for SU(2) group can grow at most polynomially in $k$. In contrast, the traces show an exponential growth in $k$ for the SO(3) group. Incorporating the well known Kashaev's conjecture \cite{kashaev1997hyperbolic}, we conjecture that the leading term of $\ln \text{Tr}[\sigma(\mathcal K \# T_{2,2})]$ and hence $\ln Z(M_{\mathcal K})$ for SO(3) gauge group, captures the hyperbolic volume of the knot complement $S^3 \backslash \mathcal{K}$ in the large $k$ limit. These are given as conjecture \eqref{SO3tracelimit} and corollary \eqref{corollary-eq} respectively.  We have performed numerical computations for torus and non-torus knots up to six crossings to validate our conjecture and proposal. 

We further propose an asymptotic form for 
$\text{Tr}[\sigma^m(\mathcal{K} \# T_{2,2})]=Z(M_{\mathcal K_m})$ for SO(3) gauge group, in the large  $k$ limit \eqref{SO3traceAsymptotic}. Using such a form, we see that the R\'enyi entropies converge to a finite number as $k \to \infty$ \eqref{limit-Renyi-proposal}. We would like to mention that currently we do not have an understanding of the geometric or topological interpretation of the large $k$ limiting values of R\'enyi entropies, and it will be worthwhile to explore this aspect in the context of SO(3) Chern-Simons theories.


\acknowledgments
SD is supported by the NSFC grants No. 12050410249 and No. 11975158. PR would like to thank SERB (MATRICS) MTR/2019/000956 funding. VKS would like to thank Rama Mishra for helpful discussion and acknowledge the hospitality of department of Mathematics, IISER, Pune (India) where this work was done during his visit as visiting fellow. BPM acknowledges the Research Grant for Faculty under IoE Scheme (number 6031).
\appendix
\section{SO(3) Chern-Simons theory and hyperbolic volumes of $S^3 \backslash \mathcal{K}$}
\label{appA}
In the appendix, we will verify the following conjecture for some of the prime knots: 
\begin{equation}
\lim_{k \to \infty} \frac{\ln\, \abs{J_{k}(\mathcal{K}\,;e^{\frac{4\pi i}{k+1}})}}{k} = \frac{\text{Vol}(S^3 \backslash \mathcal{K})}{4\pi} ~.
\end{equation}
For this, we use the equality between the colored Jones and Kashaev's invariant. In particular, setting $n=(k+1)$ with $k \in 2\mathbb{Z}$, we have:
\begin{equation}
\abs{\text{Kas}_{n}(\mathcal{K}\,;q=e^{\frac{4\pi i}{n}})} = \abs{J_{n-1}(\mathcal{K}\,;q=e^{\frac{4\pi i}{n}})} ~.
\end{equation}
This has already been discussed in section \ref{sec3} around eq.\eqref{Kas=Jones}. So, we study the asymptotics of the above Kashaev's invariant for prime knots up to six crossings using the stationary phase approximation prescribed in \cite{kashaev1997hyperbolic,hikami2003volume}. This approach works for those knots whose invariant can be written in terms of $q$-Pochhammer symbols
\begin{equation} 
(q)_{\ell} = \prod_{j=1}^{\ell} (1-q^j) ~.
\end{equation}
We consider the limit
\begin{equation} 
q=\exp(\frac{4\pi i}{n}) \quad,\quad n \to \infty ~.
\label{limittaken}
\end{equation}
The aim is to obtain the maximum summand that dominates the asymptotic of the Kashaev's invariant in the above limit. For this, the summations appearing in the Kashaev's invariant are converted into integral expressions of the following form: 
\begin{equation} 
\text{Kas}_{n}(\mathcal{K}\,;q=e^{\frac{4\pi i}{n}}) \sim \iint\ldots\int dx_1 dx_2 \ldots dx_p\,\exp(\frac{n}{4\pi i}\,V_{\mathcal{K}}(x_1,\ldots,x_p)) ~,
\label{Kash-Asymp}
\end{equation}
where the integration variable $x_i = q^{a_i}$ with $a_i$ being the summation variable in the Kashaev's invariant. The potential $V_{\mathcal{K}}$ is formally obtained by replacing each of the $q$-Pochhammer symbols appearing in the Kashaev's invariant by their asymptotic expressions. This is done by approximating the $q$-Pochhammer by a dilogarithm function as shown below:
\begin{align} 
\ln\, (q)_\ell &= \sum_{j=1}^{\ell} \ln(1- e^{\frac{4\pi i j}{n}}) \sim \frac{n}{4\pi i} \int_{0}^{\frac{4\pi i \ell}{n}} \ln(1- e^{t})\, dt  \nonumber \\
\Longrightarrow (q)_\ell &\sim \exp[\frac{n}{4\pi i}\left(\frac{\pi^2}{6}-\text{Li}_2(q^\ell)\right)] ~.
\label{qPoch-Asymp}
\end{align}
Once the potential $V_{\mathcal{K}}(x_1,\ldots,x_p)$ is known, we can apply the stationary phase approximation for the integral \eqref{Kash-Asymp} in the large $n$ limit, and obtain a saddle point solution $(x_1^0,\ldots,x_p^0)$ which satisfies:  
\begin{equation} 
\left. \frac{\partial\, V_{\mathcal{K}}(x_1,\ldots,x_p)}{\partial x_j} \right|_{(x_1,\ldots,x_p) = (x_1^0,\ldots,x_p^0)} =0,\quad \forall\,\, j=1,2,\ldots,p   ~.
\end{equation}
As a result, we can write the following limit:
\begin{equation} 
\lim_{n\to \infty}\frac{4\pi}{n}\ln \text{Kas}_{n}(\mathcal{K}\,;e^{\frac{4\pi i}{n}}) = -i\,V_{\mathcal{K}}(x_1^0,\ldots,x_p^0) ~.
\end{equation}
The real part of the right hand side or the imaginary part of the potential is expected to coincide with the hyperbolic volume of $S^3\backslash \mathcal{K}$:
\begin{equation} 
\Re[-i\,V_{\mathcal{K}}(x_1^0,\ldots,x_p^0)] = \Im[V_{\mathcal{K}}(x_1^0,\ldots,x_p^0)] = \text{Vol}(S^3 \backslash \mathcal{K}) ~.
\label{Pot-Volume}
\end{equation}
Thus, finally we can arrive at
\begin{equation} 
\lim_{n \to \infty}\frac{\ln \abs{\text{Kas}_{n}(\mathcal{K}\,;e^{\frac{4\pi i}{n}})}}{n} = \lim_{n \to \infty}\frac{\ln \abs{J_{n-1}(\mathcal{K}\,;e^{\frac{4\pi i}{n}})}}{n} = \frac{\text{Vol}(S^3 \backslash \mathcal{K})}{4 \pi} ~.
\label{LimitPotential}
\end{equation}
In the following, we will verify \eqref{Pot-Volume} and \eqref{LimitPotential} for prime knots up to six crossings. 
\subsection*{\textcolor{gray}{$\bullet$}\,\boldmath{$S^3 \backslash 3_1$}} 
The colored Jones polynomial associated with $3_1$ knot colored by an SU(2) irrep with Dynkin label $\alpha$ is given as:
\begin{equation} 
J_{\alpha}(3_1,\,q) =\sum _{j=-\alpha/2}^{\alpha/2} \left(\frac{q^{-\frac{3 \alpha ^2}{2}-\frac{5 \alpha }{2}+6 j^2-5 j+1}-q^{-\frac{3 \alpha ^2}{2}-\frac{5 \alpha }{2}+6 j^2+j}}{q^{\alpha +1}-1} \right) ~.
\end{equation}
The Kashaev's invariant for the $3_1$ knot is given as,
\begin{equation} 
\text{Kas}_{n}(3_1\,;q) =\sum_{j=0}^{n-1} (q)_j ~.
\end{equation}
We have numerically verified that:
\begin{equation}
\abs{J_{n-1}(3_1,\,e^{\frac{4\pi i}{n}})} = \abs{\text{Kas}_{n}(3_1\,;e^{\frac{4\pi i}{n}})} \quad,\quad n=1,3,5,7,\ldots ~.
\end{equation}
Using \eqref{qPoch-Asymp} and setting $x=q^j$ with $q=e^{\frac{4\pi i}{n}}$, we will have the following expression in large $n$ limit:
\begin{equation} 
\text{Kas}_{n}(3_1\,;e^{\frac{4\pi i}{n}}) \sim \int dx\,\exp(\frac{n}{4\pi i}\,V_{3_1}(x)) ~,
\end{equation}
where the potential is given as
\begin{equation}
V_{3_1}(x) = \frac{\pi^2}{6}-\text{Li}_2(x) ~.
\end{equation}
The saddle point is $x^0 = 0$ and corresponding potential is given as:
\begin{equation}
V_{3_1}(x^0) = \frac{\pi^2}{6} \quad \Longrightarrow \quad \Im[V_{3_1}(x^0)] = 0 ~.
\end{equation}
This gives the vanishing hyperbolic volume of $S^3 \backslash 3_1$ which is expected since $3_1$ is a torus knot. This verifies \eqref{LimitPotential}.
\subsection*{\textcolor{gray}{$\bullet$}\,\boldmath{$S^3 \backslash 4_1$}}
The Jones polynomial associated with $4_1$ knot colored by an SU(2) irrep with Dynkin label $\alpha$ is given as:
\begin{equation} 
J_{\alpha}(4_1,\,q) = \sum_{a=0}^{\alpha} \prod_{j=1}^a \left(q^{\frac{1}{2} (\alpha-j+1)}-q^{-\frac{1}{2} (\alpha-j+1)}\right) \left(q^{\frac{1}{2} (\alpha+j+1)}-q^{-\frac{1}{2} (\alpha+j+1)}\right)
\end{equation}
The Kashaev's invariant for the $4_1$ knot is given as \cite{kashaev1997hyperbolic},
\begin{equation} 
\text{Kas}_{n}(4_1\,;q) =\sum_{j=0}^{n-1} \abs{(q)_{j}}^2 ~.
\end{equation}
We have numerically verified that:
\begin{equation}
\abs{J_{n-1}(4_1,\,e^{\frac{4\pi i}{n}})} = \abs{\text{Kas}_{n}(4_1\,;e^{\frac{4\pi i}{n}})} \quad,\quad n=1,3,5,7,\ldots ~.
\end{equation}
Defining $x = q^{j}$ with $q=e^{\frac{4\pi i}{n}}$, and using \eqref{qPoch-Asymp}, we will have the following expression in the large $n$ limit:
\begin{equation} 
\text{Kas}_{n}(4_1\,;e^{\frac{4\pi i}{n}}) \sim \int dx\,\exp(\frac{n}{4\pi i}\,V_{4_1}(x)) ~,
\end{equation}
where the potential is given as \cite{kashaev1997hyperbolic}:
\begin{equation}
V_{4_1}(x) = \text{Li}_2(x^{-1})-\text{Li}_2(x) ~.
\end{equation}
The saddle point is $x^0=e^{-\frac{i \pi }{3}}$ at which we have:
\begin{equation}
\Im[V_{4_1}(x^0)] \approx 2.0298832128193072500424... ~.
\end{equation}
This coincides with the hyperbolic volume of $S^3 \backslash 4_1$ and verifies \eqref{LimitPotential}. 
\subsection*{\textcolor{gray}{$\bullet$}\,\boldmath{$S^3 \backslash 5_1$}}
The colored Jones polynomial associated with $5_1$ knot colored by an SU(2) irrep with Dynkin label $\alpha$ is given as:
\begin{equation} 
J_{\alpha}(5_1,\,q) =\sum _{j=-\alpha/2}^{\alpha/2} \left(\frac{q^{-\frac{5 \alpha ^2}{2}-\frac{9 \alpha }{2}+10 j^2-7 j+1}-q^{-\frac{5 \alpha ^2}{2}-\frac{9 \alpha }{2}+10 j^2+3 j}}{q^{\alpha +1}-1} \right) ~.
\end{equation}
The Kashaev's invariant for the $5_1$ knot is given as,
\begin{equation} 
\text{Kas}_{n}(5_1\,;q) =\sum_{\substack{j_1,\,j_2=0\\ 0\,\leq\, j_1+j_2\, \leq\, n-1 }}^{n-1} q^{-j_1 j_2}\, (q)_{j_1+j_2} ~.
\end{equation}
We have numerically verified that:
\begin{equation}
\abs{J_{n-1}(5_1,\,e^{\frac{4\pi i}{n}})} = \abs{\text{Kas}_{n}(5_1\,;e^{\frac{4\pi i}{n}})} \quad,\quad n=1,3,5,7,\ldots ~.
\end{equation}
Defining $x_1 = q^{j_1}$ and $x_2 = q^{j_2}$ with $q=e^{\frac{4\pi i}{n}}$, using \eqref{qPoch-Asymp} and noting that 
\begin{equation}
q^{-j_1 j_2} = e^{-\frac{4\pi i j_1 j_2}{n}} \sim e^{-\frac{n}{4\pi i}(\ln x_1) (\ln x_2)} ~,
\label{omegapower}
\end{equation}
we will have the following expression in the large $n$ limit:
\begin{equation} 
\text{Kas}_{n}(5_1\,;e^{\frac{4\pi i}{n}}) \sim \iint dx_1 dx_2\,\exp(\frac{n}{4\pi i}\,V_{5_1}(x_1,x_2)) ~,
\end{equation}
where the potential is given as
\begin{equation}
V_{5_1}(x_1,x_2) = \frac{\pi^2}{6}-\text{Li}_2(x_1 x_2) -(\ln x_1) (\ln x_2) ~.
\end{equation}
The saddle points are given as:
\begin{equation}
(x_1^0,x_2^0) = (-\tfrac{\sqrt{5}+1}{2},-\tfrac{\sqrt{5}+1}{2})\,,\, (\tfrac{\sqrt{5}-1}{2},\tfrac{\sqrt{5}-1}{2}) ~.
\end{equation}
At both of these saddle points, we find that $\Im[V_{5_1}(x_1^0,x_2^0)]=0$. This gives the vanishing hyperbolic volume of $S^3 \backslash 5_1$ which is expected since $5_1$ is a torus knot. This verifies \eqref{LimitPotential}.
\subsection*{\textcolor{gray}{$\bullet$}\,\boldmath{$S^3 \backslash 5_2$}}
The Jones polynomial associated with $5_2$ knot colored by an SU(2) irrep with Dynkin label $\alpha$ is given as:
\begin{equation} 
J_{\alpha}(5_2,\,q) =\sum_{a_1=0}^{\alpha } \sum_{a_2=0}^{a_1} \frac{(-1)^{a_2}\,q^{a_1+\frac{1}{2} a_2(5 a_2+3)}(1-q^{2 a_2+1}) \binom{a_1}{a_2}_q (q^{-\alpha};q)_{a_1} (q^{\alpha+2};q)_{a_1}}{(q^{a_2+1};q)_{a_1+1}} ~,
\end{equation}
where the following definitions are used:
\begin{equation} 
(z;q)_\ell = \prod_{j=0}^{\ell-1} (1-z q^j) \quad,\quad \binom{x}{y}_q = \frac{(q;q)_x}{(q;q)_y \, (q;q)_{x-y}} ~.
\end{equation}
The Kashaev's invariant for the $5_2$ knot is given as \cite{kashaev1997hyperbolic},
\begin{equation} 
\text{Kas}_{n}(5_2\,;q) =\sum_{j_1=0}^{n-1}\sum_{j_2=0}^{j_1}\, q^{-(j_1+1)j_2}\, \frac{(q)_{j_1}\,(q)_{j_1}}{(q)_{j_2}^{*}} ~.
\end{equation}
We have numerically checked that
\begin{equation}
\abs{J_{n-1}(5_2,\,e^{\frac{4\pi i}{n}})} = \abs{\text{Kas}_{n}(5_2\,;e^{\frac{4\pi i}{n}})} \quad,\quad n=1,3,5,7,\ldots ~.
\end{equation}
Setting $x_1 = q^{j_1}$ and $x_2 = q^{j_2}$ with $q=e^{\frac{4\pi i}{n}}$, using \eqref{qPoch-Asymp} and \eqref{omegapower}, we have the following asymptotic for the Kashaev's invariant in the large $n$ limit:
\begin{equation} 
\text{Kas}_{n}(5_2\,;e^{\frac{4\pi i}{n}}) \sim \iint dx_1 dx_2\,\exp(\frac{n}{4\pi i}\,V_{5_2}(x_1,x_2)) ~,
\end{equation}
where the potential is given as \cite{kashaev1997hyperbolic}:
\begin{equation}
V_{5_2}(x_1,x_2) = \frac{\pi ^2}{2}-2\text{Li}_2(x_1)-\text{Li}_2({x_2}^{-1})-(\ln x_1)(\ln x_2) ~.
\end{equation}
The saddle point is approximately given as: 
\begin{equation}
\left(
\begin{array}{c}
 x_1^0 \\
 x_2^0 \\
\end{array}
\right) = \left(
\begin{array}{c}
 0.3376410213776269870195456-0.5622795120623012438991821 i \\
 0.1225611668766536199752455+0.7448617666197442365931704 i \\
\end{array}
\right)
\end{equation}
at which the imaginary part of the potential is given as:
\begin{equation}
\Im[V_{5_2}(x_1^0,x_2^0)] \approx 2.82812208833078316276390... ~.
\end{equation}
This coincides with the hyperbolic volume of $S^3 \backslash 5_2$ and verifies \eqref{LimitPotential}.
\subsection*{\textcolor{gray}{$\bullet$}\,\boldmath{$S^3 \backslash 6_1$}}
The Jones polynomial associated with $6_1$ knot colored by an SU(2) irrep with Dynkin label $\alpha$ is given as:
\begin{equation}
J_{\alpha}(6_1,\,q) = \sum_{a_1=0}^{\alpha} \sum_{a_2=0}^{a_1} \left(q^{-\alpha a_1+a_2^2+a_2-a_1} \frac{(q;q)_{a_1} (q;q)_{\alpha} \left(q^{\alpha +2};q\right)_{a_1}}{(q;q)_{a_1-a_2} (q;q)_{a_2} (q;q)_{\alpha -a_1}} \right) ~.
\label{Jones61formula}
\end{equation}
The Kashaev's invariant for the $6_1$ knot is given as \cite{kashaev1997hyperbolic},
\begin{equation} 
\text{Kas}_{n}(6_1\,;q) =\sum_{\substack{j_1,\,j_2,\,j_3=0\\ j_1+j_2\,\leq\, j_3}}^{n-1}\, q^{(j_3-j_1-j_2)(j_3-j_1+1)}\frac{\abs{(q)_{j_3}}^2}{(q)_{j_1}\,(q)_{j_2}^*} \, ~.
\end{equation}
We have numerically checked that
\begin{equation}
\abs{J_{n-1}(6_1,\,e^{\frac{4\pi i}{n}})} = \abs{\text{Kas}_{n}(6_1\,;e^{\frac{4\pi i}{n}})} \quad,\quad n=1,3,5,7,\ldots ~.
\end{equation}
Setting $x_1 = q^{j_1}$, $x_2 = q^{j_2}$, $x_3 = q^{j_3}$ and using \eqref{qPoch-Asymp}, we have the following asymptotic for the Kashaev's invariant in the large $n$ limit:
\begin{equation} 
\text{Kas}_{n}(6_1\,;e^{\frac{4\pi i}{n}}) \sim \iiint dx_1 dx_2 dx_3\,\exp(\frac{n}{4\pi i}\,V_{6_1}(x_1,x_2,x_3)) ~,
\end{equation}
where the potential is given as \cite{kashaev1997hyperbolic}:
\begin{align}
V_{6_1}(x_1,x_2,x_3) = \text{Li}_2(x_1)-\text{Li}_2(x_2^{-1})+\text{Li}_2(x_3^{-1})-\text{Li}_2(x_3)+\ln(\frac{x_1}{x_3}) \ln(\frac{x_1 x_2}{x_3})-2\pi i \ln(\frac{x_1}{x_3}) ~.
\end{align}
The saddle point is approximately given as: 
\begin{equation}
\left(
\begin{array}{c}
 x_1^0 \\
 x_2^0 \\
x_3^0
\end{array}
\right) = \left(
\begin{array}{c}
 0.17384979383679558571428+1.06907189987572611850590 i \\
 0.32204184201246211276846+0.15777973787085281762619 i \\
0.27872641157714581676289-0.48341992018615350391557 i
\end{array}
\right)
\end{equation}
at which the imaginary part of the potential is given as:
\begin{equation}
\Im[V_{6_1}(x_1^0,x_2^0)] \approx 3.163963228883143983991... ~.
\end{equation}
This coincides with the hyperbolic volume of $S^3 \backslash 6_1$ and verifies \eqref{LimitPotential}.
\subsection*{\textcolor{gray}{$\bullet$}\,\boldmath{$S^3 \backslash 6_2$}}
The Jones polynomial associated with $6_2$ knot colored by an SU(2) irrep with Dynkin label $\alpha$ is given as:
\begin{align}
J_{\alpha}(6_2,\,q) &= \sum_{a_1=0}^{\alpha} \sum_{a_2=0}^{a_1} \sum_{a_3=0}^{a_2}\left( (-1)^{a_2+a_3}\, q^{\frac{1}{2}(-2 \alpha  a_1-a_2^2+2 a_1 a_2+a_2+a_3^2+a_3)} \binom{a_1}{a_2}_q\, \binom{a_2}{a_3}_q \, (q;q)_{a_2} \right. \nonumber \\
& \left. (q^{a_3+1};q)_{a_1-a_2}\, \binom{\alpha}{a_1}_q \, (q^{\alpha +2};q)_{a_1} \right) ~.
\end{align}
The Kashaev's invariant for the $6_2$ knot is given as \cite{hikami2003volume},
\begin{equation} 
\text{Kas}_{n}(6_2\,;q) =\sum_{\substack{j_1,\,j_2,\,j_3=0\\j_1\, \leq \, j_2\\ 0\,\leq\,j_1+j_3\,\leq\, n-1}}^{n-1}\, q^{-j_1(j_2+j_3+1)}\,\frac{(q)_{j_2}\,(q)_{j_2}\,(q)_{j_1+j_3}}{\abs{(q)_{j_1}}^2\,(q)_{j_2-j_1} } ~.
\end{equation}
We have numerically checked that
\begin{equation}
\abs{J_{n-1}(6_2,\,e^{\frac{4\pi i}{n}})} = \abs{\text{Kas}_{n}(6_2\,;e^{\frac{4\pi i}{n}})} \quad,\quad n=1,3,5,7,\ldots ~.
\end{equation}
Setting $x_1 = q^{j_1}$, $x_2 = q^{j_2}$, $x_3 = q^{j_3}$ and using \eqref{qPoch-Asymp}, we have the following asymptotic for the Kashaev's invariant in the large $n$ limit:
\begin{equation} 
\text{Kas}_{n}(6_2\,;e^{\frac{4\pi i}{n}}) \sim \iiint dx_1 dx_2 dx_3\,\exp(\frac{n}{4\pi i}\,V_{6_2}(x_1,x_2,x_3)) ~,
\end{equation}
where the potential is given as \cite{hikami2003volume}:
\begin{align}
V_{6_2}(x_1,x_2,x_3) = \text{Li}_2(x_1)-\text{Li}_2(x_1^{-1})+\text{Li}_2(x_1^{-1}x_2)-2 \text{Li}_2(x_2)-\text{Li}_2(x_1 x_3)-(\ln x_1) (\ln x_2 x_3)+\frac{\pi ^2}{3} ~.
\end{align}
The saddle point is approximately given as: 
\begin{equation}
\left(
\begin{array}{c}
 x_1^0 \\
 x_2^0 \\
x_3^0
\end{array}
\right) = \left(
\begin{array}{c}
 0.0904326688828523150727+1.6028830694139448442811 i \\
 -0.23270516544550685831671-1.09381177109577462363025 i \\
-0.96491338528654243537970-0.62189628352233029773000 i
\end{array}
\right)
\end{equation}
at which the imaginary part of the potential is given as:
\begin{equation}
\Im[V_{6_2}(x_1^0,x_2^0)] \approx 4.400832516123046101441... ~.
\end{equation}
This coincides with the hyperbolic volume of $S^3 \backslash 6_2$ and verifies \eqref{LimitPotential}.
\subsection*{\textcolor{gray}{$\bullet$}\,\boldmath{$S^3 \backslash 6_3$}}
The Jones polynomial associated with $6_3$ knot colored by an SU(2) irrep with Dynkin label $\alpha$ is given as:
\begin{align}
J_{\alpha}(6_3,\,q) &= \sum_{a_1=0}^{\alpha} \sum_{a_2=0}^{a_1} \sum_{a_3=0}^{a_2}\left( (-1)^{a_1+a_3}\, q^{\frac{1}{2} \left(-a_1 \left(2 \alpha +2 a_2+1\right)+a_1^2+a_3^2-2 a_2+a_3\right)} \binom{a_1}{a_2}_q\, \binom{a_2}{a_3}_q \, (q;q)_{a_2} \right. \nonumber \\
& \left. (q^{a_3+1};q)_{a_1-a_2}\, \binom{\alpha}{a_1}_q \, (q^{\alpha +2};q)_{a_1} \right) ~.
\end{align}
The Kashaev's invariant for the $6_3$ knot is given as \cite{hikami2003volume},
\begin{equation} 
\text{Kas}_{n}(6_3\,;q) =\sum_{\substack{j_1,\,j_2,\,j_3=0\\ 0\,\leq\,j_1+j_2+j_3\,\leq\, n-1}}^{n-1}\, q^{(j_1+1)(j_2-j_3)}\,\frac{\abs{(q)_{j_1+j_2+j_3}}^2\,(q)_{j_1+j_2}^*\,(q)_{j_1+j_3}}{\abs{(q)_{j_2}}^2\,\abs{(q)_{j_3}}^2 } ~.
\end{equation}
We have numerically checked that
\begin{equation}
\abs{J_{n-1}(6_3,\,e^{\frac{4\pi i}{n}})} = \abs{\text{Kas}_{n}(6_3\,;e^{\frac{4\pi i}{n}})} \quad,\quad n=1,3,5,7,\ldots ~.
\end{equation}
Setting $x_1 = q^{j_1}$, $x_2 = q^{j_2}$, $x_3 = q^{j_3}$ and using \eqref{qPoch-Asymp}, we have the following asymptotic for the Kashaev's invariant in the large $n$ limit:
\begin{equation} 
\text{Kas}_{n}(6_3\,;e^{\frac{4\pi i}{n}}) \sim \iiint dx_1 dx_2 dx_3\,\exp(\frac{n}{4\pi i}\,V_{6_3}(x_1,x_2,x_3)) ~,
\end{equation}
where the potential is given as \cite{hikami2003volume}:
\begin{align}
V_{6_3}(x_1,x_2,x_3) &= \text{Li}_2(x_2)-\text{Li}_2(x_2^{-1})+\text{Li}_2(x_3)-\text{Li}_2(x_3^{-1})+\text{Li}_2(x_1^{-1}x_2^{-1})-\text{Li}_2(x_1 x_3) \nonumber \\ &-\text{Li}_2(x_1 x_2 x_3)+\text{Li}_2(x_1^{-1}x_2^{-1}x_3^{-1})+(\ln x_1)\ln(x_2/x_3) ~.
\end{align}
The saddle point is approximately given as: 
\begin{equation}
\left(
\begin{array}{c}
 x_1^0 \\
 x_2^0 \\
x_3^0
\end{array}
\right) = \left(
\begin{array}{c}
 0.20432287067507999865586-0.97890355220475859397659 i \\
 1.6083782859756256503564+0.5587518814119368864517 i \\
0.55478836677297306027131+0.19273391491468578161798 i
\end{array}
\right)
\end{equation}
at which the imaginary part of the potential is given as:
\begin{equation}
\Im[V_{6_3}(x_1^0,x_2^0)] \approx 5.693021091281300765112... ~.
\end{equation}
This coincides with the hyperbolic volume of $S^3 \backslash 6_3$ and verifies \eqref{LimitPotential}.

\bibliographystyle{JHEP}
\bibliography{EEhyper}

\end{document}